\documentclass[fleqn]{2023SCGE}
\usepackage{graphicx}

\usepackage{mwe,tikz,ulem}
\usepackage[percent]{overpic}
\usepackage{graphicx}
\usepackage{xspace}
\usepackage[numbers]{natbib}
\usepackage[utf8x]{inputenc}

\usepackage[gen]{eurosym}
\usepackage{microtype}
\usepackage{color}
\usepackage{txfonts}
\usepackage{makeidx}
\usepackage[columns=1]{idxlayout}

\def\degr{$^\circ$}

\newcommand{\apj}{Astrophys. J.\xspace}
\newcommand{\apjl}{Astrophys. J. Lett.\xspace}
\newcommand{\apjs}{Astrophys. J. Supp. Ser.\xspace}
\newcommand{\mnras}{Mon. Not. Roy. Astron. Soc.\xspace}
\newcommand{\aap}{Astron. Astrophys.\xspace}
\newcommand{\aj}{Astron. J.\xspace}
\newcommand{\aaps}{Astron. Astrophys. Suppl.\xspace}
\newcommand{\nat}{Nature\xspace}
\newcommand{\araa}{Ann. Rev. Astron. Astrophys.\xspace}

\newcommand{\ssr}{Space Sci.\ Rev.\xspace}

\newcommand{\pasj}{PASJ\xspace}

\newcommand{\nphysa}{Nucl. Phys. A.\xspace}
\newcommand{\physrep}{Phys. Rep.\xspace}
\newcommand{\plb}{Phys. Lett. B\xspace}
\newcommand{\jcap}{J. Cosmol. Astropart. Phys.\xspace}
\newcommand{\prc}{Phys. Rev. C\xspace}
\newcommand{\prd}{Phys. Rev. D\xspace}
\newcommand{\prl}{Phys. Rev. Lett.}

\newcommand{\pasa}{Publ. Astron. Soc. Aust.\xspace}
\newcommand{\pasp}{Publ. Astron. Soc. Pac.\xspace}
\newcommand{\apss}{Astroph. Sp. Sci.}
\newcommand{\aapr}{Astron. Astroph. Rev.}

\begin{document}
\ensubject{subject}

\ArticleType{Article}
\SpecialTopic{SPECIAL TOPIC: }
\Year{ }
\Month{ }
\Vol{ }
\No{ }
\DOI{ }
\ArtNo{ }
\ReceiveDate{July 2025 (Version 2) }

\title{Dense Matter in Neutron Stars with eXTP}{extp WG1}

\author[1]{\\Ang Li}{liang@xmu.edu.cn}
\author[2]{Anna L. Watts}{}
\author[3]{Guobao Zhang}{}
\author[4]{Sebastien Guillot}{}
\author[5]{Yanjun Xu}{}
\author[6,5]{Andrea Santangelo}{}
\author[7]{\\Silvia Zane}{}
\author[5]{Hua Feng}{}
\author[5]{Shuang-Nan Zhang}{}
\author[5]{Mingyu Ge}{}
\author[5]{Liqiang Qi}{}
\author[8]{Tuomo Salmi}{}
\author[2]{\\Bas Dorsman}{}
\author[9]{Zhiqiang Miao}{}
\author[1]{Zhonghao Tu}{}
\author[10]{Yuri Cavecchi}{}
\author[11]{Xia Zhou}{}
\author[12]{Xiaoping Zheng}{}
\author[13]{\\Weihua Wang}{}
\author[12]{Quan Cheng}{}
\author[12]{Xuezhi Liu}{}
\author[12]{Yining Wei}{}
\author[14]{Wei Wang}{}
\author[1,15]{Yujing Xu}{}
\author[15]{\\Shanshan Weng}{}
\author[16,17]{Weiwei Zhu}{}
\author[18]{Zhaosheng Li}{}
\author[19]{Lijing Shao}{}
\author[6]{Youli Tuo}{}
\author[20,21]{Akira Dohi}{}
\author[18]{\\Ming Lyu}{}
\author[1]{Peng Liu}{}
\author[11]{Jianping Yuan}{}
\author[1]{Mingyang Wang}{}
\author[16]{Wenda Zhang}{}
\author[5]{Zexi Li}{}
\author[5]{Lian Tao}{}
\author[5]{\\Liang Zhang}{}
\author[22]{Hong Shen}{}
\author[23]{Constan\c{c}a Provid\^{e}ncia}{}
\author[24,25]{Laura Tolos}{}
\author[24,25]{Alessandro Patruno}{}
\author[26]{\\Li Li}{}
\author[27]{Guozhu Liu}{}
\author[28]{Kai Zhou}{}
\author[29]{Lie-Wen Chen}{}
\author[30]{Yizhong Fan}{}
\author[31,32,33]{Toshitaka Kajino}{}
\author[9,34]{\\Dong Lai}{}
\author[35]{Xiangdong Li}{}
\author[36]{Jie Meng}{}
\author[37]{Xiaodong Tang}{}
\author[38]{Zhigang Xiao}{}
\author[5]{Shaolin Xiong}{}
\author[18,39]{\\Renxin Xu}{}
\author[26]{Shan-Gui Zhou}{}
\author[40]{David R. Ballantyne}{} 
\author[41]{G. Fiorella Burgio}{} 
\author[42]{J\'er\^ome Chenevez}{} 
\author[2]{\\Devarshi Choudhury}{}
\author[43]{Anthea F. Fantina}{} 
\author[44,45]{Duncan K. Galloway}{} 
\author[46]{Francesca Gulminelli}{} 
\author[47]{\\Kai Hebeler}{} 
\author[2]{Mariska Hoogkamer}{}
\author[48]{Jorge E. Horvath}{} 
\author[2]{Yves Kini}{}
\author[49]{Aleksi Kurkela}{} 
\author[50,10]{\\Manuel Linares}{} 
\author[51,52]{J\'er\^ome Margueron}{} 
\author[47]{Melissa Mendes}{} 
\author[53]{Micaela Oertel}{} 
\author[54]{\\Alessandro Papitto}{} 
\author[55]{Juri Poutanen}{} 
\author[24,25]{Nanda Rea}{}
\author[47]{Achim Schwenk}{} 
\author[5]{Xin-Ying Song}{}
\author[47]{\\Isak Svensson}{} 
\author[56]{David Tsang}{} 
\author[8,57]{Aleksi Vuorinen}{} 
\author[58]{Nils Andersson}{} 
\author[59]{\\M. Coleman Miller}{} 
\author[60]{Luciano Rezzolla}{} 
\author[61]{Jirina R. Stone}{} 
\author[62]{Anthony W. Thomas}{} 

\address[1]{Department of Astronomy, Xiamen University, Xiamen 361005, China}
\address[2]{Anton Pannekoek Institute for Astronomy, University of Amsterdam, P.O. Box 94249, NL-1090GE Amsterdam, the Netherlands}
\address[3]{Yunnan Observatory, Chinese Academy of Sciences, Kunming 650011, China}
\address[4]{IRAP, CNRS, 9 avenue du Colonel Roche, BP 44346, F-31028 Toulouse Cedex 4, France}
\address[5]{State Key Laboratory of Particle Astrophysics, Institute of High Energy Physics, Chinese Academy of Sciences, Beijing 100049, China}
\address[6]{Institut f\"{u}r Astronomie und Astrophysik, Kepler Center for Astro and Particle Physics, Eberhard Karls Universit\"{a}t Tübingen, Sand 1, 72076 T\"{u}bingen, Germany}
\address[7]{Mullard Space Science Laboratory, University College London, Holmbury St Mary, Dorking, Surrey RH5 6NT, UK}
\address[8]{Department of Physics, University of Helsinki, P.O. Box 64, FI-00014 University of Helsinki, Finland}
\address[9]{Tsung-Dao Lee Institute, Shanghai Jiao Tong University, Shanghai 201210, China}
\address[10]{Departament de Física, EEBE, Universitat Politécnica de Catalunya, Av. Eduard Maristany 16, 08019 Barcelona, Spain}
\address[11]{Xinjiang Astronomical Observatory, Chinese Academy of Sciences, Urumqi 830011, China}
\address[12]{Institute of Astrophysics, Central China Normal University, Wuhan 430079, China}
\address[13]{Department of Physics, Wenzhou University, Wenzhou 325035, China}
\address[14]{Department of Astronomy, School of Physics and Technology, Wuhan University, Wuhan 430072, China}
\address[15]{School of Physics and Technology, Nanjing Normal University, Nanjing, 210023, Jiangsu, China}
\address[16]{National Astronomical Observatories, Chinese Academy of Sciences, A20 Datun Road, Beijing 100101, China}
\address[17]{Institute for Frontiers in Astronomy and Astrophysics, Beijing Normal University, Beijing 102206, China}
\address[18]{Department of Physics, Xiangtan University, Xiangtan, Hunan 411105, China}
\address[19]{Kavli Institute for Astronomy and Astrophysics, Peking University, Beijing 100871, China}
\address[20]{RIKEN Pioneering Research Institute (PRI), Astrophysical Big Bang Laboratory (ABBL), Wako, Saitama, 351-0198 Japan}
\address[21]{RIKEN Center for Interdisciplinary Theortical and Mathematical Sciences (iTHEMS), RIKEN 2-1 Hirosawa, Wako, Saitama 351-0198, Japan}
\address[22]{School of Physics, Nankai University, Tianjin 300071, China}
\address[23]{CFisUC, Department of Physics, University of Coimbra, 3004-516 Coimbra, Portugal}
\address[24]{Institute of Space Sciences (ICE, CSIC), Campus UAB, Carrer de Can Magrans, 08193, Barcelona, Spain}
\address[25]{Institut d'Estudis Espacials de Catalunya (IEEC), 08860 Castelldefels (Barcelona), Spain}
\address[26]{Institute of Theoretical Physics, Chinese Academy of Sciences, Beijing 100190, China}
\address[27]{Department of Modern Physics, University of Science and Technology of China, Hefei, Anhui 230026, China}
\address[28]{School of Science and Engineering, The Chinese University of Hong Kong, Shenzhen (CUHK-Shenzhen), Guangdong, 518172, China}
\address[29]{School of Physics and Astronomy, Shanghai Jiao Tong University, Shanghai 200240, China}
\address[30]{Key Laboratory of Dark Matter and Space Astronomy, Purple Mountain Observatory, Chinese Academy of Sciences, Nanjing 210023, China}
\address[31]{School of Physics and International Research Center for Big-Bang Cosmology and Element Genesis, Beihang University, Beijing 100083, China}
\address[32]{Center for Nuclear Study, The University of Tokyo, RIKEN campus, 2-1 Hirosawa, Wako, Saitama 351-0198, Japan}
\address[33]{National Astronomical Observatory of Japan 2-21-1 Osawa, Mitaka, Tokyo 181-8588, Japan}
\address[34]{Department of Astronomy, Cornell Center for Astrophysics and Planetary Science, Cornell University, Ithaca, NY 14853, USA}
\address[35]{Department of Astronomy, Nanjing University, Nanjing 210046, China}
\address[36]{State Key Laboratory of Nuclear Physics and Technology, School of Physics, Peking University, Beijing 100871, China}
\address[37]{Institute of Modern Physics, Chinese Academy of Sciences, Lanzhou 730000, China}
\address[38]{Department of Physics, Tsinghua University, Beijing 100084, China}
\address[39]{Department of Astronomy, School of Physics, Peking University, Beijing 100871, China}
\address[40]{School of Physics and Center for Relativistic Astrophysics, 837 State St NW, Georgia Institute of Technology, Atlanta, GA 30332, USA} 
\address[41]{INFN Sezione di Catania and Dipartimento di Fisica, Universit\'a di Catania, Via Santa Sofia 64, 95123 Catania, Italy} 
\address[42]{DTU Space, Technical University of Denmark, Elektrovej 327-328, DK-2800 Kongens Lynby, Denmark} 
\address[43]{Grand Acc\'el\'erateur National d'Ions Lourds (GANIL), CEA/DRF - CNRS/IN2P3, Boulevard Henri Becquerel, 14076 Caen, France} 
\address[44]{School of Physics \& Astronomy, Monash University, Melbourne VIC 3800, Australia} 
\address[45]{Institute for Globally Distributed Open Research and Education (IGDORE)} 
\address[46]{Université de Caen Normandie, ENSICAEN, CNRS/IN2P3, LPC Caen UMR6534, F-14000 Caen, France} 
\address[47]{Technische Universit\"at Darmstadt, Department of Physics, 64289 Darmstadt, Germany} 
\address[48]{Astronomy Department, IAG-USP. R. do Matão 1226, 05508-090 São Paulo SP, Brazil} 
\address[49]{Faculty of Science and Technology, University of Stavanger, 4036 Stavanger, Norway} 
\address[50]{Institutt for Fysikk, Norwegian University of Science and Technology, Trondheim, Norway} 
\address[51]{Univ Lyon, Univ Claude Bernard Lyon 1, CNRS/IN2P3, IP2I Lyon, UMR 5822, F-69622, Villeurbanne, France} 
\address[52]{International Research Laboratory on Nuclear Physics and Astrophysics, Michigan State University and CNRS, East Lansing, MI 48824, USA} 
\address[53]{Laboratoire Univers et Th\'eories, CNRS, Observatoire de Paris, Universit\'e PSL, Université Paris Cité, 5 place Jules Janssen, 92195 Meudon, France} 
\address[54]{Osservatorio Astronomico di Roma, Via Frascati 33, 00078 Monte Porzio Catone (RM), Italy} 
\address[55]{Department of Physics and Astronomy, FI-20014 University of Turku, Finland} 
\address[56]{Department of Physics, University of Bath, Claverton Down, Bath BA1 1AL, UK} 
\address[57]{Helsinki Institute of Physics, P.O.~Box 64, FI-00014 University of Helsinki, Finland} 
\address[58]{Mathematical Sciences and STAG Research Centre, University of Southampton, Southampton SO17 1BJ, UK} 
\address[59]{Department of Astronomy and Joint Space-Science Institute, University of Maryland, College Park, MD 20742, USA} 
\address[60]{Institut f\"{u}r Theoretische Physik, Goethe Universit\"at, Max-von-Laue-Str. 1, 60438 Frankfurt am Main, Germany} 
\address[61]{Department of Physics (Astrophysics), University of Oxford, Oxford OX1 3RH, UK}  
\address[62]{CSSM and ARC Centre of Excellence for Dark Matter Particle Physics, Department of Physics, University of Adelaide, Adelaide SA 5005, Australia} 


\AuthorMark{ }


\AuthorCitation{}


\abstract{
In this White Paper, we present the potential of the 
\textit{enhanced X-ray Timing and Polarimetry} (eXTP) mission to constrain the equation of state of \textit{dense matter} in neutron stars, exploring regimes not directly accessible to terrestrial experiments. 
By observing a diverse population of neutron stars — including isolated objects, X-ray bursters, and accreting systems — eXTP's unique combination of timing, spectroscopy, and polarimetry enables high-precision measurements of compactness, spin, surface temperature, polarimetric signals, and timing irregularity.
These multifaceted observations, combined with advances in theoretical modeling, pave the way toward a comprehensive description of the properties and phases of dense matter from the crust to the core of neutron stars.
Under development by an international Consortium led by the Institute of High Energy Physics of the Chinese Academy of Sciences, the eXTP mission is planned to be launched in early 2030.
}

\keywords{dense matter, equation of state, X-rays, neutron stars}

\PACS{26.60.Kp, 95.55.Ka, 97.60.Jd}

\maketitle


\begin{multicols}{2}

\section{Introduction 
}
\label{Sec:Intro}

The \textit{enhanced X-ray Timing and Polarimetry} (eXTP) is a science mission primarily designed to study laws of physics under extreme conditions of density (this paper), gravity~\cite{WP-WG2}, magnetism~\cite{WP-WG3} and the potential time-domain and multi-messenger transients~\cite{WP-WG4} as well as observatory science~\cite{WP-WG5}. eXTP is currently scheduled to launch in early 2030. 

In the new baseline design, the scientific payload of eXTP consists of three main instruments: the Spectroscopic Focusing Array (SFA), the Polarimetry Focusing Array (PFA) and the Wide-band and Wide-field Camera (W2C). 
The SFA, PFA, and W2C provide complementary observational capabilities on board eXTP to constrain the equation of state (EOS) of dense matter in neutron stars (NSs).
We briefly summarize and benchmark their key instrumentation properties—focusing on effective area, background, and polarimetric sensitivity—against other current and planned missions relevant to dense matter research,
and refer to~\cite{WP-main} for a detailed description.
With the large effective area of eXTP, high time resolution, and combined spectral, timing, and polarization capabilities, precise measurements of NS properties hold promise in constraining the NS EOS and the composition of dense matter in its interior~\cite{lattimer2007neutron,Watts:2016uzu,ascenzi2024neutron}. 

The SFA consists of five SFA-T (where T denotes Timing) X-ray focusing telescopes covering the energy range $0.5$--$10\, \mathrm{keV}$, featuring a total effective area of $2750\,{\rm cm^2}$ at $1.5\, \mathrm{keV}$ and $1670\,{\rm cm^2}$ at $6\,\mathrm{keV}$. The designed angular resolution of the SFA is $\le 1^\prime$ (HPD) with a $18^{\prime}$ field of view (FoV). The SFA-T are equipped with silicon-drift detectors (SDDs), which combine good spectral resolution ($\sim$ 180~eV at 1.5~keV) with very short dead time and a high time resolution of $10\,{\mu\mathrm {s}}$. They are therefore well-suited for studies of X-ray emitting compact objects at the shortest time scales. 
The SFA array also includes one unit of the SFA-I (where I signifies Imaging) telescope equipped with pn-CCD detectors (p-n junction charged coupled device) to enhance imaging capabilities, which would supply strong upper limits to detect weak and extended sources. The expected FoV of SFA-I is $18^\prime \times 18^\prime$. Therefore, the overall sensitivity of SFA could reach around $3.3\times 10^{-15}\,{\rm ergs\,cm^{-2}\,s^{-1}}$ for an exposure time of $1\, \mathrm{Ms}$—ideal for constraining faint persistent or extended emission. 
Since it is not excluded that the SFA might in the end include six SFA-T units, simulations presented here have taken this possibility into consideration. 

The PFA features three identical telescopes, with an angular resolution better than $30^{\prime\prime}$ (HPD) in a $9.8^{\prime} \times 9.8^{\prime}$ FoV, and a total effective area of $250\,{\mathrm{ cm^{2}}}$ at $3\, \mathrm{keV}$ (considering the detector efficiency). Polarization measurements are carried out by gas pixel detectors (GPDs) working at 2--8\,keV with an expected energy resolution of 20\% at 6\,keV and a time resolution better than $10\,{\mathrm {\mu{s}}}$ \cite{2001Natur.411..662C, 2003NIMPA.510..176B,2007NIMPA.579..853B,2013NIMPA.720..173B,eXTP2019}. The instrument reaches an expected minimum detectable polarization (MDP) at $99\%$ confidence level ($\mathrm{MDP}_{99}$) of about $2\%$ in $1\,\mathrm {Ms}$ for a milliCrab-like source.

The W2C is a secondary instrument of the science payload, featuring a coded mask camera with a FoV of approximately {1500} {square degrees} (Full-Width Zero Response, FWZR). The instrument achieves a sensitivity of $4\times 10^{-7}\,{\rm ergs\, cm^{-2}\,s^{-1}}$ (1\,s exposure) across the 10–600\,keV energy range, with an angular resolution of $20^{\prime}$ and an energy resolution better than $30\%$ at 60\,keV.

The combined capabilities of eXTP's three major payloads—
the SFA for high-throughput spectroscopy, the PFA for sensitive X-ray polarimetry, and the W2C for transient monitoring
—uniquely positions the mission to deliver next-generation constraints on the EOS of NSs across multiple observational messengers and timescales (see Table I in~\cite{WP-main} for a benchmark comparison of eXTP's instrument capabilities against current and planned X-ray missions):
\begin{itemize}
  \item SFA combines high effective area with focusing optics, good spectral resolution ($\sim$180 eV), and microsecond timing—surpassing the Rossi X-ray Timing Explorer (RXTE) and the Neutron star Interior Composition Explorer (NICER) in soft X-ray sensitivity.
  \item PFA opens a new observational window for dense matter studies by enabling phase-resolved polarimetry of surface hotspots, aiding mass-radius inference via geometric constraints.
  \item W2C offers continuous hard X-ray monitoring ($10$--$600\,\mathrm{keV}$) over a very wide FoV, enabling detection of rare events like superbursts, long/intermediate X-ray bursts, and glitch-related transients with high cadence and sensitivity.
\end{itemize}

Through the timely efforts enabled by eXTP, we will be able to advance the understanding of several key questions central to the long-standing challenge of determining the EOS of dense matter:
\begin{itemize}
\item How does pressure depend on density, composition and temperature, that is, the EOS, and how does this influence the properties of NSs?

\item What is the observational evidence of the presence of heavy baryons or deconfined quarks at the extreme densities in NS cores, and what are the implications?
\item What is the role of superfluidity in regulating thermal evolution and glitch behavior?
\item How does nuclear pasta vary throughout the deep layers of NS crusts?
\item What is the maximum possible rotation speed of a NS?
\end{itemize}
\vspace{0.5cm}

After a general introduction to dense matter, pulsars, and the NS structure, the following sections outline the various techniques that eXTP will use to measure the dense matter EOS, and explore its expected performance in more detail. Our study emphasizes the full thermodynamics of NS matter to understand its thermal and dynamic evolution. We tactically adopt multi-wavelength observations to study dense matter and approach the key questions in a multi-scale, multi-messenger manner. 

\subsection{State of the art of dense matter
} 
\label{Intro:dense}

NSs are stellar objects with the highest baryonic densities in their interior~\cite{Glendenning,lattimer2004physics}. They are therefore expected to allow us to learn about the high-density, low- and medium-temperature regime of the Quantum Chromodynamics (QCD) phase diagram~\cite{blaschke2010constraints}. Understanding the properties and phases of QCD matter is one of the most fundamental and exciting challenges in nature and is at the heart of many efforts from both theory and experiments~\cite{Fukushima:2013rx}. 
The EOS of QCD matter describes the behavior of strongly interacting matter under extreme conditions, found in different scenarios ranging from cosmology and astrophysics~\cite{Rezzolla2018} to heavy-ion collisions (HIC)~\cite{luo2022properties}. 
Moreover, reliable experimental and observational data interpretation is based on state-of-the-art theoretical models for extreme QCD matter.

The regime relevant for NS interiors lies in a region of the QCD phase diagram where the theory is strongly coupled and non-perturbative. 
First-principles lattice QCD provides robust results at zero baryon chemical potential~\cite{2015IJMPE..2430007D}, but fails at finite density due to the sign problem. 
Extrapolations to finite baryon chemical potential $\mu_B$ are only reliable at small to moderate $\mu_B/T$~\cite{2021PhRvL.126w2001B}, though lattice simulations of theories where the sign problem can be avoided, such as QCD at $\mu_B=0$ but at finite isospin chemical potential or phase quenched QCD, can provide rigorous constraints to the EOS at large $\mu_B$ \cite{Moore:2023glb,Fujimoto:2023unl}.

At extremely high densities, beyond those realized in NSs, perturbative QCD (pQCD) calculations become applicable and provide EOS constraints through stability and asymptotic consistency requirements~\cite{Komoltsev:2021jzg,Gorda:2022jvk,somasundaram2023perturbative}. 
These calculations have reached nearly complete Next-to-Next-to-Next-to-Leading order (N3LO) precision~\cite{Gorda:2021znl,Gorda:2023mkk}, with systematic uncertainties from missing higher-order terms quantified via Bayesian methods~\cite{Gorda:2023usm}.

At lower densities, up to 1--2 times the saturation density, the properties of matter can be calculated from first-principles calculations based on chiral effective field theory (EFT) of QCD~\cite{Hebeler:2013nza,Tews:2012fj,Lynn:2015jua,Drischler:2017wtt,Carbone:2019pkr,Keller:2022crb}.
In chiral EFT, the strong interactions between nucleons are given by long-range pion exchanges and short-range contact interactions, which can be organized in a systematic expansion in powers of $Q/\Lambda_b$, where $Q$ is a typical momentum in the system and $\Lambda_b \sim 500$-$600$\,MeV is the breakdown scale~\cite{Epelbaum:2008ga,Machleidt:2011zz}. 
Up to N3LO, it enables uncertainty-quantified calculations of neutron-rich matter across multiple many-body frameworks~\cite{Hammer:2012id,Hebeler:2015hla,Hebeler:2020ocj,Hergert:2020bxy,Hebeler:2009iv,Tews:2012fj,Carbone:2013rca,Hagen:2013yba,Coraggio:2014nva,Wellenhofer:2014hya,Wellenhofer:2015qba,Lynn:2015jua,Drischler:2015eba,Ekstrom:2017koy,Lu:2019nbg,Keller:2020qhx,Marino:2024tfp}.

Other model approaches include the use of the Dyson-Schwinger equation~\cite{Chen:2011my} and the Nambu–Jona-Lasinio model (NJL) \cite{Hatsuda:1994pi,Buballa:2003qv, Gholami:2024diy}.
An extended NJL description of hadronic matter has also been successfully used to describe NS matter \cite{Wei:2015aep,Pais:2016dng,Marquez:2024bzj}.

It is known that dense matter within NSs exhibits significant quantum many-body correlation effects in the core regions, such as the strong Landau damping of nucleon quasi-particles and the renormalization of nucleon masses and Fermi velocities~\citep{abrikosov2012methods,fetter2012quantum,Coleman}. 
These effects significantly impact the NS EOS and cannot be captured by mean-field or weak-coupling perturbative approaches~\citep{serot1986advance}, highlighting the need for non-perturbative methods. The Dyson-Schwinger equation offers a promising framework, allowing unified treatment of the EOS and superfluid transition temperature~\citep{2024PhRvC.110c5810Z}.
Most thermal evolution models (see Sect.\ref{Sec:Nscool}) assume a Fermi liquid with a sharp Fermi surface and linear specific heat. In this picture, the direct Urca (dUrca) process is only allowed above a threshold proton fraction~\citep{Lattimer1991,Pethick1992_RMP64-1133,Prakash1992}. However, in the dense NS core, strong interactions and quantum fluctuations may induce non-Fermi liquid behavior~\citep{Sedrakian_PRL_24}, characterized by a blurred Fermi surface and significant many-body corrections to nucleon mass~\citep{Coleman}. This can qualitatively alter neutrino emissivity, heat capacity, and thermal conductivity, and may lower the threshold for dUrca processes.
Cooling in accreting NS crusts also depends sensitively on nuclear structure inputs~\citep{LJWang_2018_PRC_GT,Javier2011PRL}, while strong magnetic fields modify the electron chemical potential and the number of Landau levels, further affecting Urca rates~\citep{Famiano_2022_ApJ}.

Energy-density functional theory (EDFT)–based approaches have also been widely used to compute the EOS of NS cores. These methods use density functionals whose parameters are typically calibrated using a wide range of experimental nuclear data or microscopic calculations of neutron matter. Common implementations include nonrelativistic models based on Skyrme or Gogny interactions, as well as the relativistic mean-field theory (RMFT); see, e.g., \cite{Oertel2017,FiorellaBurgio:2018dga} for reviews. EDFT has achieved remarkable success in the theoretical description of finite nuclei \cite{Chen2014,Liliani2021} involving densities that are still much lower than the nucleon density in the core region of NS.
Another approach to the problem of nuclear binding involves working at the quark level, to self-consistently calculate the change in the internal structure of nucleons and hyperons immersed in the strong scalar mean fields that arise in dense matter~\cite{Guichon:1987jp}, called the quark-meson coupling model (QMC) or the quark mean-field (QMF) model~\cite{Guichon:1995ue,2018PrPNP.100..262G,Stone:2019blq,Guichon2024}. With a set of five parameters characterizing the coupling of mesons to the light quarks, a non-relativistic reduction of this model yields reliable predictions for nuclear binding energies, root-mean-square radii and nuclear deformations across the periodic table~\cite{Thomas:2022dnj,Martinez:2020ctv}. 

As a relatively different approach, holographic QCD describes non-perturbative QCD by mapping it to classical gravity with one higher spacetime dimension, thanks to the establishment of the holographic duality~\cite{Liu:2020rrn}. It has proven to be a useful tool in studying various aspects of QCD matter, including the spectrum, transport properties, QCD phase transitions, thermodynamic and transport properties, and out-of-equilibrium dynamics; for reviews and applications, see~\cite{2022EPJC...82..282J, Tootle2022,2022CoTPh..74i7201C,2023arXiv230703885R,Hoyos:2021njg,CruzRojas:2024etx}. Compared with other effective approaches, it can describe different phases within the same framework and provide a well-understood technique to treat dynamical processes at strong coupling. Nevertheless, a unified holographic QCD theory from a top-down approach has not yet been constructed. The holographic approach is complementary to other methods, see e.g.~\cite{2022PhRvX..12d1012D,2022PhRvD.106l1902C,2022PhRvL.129h1601J}.

\subsection{Observational properties of pulsars and other NSs
}
\label{Intro:pulsar}

Pulsars are fast-rotating, strongly magnetized NSs that emit electromagnetic radiation from the magnetosphere (inside or outside the light cylinder) or from the NS surface. The rotation of the NS then may cause the emission to appear pulsed to a distant observer, depending on the magnetic field inclination and the observer viewing angles. 

\begin{figure}[H]
\centering
\includegraphics[width=0.5\textwidth]{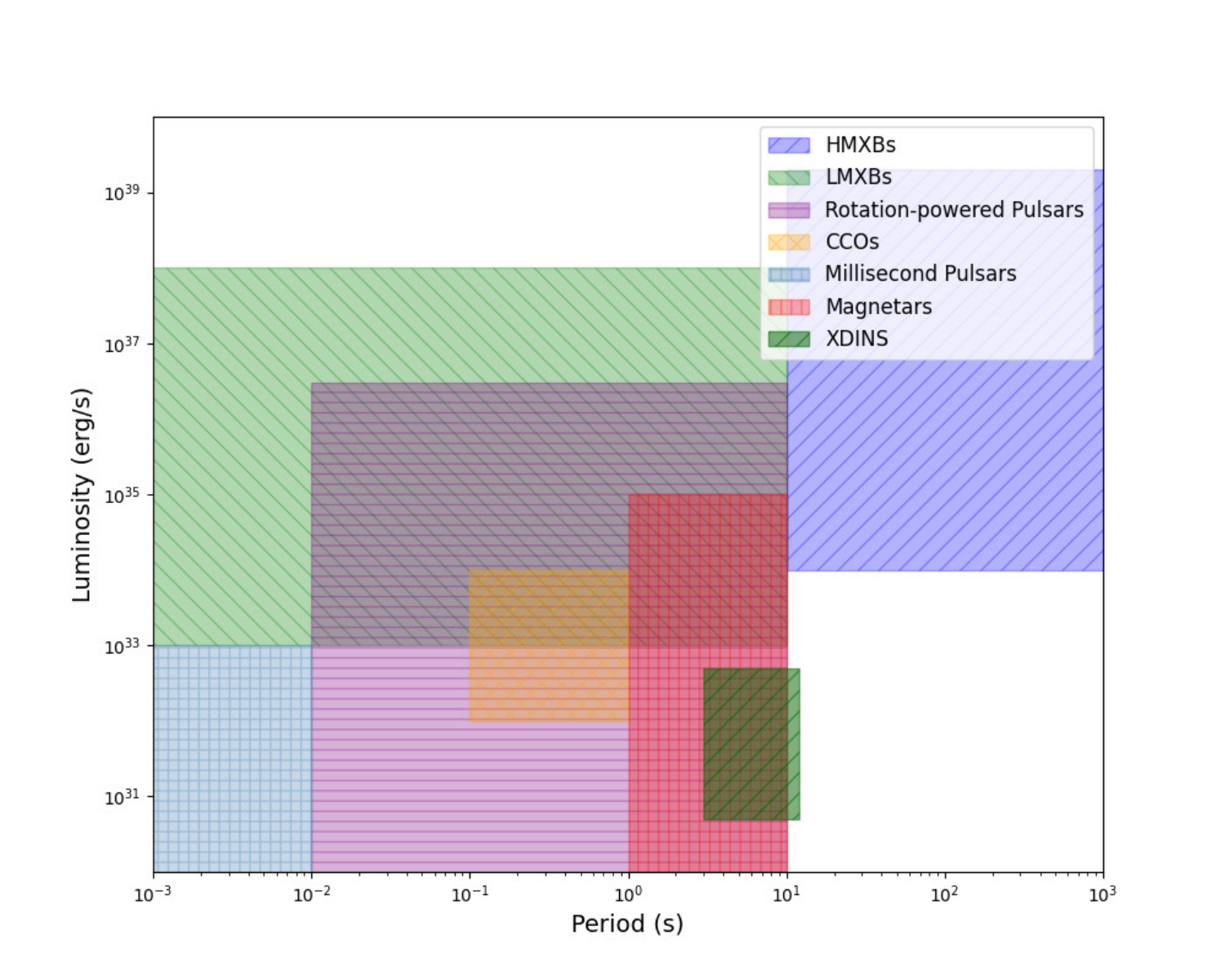}
\caption{Diagram of X-ray luminosity versus spin period. 
eXTP probes a diverse range of X-ray sources, e.g., high-mass X-ray binaries (HMXBs), low-mass X-ray binaries (LMXBs), rotation-powered pulsars, central compact objects (CCOs), X-ray dim isolated NSs (XDINs), millisecond pulsars, and magnetars, with the design and optimization of its payload driven by a set of primary goals, including the study of the dense matter EOS.
}
\label{fig:psr-type}
\end{figure}

The first pulsar was discovered in the radio band in 1967~\cite{hewish1968}; at present more than 4000 pulsars have been discovered with multiple telescopes, from radio to X-ray and gamma rays~\cite{smith2023,han2025}. The known pulsars have rotation periods $P_{\rm spin}$ distributed from about 1.4~ms to 20~s~\cite{hessel2006,manch2005}. Millisecond pulsars (MSPs, with $P_{\rm spin}\sim 1.4$--$30$ ms) are old NSs that have been spun up (after spinning down to periods of order $\sim$seconds) by accretion of material — and angular momentum — from a companion star in a process called recycling~\citep{Alpar82,Radhakrishnan82,Bhattacharya91}. MSPs have comparatively weak magnetic fields~\cite{lorim2008}, typically $10^{8}$--$10^{9}$~G, in contrast to the $\sim 10^{11}$--$10^{13}$~G typical of young NSs~\cite{manch2004}, where the reduction in field strength could be related to the accretion needed to spin-up the star \cite{konar1997}. MSP surfaces are expected to have nearly pure composition, of the lightest accreted elements, because heavier elements sink quickly to the interior~\cite{Bildsten1992}. Pulsars emit radiation across multiple wavelengths; accordingly, they can be broadly classified into two categories: radio pulsars, which emit primarily in the radio band, and high-energy pulsars, which radiate in the X-ray and gamma-ray bands. More than 300 high-energy pulsars have been discovered so far, while relatively few pulsars have been detected in the infrared, optical, and ultraviolet bands. Fig. \ref{fig:psr-type} presents a simple sketch that illustrates the different subcategories of X-ray pulsars visible to eXTP that we will discuss in what follows.

Another, more physically-based pulsar classification consists of dividing them according to the mechanism powering their radiation: rotation-powered pulsars (RPPs), magnetars (magnetic-powered pulsars), and accretion-powered pulsars in binary systems. RPPs have been generally characterized as radio pulsars, but they can also be some of the brightest X-ray and gamma-ray sources in the sky. For example, 231 X-ray counterparts of RPP were recently identified~\cite{Xu2025} by cross-correlating their timing positions with the {\it Chandra} and {\it XMM-Newton} catalogs, including 98 normal pulsars (NPs) and 133 MSPs. Their X-ray luminosity ($L_{\rm X}$) is positively correlated with both the spin-down power ($\dot{E}$) and the light cylinder magnetic field ($B_{\rm lc}$), supporting the outer-gap model of high-energy emission. However, because of low flux levels and/or low time resolution of data from {\it Chandra} (typically $\sim 3.2$~s) or {\it XMM-Newton} (73.4~ms for the full frame mode), X-ray pulsations have been found only in 51 NPs and 18 MSPs. eXTP-SFA has a large effective area and high time resolution, making it highly suitable for searching for X-ray pulsations from more RPPs, and to model the X-ray pulse profiles to determine the masses and radii of NSs (see Sect.~\ref{Sec:Pulse}).

Magnetars are isolated NSs that are largely powered by strong magnetic fields ranging from $10^{13}$--$10^{15}$~G~\cite{Duncan1992}. Magnetars have spin periods of $\sim 0.1 $--$ 12$~s and X-ray luminosities as high as $\sim10^{36}\,\rm erg\ s^{-1}$~\cite{Mereghetti2008, kaspi2017, Esposito2021, Rea25}. Initially, they were identified as a distinct class of pulsars, with rotational energy typically much less than their X-ray luminosity and relatively slow rotation. This led to the identification of magnetic energy as the main power of both their persistent emission and their different types of bursts \citep[see][and references therein]{belob2009,Vigano2013}. Observationally, anomalous X-ray pulsars (AXPs) and soft gamma-ray repeaters (SGRs) are now also classified as a single magnetar class. In general, their X-ray spectra can be characterized by one or two thermal components deriving from hot spots on their surface ($kT\sim 0.1$--$1$ keV) and in some cases with one or two power-law components extending to higher energies (in some cases up to $\sim 200$ keV \cite{Kuiper2006}). The non-thermal component is commonly attributed to resonant cyclotron scattering of the seed surface thermal photons by non-relativistic electrons in the magnetar magnetospheres, or by a relativistic electron component or, in the case of hundred-keV spectra, by a relativistic electron component \cite{Lyutikov06, Nobili08}. Magnetars show a large variety of flaring activity from their long-term outbursts lasting months-years \cite{Cotizelati2018}, to energetic bursts and flares of magnetospheric origin. The rare and hyper-energetic giant flares from SGRs can have X-ray luminosities up to $\sim 10^{44}$--$10^{47} \rm erg\ s^{-1}$ with a duration of $\sim 1$--$100$ s. This distinctive feature may reflect the interaction between the ultra-strong magnetic field and the NS crust or interior.

X-ray Dim Isolated NSs (XDINs) are close-by isolated NSs possibly associated with old magnetars with $>$1\,Myr and fields of $\sim10^{13}$\,G. They can be well described by blackbody-like spectrum or magnetised atmosphere models, corresponding to thermal emission from the surface, possibly with a phase-dependence if the surface temperature distribution is not uniform.Furthermore, some XDINS also exhibit broad and/or phase-dependent spectral (absorption) lines, thought to be cyclotron resonance features originating close to the surface \citep{Borghese2025}.

Central compact objects (CCOs) are compact stars associated with their supernova remnants and are thought to be isolated NSs~\cite{deluca2008}. Some of them present small hot spots detectable as X-ray pulsations (e.g., a region of $R\sim 1$~km and $kT\sim 0.5$~keV). Such hot spots may require a crustal field up to magnetar strength~\cite{Shabaltas2012}. However, the timing properties of CCOs suggest a relatively low magnetic field ($<10^{12}$~G)~\cite{Halpern2010}. Their young age, hot crusts and low dipolar field might be explained in terms of large episodes of fall-back accretion after the supernova, at birth.

Recently, other source classes have shown magnetar flares and outbursts, starting from typical RPPs \cite{Gavriil2008} and the CCO in RCW103, which has a measured spin period of $\sim6.4$\,hr, reachable only via a propeller phase during supernova fall-back accretion. Magnetar emission has then been observed in nearly all classes of isolated NSs, probably powered by non-dipolar field components even in objects with low dipolar fields including some old magnetars \cite{Rea2010,Rea2012} or CCOs \cite{deluca2008,Dai2016, Rea2016}.

Accretion-powered pulsars appear when NSs in X-ray binary systems accrete matter from a companion star and their magnetic field is strong enough to channel the material onto the magnetic poles.
X-ray binaries can be further classified as high-mass X-ray binaries (HMXBs) or low-mass X-ray binaries (LMXBs), according to the donor star masses. 
Most X-ray pulsars are observed in HMXBs. Based on the spectral type of the donor star, NS HMXBs are also classified as supergiant X-ray binaries or Be/X-ray binaries (BeXBs)~\citep[e.g.][]{yang2024}. A majority of the HMXB systems are known to be BeXBs with young optical companions of spectral type O or B and they show outburst activity in X-ray \cite{Caballero2012,stella1986,Okazaki2001}.

In LMXBs, where the companion has a mass of $\lesssim 2M_\odot$, the majority of accreting NSs have weak magnetic fields ($\lesssim 10^8$~G), insufficient to strongly influence accretion, and are non-pulsating, but there are also sources observed as accreting millisecond X-ray pulsars (AMXP).

LMXBs may also show recurrent type-I X-ray bursts, which are attributed to unstable thermonuclear burning on the surface of an accreting NS, with typical durations of $\sim 10$--$100 $~s.

Most importantly, the spectra associated with the bursts can be used to constrain the properties of NSs~\cite{Ozel2016}. Additionally, the burst light curves may occasionally show pulsations called burst oscillations which are associated with inhomogeneous emission from the surface and can also be used to put constraints on the NS EOS \citep{2012ARA&A..50..609W}.

As ``cosmic lighthouses'', pulsars serve as extreme laboratories where all four fundamental forces of nature — strong, weak, electromagnetic and gravitational — coexist.
In the following sections we will see how eXTP observations of the various kinds of NSs can help understanding the physics of the NS interiors and their extreme magnetic field environments.

\subsection{Matter in NS interiors
}
\label{Intro:tools}

According to current theories, a (cold catalysed) NS can be broadly divided into two main regions, lying below a thin atmosphere (see below in Fig.~\ref{fig:structure}):
the crust and the core, with the dividing density being approximately half the nuclear saturation density~\cite{Haensel2007, 2008LRR....11...10C}.
Most approaches to construct the NS EOS have employed ad-hoc matching procedures of different EOSs for the crust and the core, computed using different models.
On the other hand, recent efforts have been devoted to develop a unified treatment of the crust and core (e.g., \cite{2001A&A...380..151D,2015A&A...584A.103S,2013A&A...559A.128F,Gulminelli:2015csa,Pearson:2018tkr,Scurto:2024ekq}; see the \texttt{CompOSE} database \cite{CompOSECoreTeam:2022ddl} for a collection). 
Indeed, it has been shown that a non-uniform treatment of the inner crust-core EOS can lead to uncertainties as large as $\sim 5$\% in the radius estimation of medium and low-mass NSs~\cite{Fortin:2016hny,Pais:2016xiu,Suleiman:2021hre}.

\subsubsection{The crust and nonuniform nuclear matter}
\label{Sec:Crust}

The crust is the outermost layer, extending from the surface to a depth where the density reaches about half the nuclear saturation density, at which point heavy nuclear clusters tend to dissolve into a uniform nucleon liquid.
The crust is divided into the outer crust and the inner crust.
The outer crust is composed of fully ionised atoms immersed in a charge-compensating electron background,
while the inner crust contains a mixture of nuclei, free neutrons, and electrons. 
The outer crust is mainly composed of atomic nuclei such as iron (Fe), nickel (Ni), and other heavy elements~\cite{Baym:1971pw,Haensel:1993zw,Ruester:2005fm,Pearson:2018tkr,Antic2020,Antic2020a}. These elements are arranged in a crystalline lattice, with electrons occupying the interstitial spaces.
As the density increases, the nuclei become more neutron-rich, and free neutrons begin to appear, marking the transition to the inner crust. 
Recently, microscopic calculations have also shown that it is possible to have proton drip in addition to neutron drip in the non-uniform phase toward the crust-core transition~\cite{Keller:2024snt}.

The composition of the inner crust is influenced by the nuclear symmetry energy and its density dependence, since the nuclei become heavier and more neutron-rich with increasing density.
Non-spherical nuclei, collectively known as ``pasta phases'', may appear in the inner crust as the density approaches the phase transition to uniform matter~\cite{Ravenhall:1983uh}.
The appearance of nuclear pasta is mainly caused by the competition between the surface and Coulomb energies of heavy nuclei. As a result, the stable nuclear shape in non-uniform matter can change from droplet to rod, slab, tube, and bubble, among others, with increasing density~\cite{2015PhRvC..91a5807B}. 
Over the past decades, the properties of the inner crust and of pasta phases have been studied using various methods, such as the liquid-drop model~\cite{Ravenhall:1983uh,2001A&A...380..151D,Carreau:2019zdy,Thi:2021hai,Balliet:2020nsh} and the Thomas-Fermi
approximation~\cite{Maruyama:2005vb,Avancini:2008zz,2015PhRvC..91a5807B,Shchechilin:2023erz}. 
Generally, the Wigner-Seitz approximation with typical geometric shapes of nuclear pasta is used to simplify the calculations.
For a more realistic description, there are some studies that have not explicitly assumed any geometric shape and performed fully three-dimensional calculations for nuclear pasta based on the Thomas-Fermi approximation and molecular dynamics 
methods~\cite{Horowitz:2004yf,Watanabe:2004tr,Okamoto:2013tja,Schneider:2013dwa} (see e.g., \cite{Oertel2017,Blaschke:2018mqw,FiorellaBurgio:2018dga} for a review). 
A recent work includes a fully self-consistent three-dimensional Hartree–Fock+BCS calculation, which provides a comprehensive survey of the shape parameter space in the pasta phase of the cold NS~\cite{Newton2022}. 
It has been shown that nuclear symmetry energy and its slope could
significantly affect the pasta phase structure and crust-core transition
of NSs~\cite{Ducoin:2011fy,Newton:2011dw,Newton2013_ApJL779-L4}. A strong magnetic field could also have non-negligible effects, leading to an increase in the extension of the crust and the charge content of the clusters \cite{Wang:2022sxx}.
The pasta phases in the inner crust of NSs play an important role in the determination of the neutrino scattering rates during proto-NS evolution~\cite{Horowitz:2004yf,Sonoda:2007ni,Lin:2020nxy}. Additionally, the elastic properties of nuclear pasta are closely related to the search for gravitational waves from NSs, supernovae, and NS mergers (e.g., \cite{Gearheart:2011qt}). 

Understanding the composition of the NS crust is crucial for the interpretation of observational data~\cite{2008LRR....11...10C}. These models are essential for understanding the behavior of dense matter and for interpreting multi-messenger observations, such as NS cooling~\cite{Yakovlev2004} (addressed in Sect.~\ref{Sec:Crustcool} and Sect.~\ref{Sec:Nscool}), pulsar glitches~\cite{2022RPPh...85l6901A} (addressed in Sect.~\ref{Sec:Timing}), and r and i-mode oscillationsn in NSs~\cite{2001IJMPD..10..381A,2012PhRvL.108a1102T} (see related discussions in the WG4 white paper~\cite{WP-WG4}).

\subsubsection{The core and uniform nuclear matter}
\label{Sec:Core}

The composition and EOS of the NS core can in principle be determined using the fact that NSs are charge-neutral objects, whose equilibrium configuration is governed by weak-interaction processes. However, although it is responsible for most of the mass of the star, the nature of matter in the core is presently unknown~\cite{Glendenning,Haensel2007,Baldo2016,Oertel2017,Lattimer2021,Blaschke:2018mqw}.

The behavior of EOS around the nuclear saturation density can be determined by nuclear experiments\footnote{These constraints have theoretical uncertainties due to model extractions and/or extrapolations to the neutron-rich conditions not attainable by the experiments.}. 
The nuclear saturation density and the binding energy at that density for symmetric nuclear matter (equal number of neutrons and protons where the Coulomb interaction has been switched off) can be inferred from measurements of the distributions of the nuclear charge density distributions~\cite{angeli2013table} and masses of nuclei~\cite{huang2021ame,wang2021ame}. Another important parameter, incompressibility at nuclear saturation density, is determined primarily by isoscalar giant monopole resonances in medium and heavy nuclei~\cite{Blaizot:1980tw,youngblood1999incompressibility,Piekarewicz:2003br,Khan:2012ps,Stone:2014wza}.

The EOS parameters for asymmetric nuclear matter, such as the symmetry energy or slope of the symmetry energy ($L$) around saturation density, can be extracted from experiments involving isospin diffusion measurements and pygmy resonances, isobaric analog states, production of pions and kaons in HIC or data on the neutron skin thickness of heavy nuclei~\citep[see reviews e.g.,][and references therein]{li2014topical,Baldo2016}. 
Note that previous nuclear experiments tested matter below or close to saturation density~\cite[e.g.,][]{danielewicz2002determination,stone2007skyrme}, whereas HIC probes matter with similar numbers of neutrons and protons up to 2--3 times nuclear saturation density~\cite{Stone2017,Stone2024} 
and temperatures well above zero~\cite[e.g.,][]{qin2012laboratory,bougault2020equilibrium}. Thus, in order to describe NS matter, models need to be extended to high baryon density in an extremely isospin asymmetric environment at low temperatures~\cite[see discussion in e.g.,][]{li2014topical,Baldo2016,horowitz2014way,lattimer2016equation}. 
Nevertheless, it is not obvious whether the information on the nuclear EOS from high-energy HICs can be related to the physics of NS interiors. Methods to integrate diverse data from nuclear experiments and multi-messenger astrophysics to advance the study of dense matter are discussed in Sect.~\ref{Sec:MM}.

Using various theories detailed in Sect.~\ref{Intro:dense} to describe dense baryonic matter, it has been shown that the conditions inside NSs could allow for the appearance of exotic (non-nucleonic) degrees of freedom, including hyperons, $\Delta$ baryons, as well as deconfined quark matter. 
At large densities in the centers of massive NSs, exotic quark phases such as color superconducting phases have also been predicted. 
Reviews of EOS physics, including exotic degrees of freedom, can be found in e.g.,~\cite{Oertel2017,Tolos:2020aln,Burgio:2021vgk}.

Many studies have been carried out with the aim of determining the EOS of strongly interacting matter in a model-independent approach constrained by observations and possibly by well-accepted ab initio calculations, both chiral EFT calculations at nuclear densities~\cite{Hebeler:2013nza,Tews:2012fj,Lynn:2015jua,Drischler:2017wtt,Carbone:2019pkr,Keller:2022crb} and pQCD at high densities~\cite{Kurkela:2009gj,Gorda:2021znl,Gorda:2023mkk}. 
In order to cover the whole phase space connecting the low-density to the high-density constraints, several interpolation schemes have been implemented based on agnostic descriptions of the EOS, both parametric, including polytropes~\cite{Read:2008iy,Hebeler:2010jx,Hebeler:2013nza,Kurkela:2014vha,Annala:2017llu,Most2018,Annala:2019puf,OBoyle:2020qvf}, spectral representations~\cite{Lindblom:2010bb,LIGO2018}, speed of sound~\cite{Alford:2013aca,Greif:2018njt, Tews:2018iwm,Annala:2019puf,Somasundaram:2021clp, Altiparmak:2022bke,Ecker:2022, Ecker:2022b,Annala:2021gom,Annala:2023cwx}, metamodeling based on a Taylor expansion around saturation density~\cite{Margueron:2017eqc,Li:2021thg}, and non-parametric descriptions~\cite{Landry:2020vaw,Essick2020,Gorda:2022jvk}. 
In addition, machine learning frameworks have been applied to infer the EOS from NS observations~\cite{Fujimoto_2020,morawski2020neural,Fujimoto_2021,Soma:2022qnv,Soma:2022vbb,Carvalho:2023ele,Krastev:2023fnh,Carvalho:2024kgf,Fujimoto:2024cyv,Carvalho:2024bxv,Somasundaram:2024ykk}. See more discussions in Sect.~\ref{Sec:MM} for studies in the framework of machine learning with physical priors.

Although very general, these approaches have the limitation of being unable to directly determine the internal composition of NSs. In particular, the identification of the presence of a deconfined phase is generally proposed by analyzing the approach to the limit of approximative conformal symmetry, a known property of weakly coupled quark matter that is severely broken in nuclear matter \cite{Annala:2019puf,Annala:2017llu}. Since the sound velocity cannot effectively identify the composition of matter at the intermediate densities relevant to compact stars~\cite{2022PhRvD.106a4014S}, attempts are being made to directly link NS observations and microphysically driven investigations of the EOS across different density regimes, following the description of the RMFT~\cite{Malik:2022jqc,2023ApJ...943..163Z} or Skyrme energy density functional~\cite{2020PhRvC.101c4303Z,Beznogov:2023jqp}, possibly together with a phase transition to deconfined matter within the NJL model \cite{Albino:2024ymc}. 
This approach supports quantitative studies of the EOS in different density regimes, inferring the values of unresolved nuclear saturation and phase transition properties, especially with the increasing accumulation of robust NS observables. 
Problems remain in combining the effective interactions with realistic ones to describe the properties of dense matter~\cite{2024Univ...10..226V}.

As pointed out in~\cite{Alford:2004pf} and several later studies, observations are still unable to identify the composition of NSs, in particular to confirm or exclude the presence of exotic degrees of freedom. Recent Bayesian inference calculations confirm this degeneracy of stellar properties when comparing nucleonic with hyperonic and hybrid NSs~\cite{Malik:2022jqc,Sun:2022yor,Malik:2023mnx,Huang:2024rvj,2022MNRAS.515.5071M}. Studies considering quark degrees of freedom also show that the available NS observations are compatible with both hybrid stars or pure quark stars~\cite{Alvarez-Castillo:2016oln, Montana2018, Albino:2024ymc,2018PhRvD..97h3015Z}, while model-independent calculations indicate a rapid conformalization of the system close to the central densities of maximally massive NSs, pointing towards the possible presence of quark matter therein \cite{Annala:2019puf,Annala:2023cwx}. In the future, observations with much smaller associated uncertainties, particularly in radius and moment of inertia measurements~\cite{2024Univ...10..160H,2024PhRvD.110d3013W}, along with the potential discovery of sub-millisecond spinning NSs and more quantitative comparisons of their thermal and glitching properties, may allow us to distinguish between the different scenarios and resolve the current degeneracy.

NSs, due to their high densities and strong gravity, are promising astrophysical laboratories for probing the properties of dark matter~\cite{Kouvaris2008,Leung2011,Tolos:2015qra,Bell:2020obw}. Although dark matter constitutes about 27\% of the universe’s energy budget—far exceeding the $\sim$5\% from ordinary matter—its fundamental nature remains unknown, and direct detection efforts have so far been inconclusive.
If dark matter accumulates in NSs, it could significantly alter observable properties such as the mass-radius relation~\cite{Narain2006,Leung2011,Li2012,Tolos:2015qra}, tidal deformability~\cite{Ellis2018,Nelson2019,Dengler:2021qcq,Husain:2021hrx,Karkevandi2024,Barbat:2024yvi}, and evolutionary processes~\cite{Goldman1989,Kouvaris2008,PerezGarcia2010,PerezGarcia2012,Husain:2023fwb}. Depending on its distribution, dark matter may form a compact core or an extended halo, leading to contrasting effects: a dark matter halo tends to increase mass and deformability, while a dark matter core can reduce the maximum mass and shrink the radius~\cite{Miao22}. These signatures offer a valuable avenue for constraining dark matter through NS observations.

In the context of X-ray observations, 
the strong gravitational field of a NS bends the X-rays emitted from its surface. 
Tracking this deflected light allows for the reconstruction of the external spacetime of the NS, thereby constraining its mass and radius (see discussion in Sect.~\ref{Sec:Pulse}). 
This technique has been employed by NICER \cite{Gendreau16} and is expected to play a crucial role in future missions such as eXTP. 
However, if the NS is surrounded by a dark matter halo, an additional gravitational potential will act on the light emitted from its baryonic surface, leading to modifications in the pulse profile. 
Studies have shown that these modifications can be significant, reaching up to 10\% for a dense dark matter halo~\cite{Miao22,Shawqi24,Liu2024prd}. 
In such cases, measurements based on the assumption of purely baryonic matter may no longer be accurate. To obtain reliable results, both baryonic matter and dark matter must be considered simultaneously in the modeling~\cite{Rutherford23,Rutherford24}. 
Such a strategy may not only allow for a more precise determination of the NS baryonic radius but could also indirectly constrain the mass, interaction strength, and distribution of dark matter particles, providing valuable insights into the fundamental nature of dark matter.

Thus, NSs are invaluable astrophysical laboratories, providing insight into dense nuclear matter, exotic phases, and dark matter.
The capabilities of eXTP will enable us to effectively constrain the EOS of matter under extreme densities and low temperatures, conditions that cannot be replicated in laboratory experiments. Upcoming high-precision observations~\cite{2019SCPMA..6229503W,Cruise2025,yuan2025science} together with advanced theoretical modeling will be crucial to solving many fundamental open questions and deepening our understanding of the most extreme states of matter in the Universe.

\section{Pulse profile modeling}
\label{Sec:Pulse}

Pulse profile modeling (PPM) exploits the variation in brightness and spectrum as a function of the rotational phase to extract properties of either the emission or the NS itself. In the context of constraints on dense matter, PPM is used to model the emission of MSPs (rotation- or accretion- powered) \citep{Poutanen03}: more specifically the emission originating close to the NS surface, i.e., the emission affected the most by the strong NS surface gravity. As this is related to the NS compactness, PPM facilitates the measurement of the NS mass and radius, among other properties. This requires not only high quality pulse profile data (e.g., with order 10--100~$\mu$sec timing resolution and $\sim$100~eV spectral resolution in the soft X-ray band) but also the ability to simulate pulse profiles using an adequate physical model (to form the basis for likelihood evaluations). Model ingredients include the physics of relativistic ray-tracing in the space-time of a rapidly rotating NS for a given mass and radius, a prescription for the possible properties of the X-ray emitting hotspots, an atmospheric beaming function, the effects of interstellar absorption and the instrument response (taking into account calibration uncertainty). Other relevant parameters will include the distance, pulsar inclination and any non-source background. Priors must be chosen for all of the model parameters. For a detailed overview of the simulation and inference process involved in PPM, see \cite{Bogdanov19b,Bogdanov21} and references therein.

In the subsections that follow, we discuss PPM for rotation-powered MSPs and accretion-powered MSPs, two major target source classes for eXTP that are expected to deliver good constraints on the EOS. While not discussed in depth in this white paper, PPM for magnetars - where polarimetry can help to break degeneracies associated with the strong magnetic field (see \cite{WP-WG3}) - may also provide EOS constraints. 

\begin{figure}[H]
\begin{center}
\includegraphics[width=0.99\linewidth]{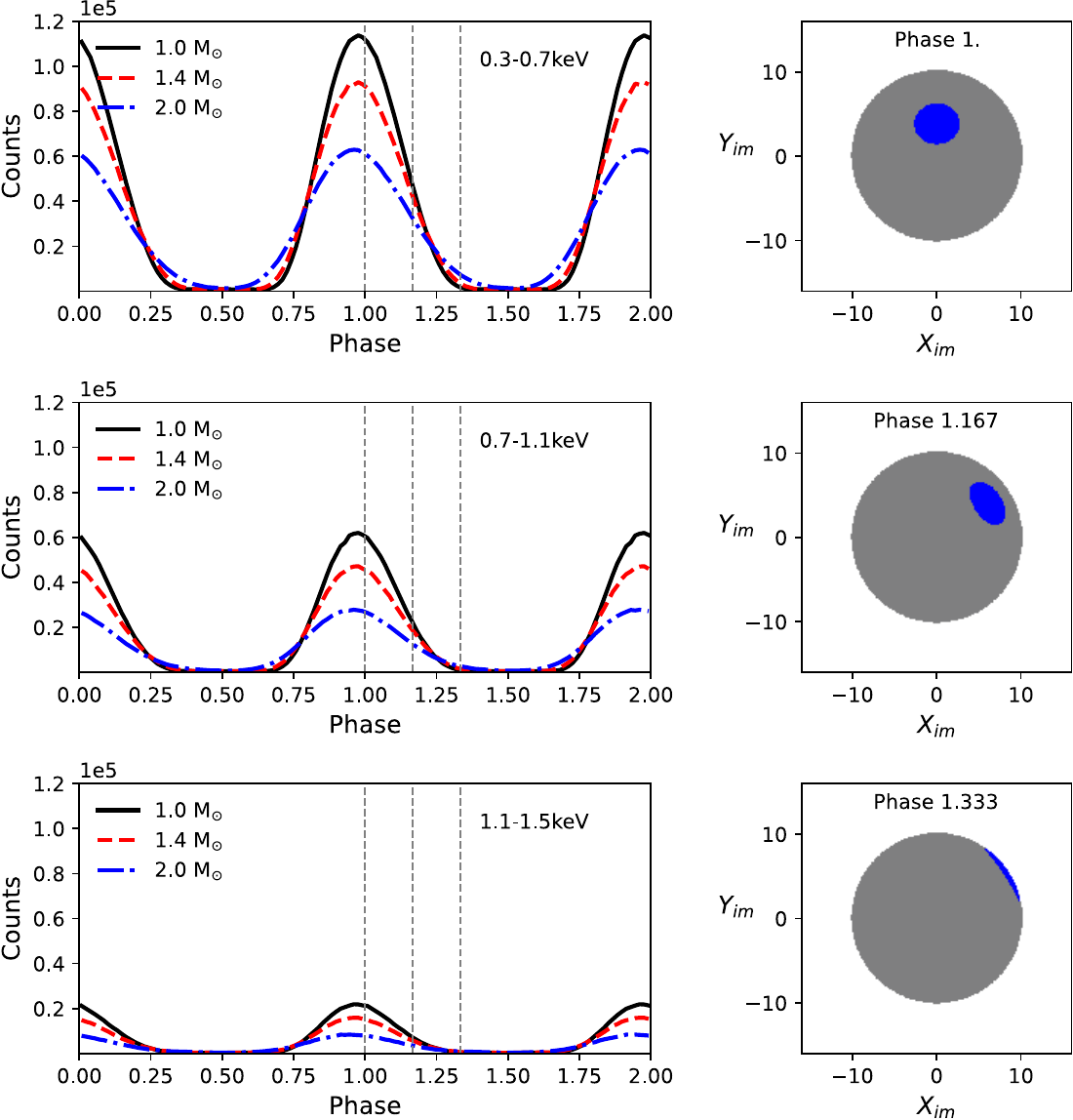}
\end{center}
\caption{Simulated pulse profiles observed by eXTP-SFA for NSs with varying gravitational masses, demonstrating the sensitivity of pulse profiles to stellar compactness. The effective exposure of the simulation is 1~Ms. The other parameters of the simulation \citep{Qi25} include an equatorial radius of 12~km, a spin frequency of 200~Hz, a hydrogen column density of 1~$\times$~$10^{20}~$cm$^{-2}$, a distance of 500~pc, and an inclination of 90\degr. A single-temperature circular hotspot (angular radius = 20\degr, temperature = 0.1 keV) is placed at 60\degr\ colatitude. Two rotational cycles are plotted for clarity. Right panels show the corresponding gravitational lensed geometry of the hotspot at rotational phases marked by vertical dashed lines in the pulse profiles. 
}
\label{fig:msp_illustration}
\end{figure}

\subsection{Rotation-powered MSPs}
While most MSP pulsations are detected in the radio and gamma-ray bands, a handful are also seen in the X-ray band with a characteristic {\it pulse profile}: either narrow and single/double peaked for magnetospheric emission \citep[e.g.,][]{Rowan2020}, or broad and sometimes quasi-sinusoidal for thermal emission originating from the surface \citep[e.g.,][]{Bogdanov19a}. In the latter, as used for PPM, the thermal photons propagate through the gravitational potential well of the star on its way to the observer - picking up the imprints of various relativistic effects that can be used to infer pulsar mass and radius (see Fig.~ \ref{fig:msp_illustration}). The X-ray thermal emission of MSPs generally comes from small regions at the magnetic poles that are heated by return currents from the magnetosphere \cite{Ruderman75,Arons81,Harding01}. 

\begin{figure}[H]
\begin{center}
\includegraphics[width=0.99\linewidth]{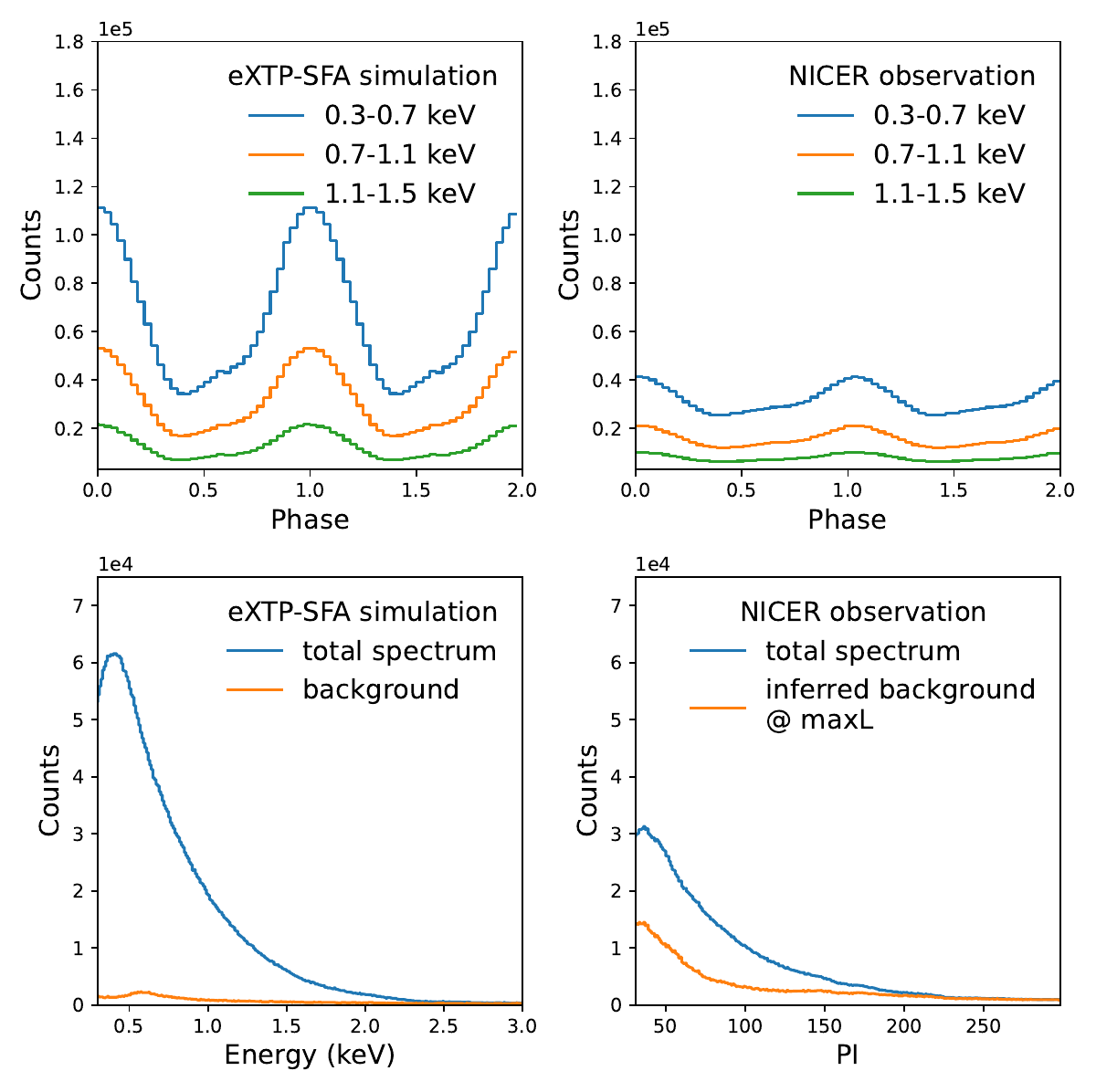}
\end{center}
\caption{Simulated energy-resolved pulse profiles of PSR~J0437$–$4715 observed by eXTP-SFA (32 phase bins), compared with NICER observations. The simulation adopts an effective exposure of 1.328 Ms, matching the NICER observation time. Two rotational cycles are plotted for clarity. The simulation uses the best-fit \textbf{CST-PDT} model parameters derived from NICER observations \citep[as determined by][]{Choudhury24}. In the lower panels, the comparison between the total energy spectrum and the the background estimation (or the inferred background at maximum likelihood) is plotted for eXTP-SFA and NICER, respectively. Additionally, ray-tracing simulations confirm that AGN contamination within the field of view has a negligible effect on PSR~J0437$–$4715 for eXTP-SFA. 
}
\label{fig:J0437_300ks_simulation}
\end{figure}

\subsubsection{State of the art - MSP PPM with NICER}

NICER is a soft X-ray telescope operating on the International Space Station since 2017. Its primary science goal is to explore the EOS of NSs by collecting high quality pulse profile data sets of MSPs to be used for PPM. 
The NICER MSPs are also radio pulsars: radio timing provides the spin ephemeris necessary to generate the X-ray pulse profile, and other informative priors such as distance or, if the source is in a binary, the mass and inclination \cite{Fonseca21,Reardon24}. Constraints on background can come either from NICER background models (e.g., \cite{Remillard22}) or indirectly via joint fitting of (phase-averaged) data for the same source from other X-ray observatories with imaging capabilities (e.g., XMM-Newton, with a well constrained background, enabling a good estimate of the number of photons coming from the source in the XMM-Newton data set).

To date, PPM using NICER data has yielded results for five MSPs: PSR~J0030+0451, for which the mass and inclination are not known a priori \cite{Riley19,Miller19,Bogdanov19a,Salmi23,Afle23,Vinciguerra23,Vinciguerra24}; the $\sim$2.1$M_\odot$ pulsar PSR~J0740+6620 \cite{Riley21,Miller21,Wolff21,Salmi22,Salmi23,Salmi24a,Dittmann24,Hoogkamer25}; the $\sim$1.4$M_\odot$ pulsar PSR~J0437$-$4715 \cite{Choudhury24}; PSR~J1231$-$1411, which has only a weakly-informative prior on mass and inclination \cite{Salmi24b,Qi25}; and the $\sim 1.4 M_\odot$ pulsar PSR J0614-3329~\cite{Mauviard25}. In support of this effort, extensive work has gone into developing and testing PPM pipelines \cite{Bogdanov19b,Bogdanov21,Riley23,Choudhury24rt}. The tightest constraints on radius so far (for PSR~J0437$-$4715, \cite{Choudhury24}) are at the $\sim\pm7$\% level (68\% credible interval). The results are being used to place constraints on both the dense matter EoS \cite[see, e.g.,][ for recent analyses that include PSR~J0437$-$4715]{Rutherford24a,Huang:2024rvj,Kurkela24,LiJJ25} and the magnetic field geometry of the MSPs~\cite{WP-WG3}. 

\begin{figure}[H]
\begin{center}
\includegraphics[width=1\linewidth]{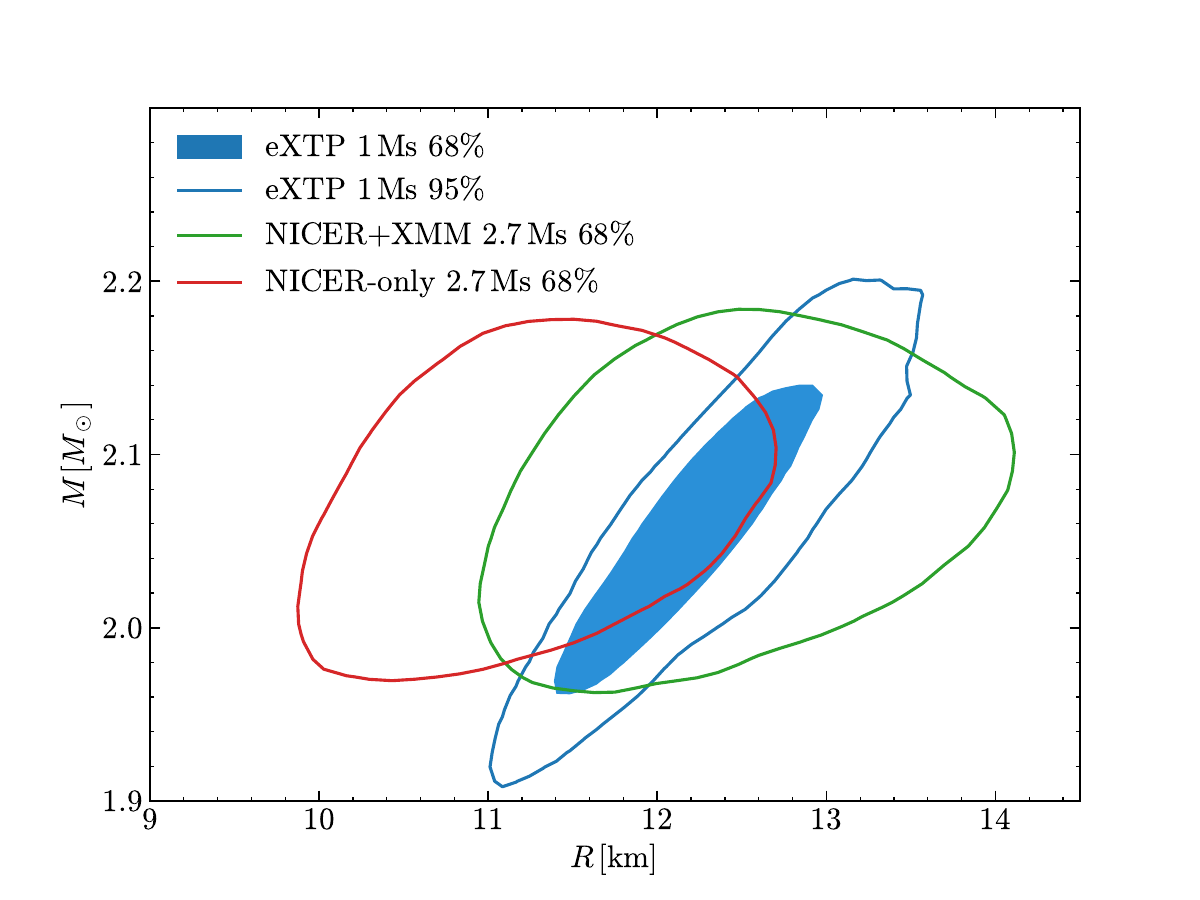}
\end{center}
\caption{2-D marginalized posteriors of mass and radius of PSR~J0740+6620, reconstructed from the simulated pulse profiles observed by eXTP with an effective exposure time of 1\,Ms. The \textbf{ST-U} model is employed on the synthetic data with an upper ($\mathcal{B}+3\sqrt{\mathcal{B}}$) and a lower background limit (max(0, $\mathcal{B}-3\sqrt{\mathcal{B}}$)) in the background-marginalized likelihood function. For comparison, the inferred results of \cite{Salmi24a}, both with and without XMM-Newton data (which illustrates the effect of including background constraints), are also shown. The current background level estimation of eXTP-SFA is nearly 3 times lower than that of NICER for PSR~J0740+6620. Note that by 2030, the radius precision is expected to further increase due to anticipated improvement in the mass prior from radio observations (see discussion in Sect.~\ref{mass}).}
\label{fig:J0740_MR_simulation}
\end{figure}

\subsubsection{MSP PPM with eXTP}

\textit{NICER} has a peak effective area at 1 keV of 1800 cm$^2$. The \textit{eXTP} SFA will be a factor of $\sim$2.5 larger below 2~keV (the energy range of interest for MSP PPM), with much lower background, enabling improved constraints on mass and radius for the NICER MSPs \cite{Lo13,Psaltis14,Bootsma25} and the collection of data sets large enough to enable PPM for MSPs that are too faint to be studied with NICER \cite{Guillot2019}. Figs. \ref{fig:J0437_300ks_simulation} and \ref{fig:J0740_MR_simulation} show simulated pulse profiles and the level of constraints achievable within $\sim$1 Ms exposure times with eXTP for PSR~J0437$–$4715 and PSR~J0740+6620, compared to what is possible with current NICER data. Of greatest interest are nearby MSP binaries with precise measurements of the NS mass from radio pulse timing (see discussion in Sect.~\ref{mass}), since this reduces one source of uncertainty. In addition to the pulsars mentioned above already being studied by NICER, these include PSR~J1909$-$3744 \citep[$M=1.57^{+0.020}_{-0.019}$ $M_{\odot}$;][]{Agazie23}, PSR~J0751$+$1807~\citep[$M=1.64\pm0.15$ $M_{\odot}$; ][]{Desvignes16}, PSR~J2222$-$0137~\citep[$M=1.831\pm0.010$ $M_{\odot}$; ][]{Guo21}, and PSR~J1614$-$2230~\citep[$M=1.937\pm0.014$ $M_{\odot}$; ][]{Agazie23}.

With eXTP, PPM modeling will be possible for a sample of at least 10 sources dominated by thermal radiation, under a reasonable observing strategy and within the anticipated mission lifetime. These sources, spanning a mass range of 1.2–2.1 M$_{\odot}$, will enable eXTP to map the EOS over a broad range of central densities well above nuclear saturation density — particularly for massive NSs such as PSR~J0740+6620, as illustrated in Fig.~\ref{fig:J0740_MR_simulation}.

\subsection{Accretion-powered MSPs} 
\label{Sec:amp}

For AMXPs, hot spots form as matter accreting via a disk is channeled onto the star surface by the NS magnetic field. The X-ray emission from these sources therefore includes a pulsed component from the hot spots (likely scattered by the material in the accretion funnel, see below), together with emission from the rest of the stellar surface (heated as the accreting material flows over the rest of the star) and the accretion disk itself. Given suitable models for the emission from the various components, AMXP pulsations are a viable target for PPM \cite{Poutanen03,Salmi18}. 

The radiation emitted from AMXPs is expected to be linearly polarized because the opacity is dominated by electron scattering \cite{Viironen2004}. And indeed polarized radiation was recently discovered from the AMP SRGA~J144459.2$-$604207 \cite{Papitto2025} by the currently active Imaging X-ray Polarimetry Explorer (IXPE, \cite{Weisskopf2022}). This means that additional constraints on the hot spot localization and NS inclination can be obtained from modeling the X-ray polarization signal, which can break the degeneracy between the geometrical parameters and the NS mass and radius. eXTP will significantly enhance the sensitivity of polarization measurements and the achievable parameter constraints. 

The technique used to model the observations is based on the rotating vector model, where the relativistic effects and the oblate shape of the star can be accounted for \cite{Poutanen2020RRM,Loktev2020}.
In addition, the emission of the polarized radiation can be modeled using a slab of hot electrons up-scattering the soft photons coming from the NS surface \cite{Bobrikova2023}. 

\subsubsection{Geometric constraints from polarimetry}
\label{Sec:amp_pol}

\begin{figure}[H]
\begin{center}
\includegraphics[width=1\linewidth]{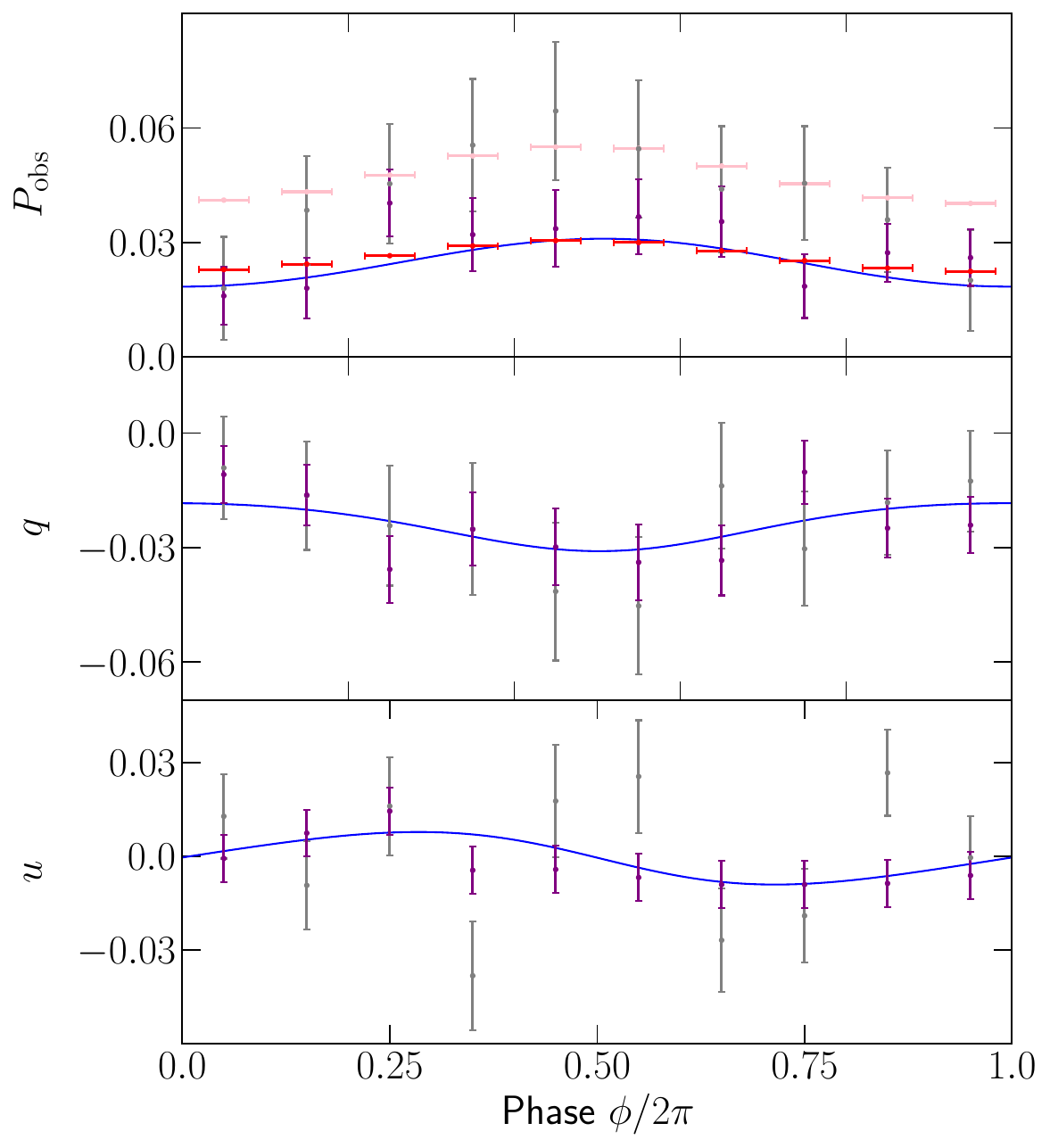}
\end{center}
\caption{
Synthetic polarization degree ($P_{\mathrm{obs}}$) and normalized PFA Stokes $q$, $u$ data. 
Scenario B of \cite{Salmi25} is taken as the input model. 
The blue curves show the model curve using the injected parameters, the purple dots and error bars show the synthesized observed PFA data, and the red bars show the minimum detectable polarization (MDP) values for the corresponding phase points. 
The corresponding IXPE data and MDP values are shown with gray and pink bars, respectively.
}
\label{fig:AMP_polarization_data}
\end{figure}

\begin{figure}[H]
\begin{center}
\includegraphics[width=1\linewidth]{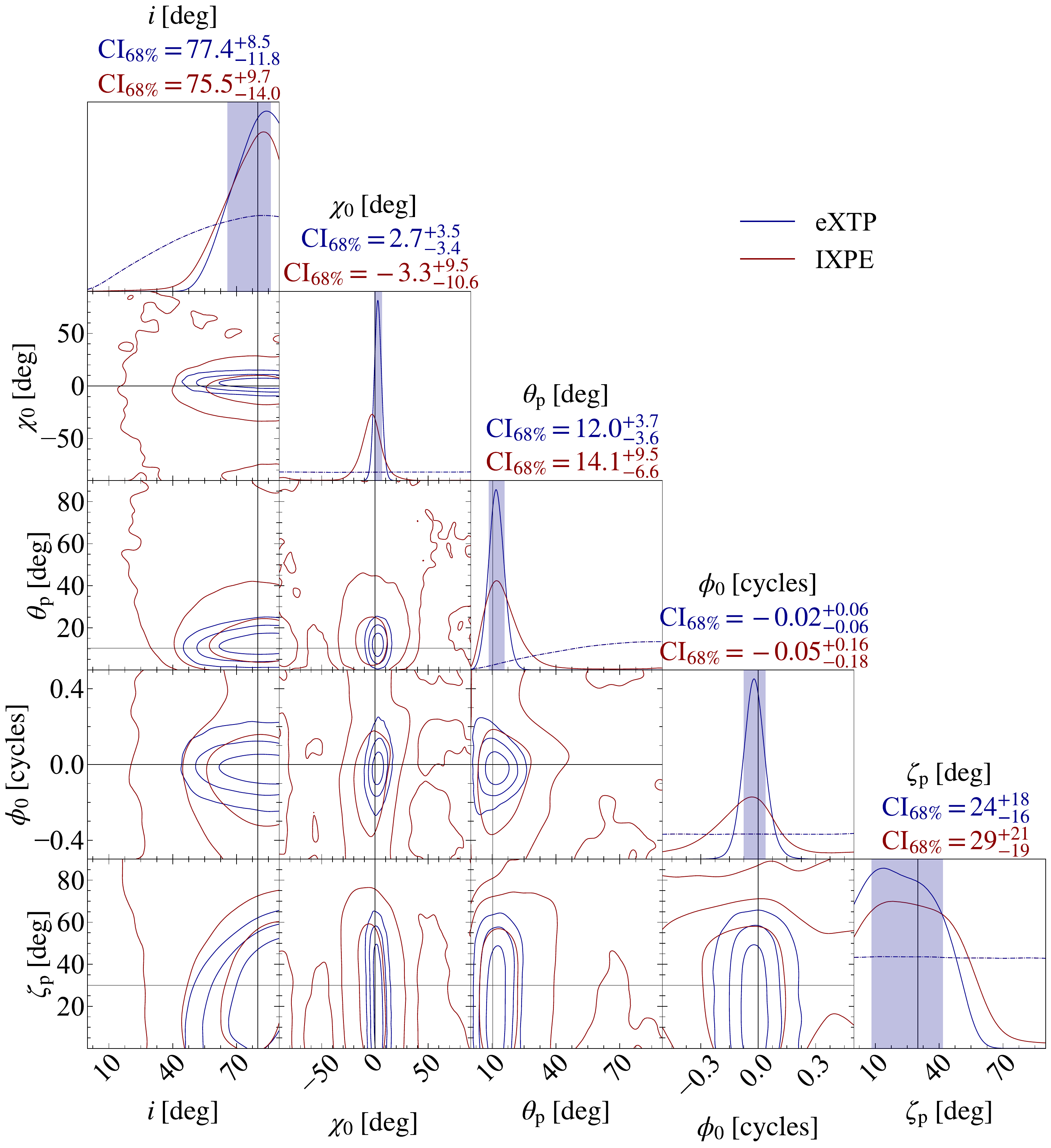}
\end{center}
\caption{Posterior distributions of NS inclination $i$, spin axis position angle $\chi_{0}$, hot spot colatitude $\theta_{\mathrm{p}}$, phase zero $\phi_{0}$, and hot spot angular radius $\zeta_{\mathrm{p}}$ when fitting simulated Stokes $q$ and $u$ data from eXTP or IXPE. 
The true values are shown with thin-solid lines (matching Scenario B from \cite{Salmi25}), and the dash-dotted curves represent the prior distributions.
The 1D credible intervals contain 68.3\% of the posterior mass, and the 2D contours contain 68.3\%, 95.4\% and 99.7\% of the posterior mass.
}
\label{fig:AMP_polarization_posterior}
\end{figure}

Polarization simulations and parameter inference predictions for IXPE were recently studied in \cite{Salmi25}. 
Using the same pipeline (adapted to eXTP) and their scenario B (that is closest to the true observations of SRGA~J144459.2$-$604207, with a polarization degree of a few per cent) fitting Stokes $q$ and $u$ data, we simulate data with a 600 ks effective observation time (shown in Fig. \ref{fig:AMP_polarization_data}) and parameter constraints for the PFA instrument of eXTP.
The resulting posterior distributions for the geometry parameters are shown in Fig. \ref{fig:AMP_polarization_posterior} when the prior of the hot spot colatitude $\theta_{\mathrm{p}}$ is limited between 0 and 90\degr. 
Significant improvements over the IXPE constraints are observed with the eXTP simulated data, particularly in $\theta_{\mathrm{p}}$, where the 68\% credible interval shrinks by more than half.

\begin{table*}
    \renewcommand\arraystretch{2}
\centering     
\caption{Precision and Systematics of Pulse Profile Observations for NS Mass-Radius Constraints}
\label{tab:techniques_precision}
\footnotesize
\begin{tabularx}{\textwidth}{XXXXXX}
\toprule
Observational Technique & Instrumental Mode & Required Effective Exposure & Expected Precision on M-R & Leading Systematic Limits \\
\midrule
Pulse profile: non-accreting MSP & SFA long-term pointed observation & 1–2 Ms per source & Radius: $\pm$5–10\%; Mass: $\pm$5–10\% & Background, hot spot configuration, mass prior constraint \\
Pulse profile: accreting MSP & SFA and PFA long-term pointed observation & 100–600 ks per source & Radius: $\pm$5–10\%; Mass: $\pm$10–15\% & Hot spot and accretion disk modelling, accretion emission variability \\
\bottomrule
\end{tabularx}
\end{table*}

\subsubsection{Combined constraints}

\begin{figure}[H]
  \centering
    \includegraphics[width=1\linewidth]{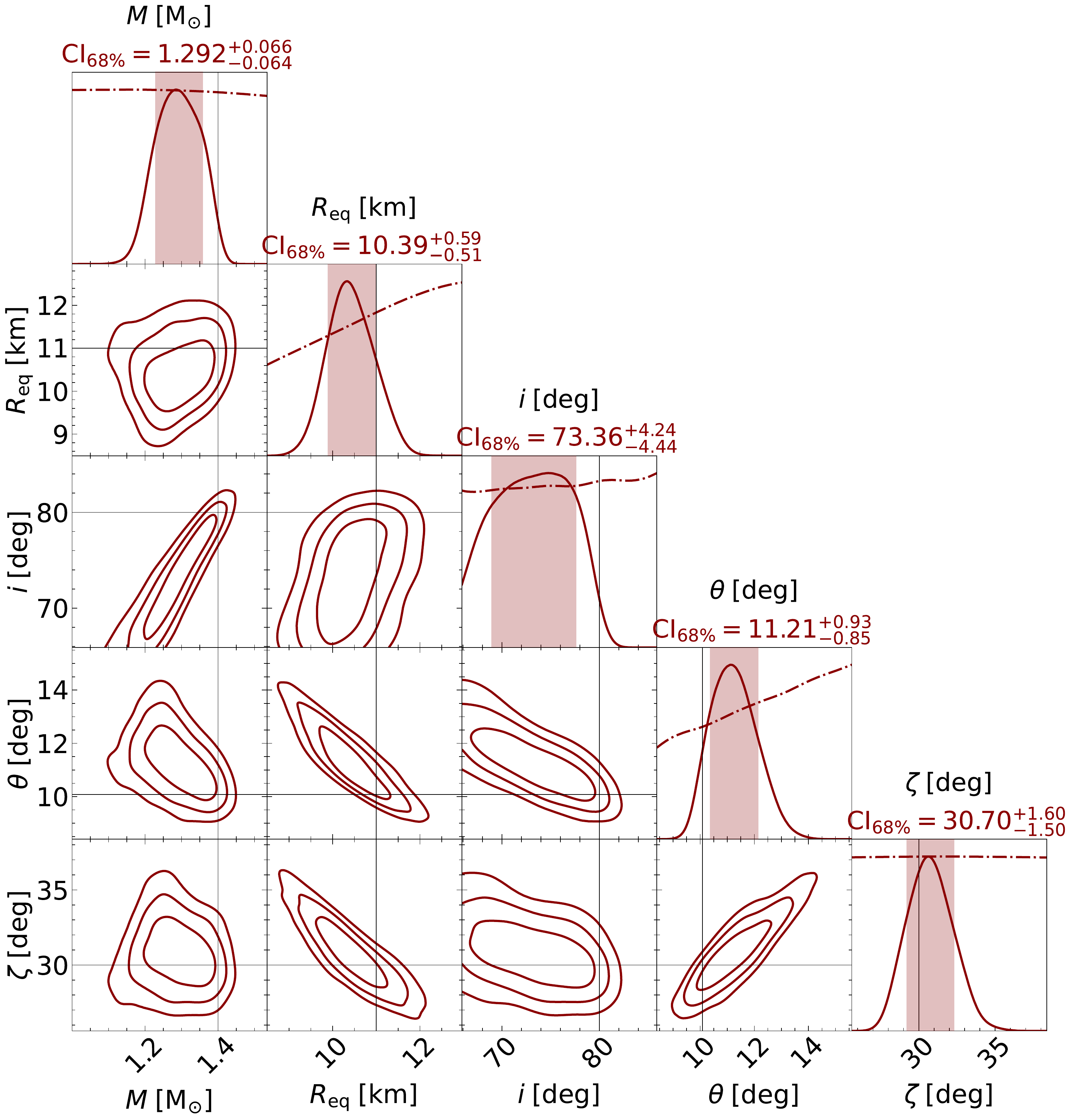}
    \caption{Posterior distributions of mass $M$, equatorial radius $R_{\rm eq}$, $i$, $\theta_{\mathrm{p}}$, and $\zeta_{\mathrm{p}}$ obtained by fitting the simulated eXTP-SFA pulse profile. Here, the 68.3\% intervals of $i$ and $\theta_{\mathrm{p}}$ resulting from the eXTP polarization analysis (Fig. \ref{fig:AMP_polarization_posterior}) were taken as a prior. 
    }
    \label{fig:combined_constraints}
\end{figure}

The constraints on geometry derived from the polarimetric data for the AMXPs can then be used together with PPM on other (non-polarimetry) data sets. Fig. \ref{fig:combined_constraints} shows simulated constraints for 100 ks of exposure time for eXTP-SFA for the scenario explored in Sect.~\ref{Sec:amp_pol}, but using the geometric constraints derived from the eXTP-PFA analysis as priors. This analysis uses the AMXP PPM pipeline of \cite{Dorsman25}, which incorporates an appropriate atmosphere model \cite{Bobrikova2023} and model for emission from the accretion disk. Compared to analysis of the same scenario for simulated NICER data, the constraints on mass and radius are much tighter with eXTP-SFA: the 68\% credible interval on mass has reduced from $\sim$0.20 to $\sim$0.13 M$_{\odot}$, and the 68\% credible interval on radius has reduced from 0.19 to 0.11 km. The improvement is in part due to larger effective area, and in part due to polarization constraints. This is a conservative estimate for eXTP with an effective observation time here of only 100 ks, compared to the simulation in \cite{Dorsman25} of 130 ks.
Table \ref{tab:techniques_precision} summarizes the quantitative performance of PPM techniques in constraining NS mass and radius.

\begin{figure*}
\centering
\includegraphics[width=0.95\textwidth,,height=0.5\textwidth]{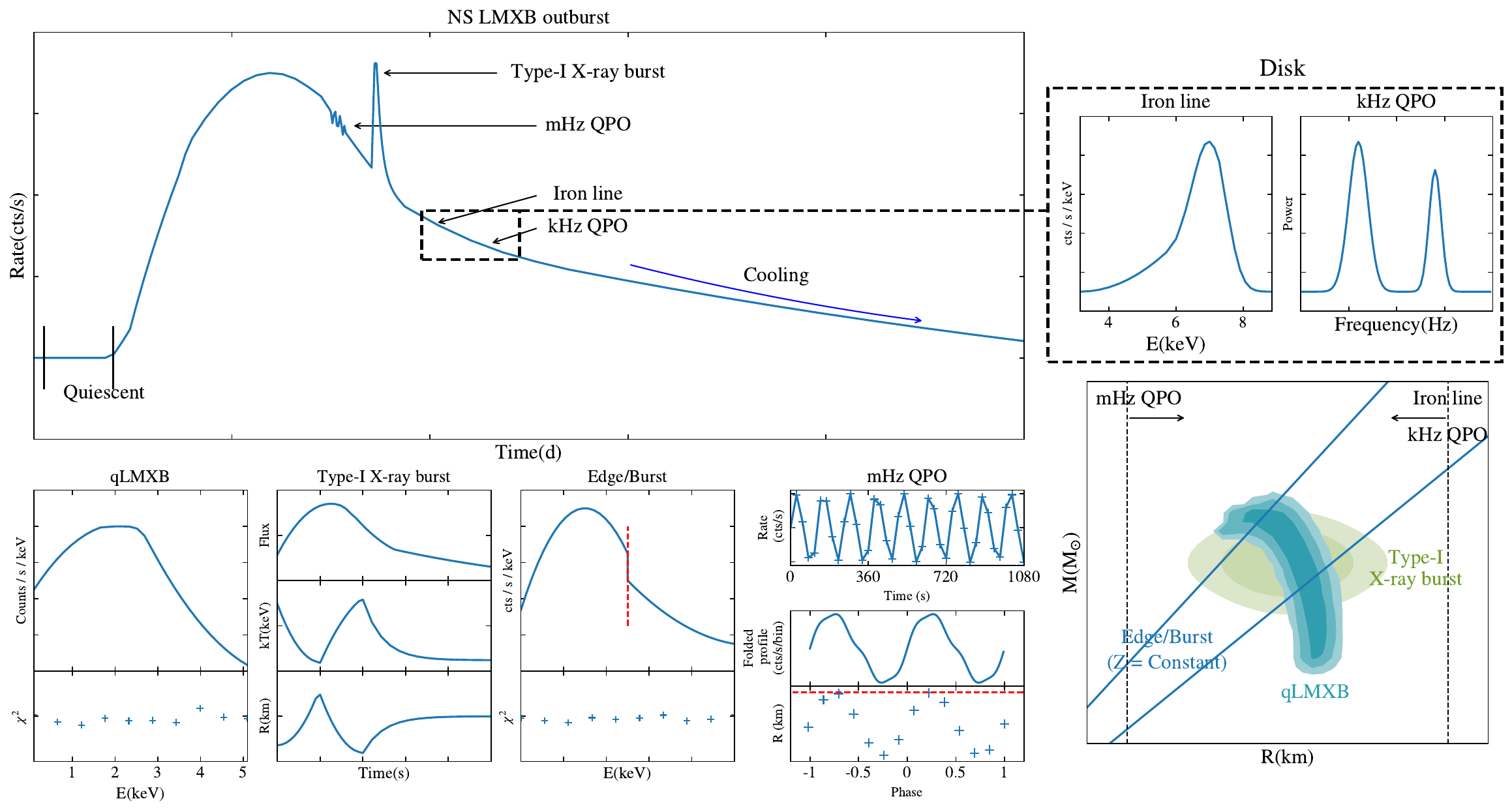}
\caption{This schematic illustrates key observational and temporal features of a NS-LMXB system during outburst and quiescent phases. Upper Left Panel: Depicts the X-ray outburst phase, highlighting prominent phenomena such as type I (thermonuclear) X-ray bursts, mHz QPOs, and kHz QPOs. Lower Left Panel: Shows detailed observational features tied to emission from the NS. Subfigures (left to right) include: The X-ray spectrum of a quiescent NS-LMXB. Light curves, fitted blackbody temperature, and radius evolution from a photospheric radius-expansion (PRE) type I X-ray burst. 
Absorption edges observed during bursts and mHz QPO variability. Top Right Panel: Displays accretion disk features, including broadened Fe K$_\alpha$ emission lines and kHz QPOs. Lower Right Panel: Emphasizes how combined timing, spectral, and thermal datasets synergistically constrain the NS mass and radius.
}
\label{fig:lmxb}
\end{figure*}

\section{Other-methods: spectroscopy, polarimetry and flux evolution}
\label{Sec:Spectr}

In NS X-ray Binary (NS-XRB) systems, either LMXBs or HMXBs, the companion star transfers material onto the NS via Roche-lobe overflow or stellar winds. As the accreted material spirals inward, it heats up due to viscous dissipation and emits X-rays. These systems provide a natural laboratory for studying the physics of NSs and offer critical insights into fundamental astrophysical phenomena, such as strong gravity, nuclear processes, and accretion dynamics.

X-ray outbursts are sudden increases in the X-ray luminosity of NS-XRBs. They typically occur when there is a temporary enhancement of the mass transfer rate from the companion star. These events can last from days to months and 
can evolve from initially ``hard'' spectral states to ``soft'' and back again~\citep{van_der_Klis1994, Hasinger1989}. NS-LMXBs exhibit distinct observational and temporal features during both outburst and quiescent phases. Fig. \ref{fig:lmxb} illustrates these key features, which will be discussed in detail in the following subsections.

In the quiescent state of NS-XRBs, the radiation from the NS surface may dominate the emission in the soft X-ray band. This thermal emission can be well described by a NS atmosphere model, which can then be used to measure the NS mass and radius. This approach will be discussed in detail in Sect.~\ref{Sec:qLMXB}.
A similar method can be applied to isolated NSs, which lack strong magnetic fields and accretion-related complexities. The absence of these factors simplifies the interpretation of their spectra, reducing uncertainties associated with radiative transfer in extreme magnetic environments or accretion physics. This approach will be discussed in detail in Sect.~\ref{Sec:CCO}.

Thermonuclear X-ray bursts \citep[e.g.,][]{Lewin93, Galloway2021ASSL}, also known as type I X-ray bursts, result from runaway thermonuclear burning on the surface of a NS in a LMXB system. The observational properties of type I X-ray bursts — such as their spectra, light curves, and peak luminosities — can be used to infer the NS mass and radius. This approach will be discussed in detail in Sect.~\ref{Sec:burst_cooling}.

During some type I X-ray bursts, X-ray spectra exhibit absorption edges at specific energy levels \citep[e.g.,][]{int_Zand2010A&A, Kajava2017}. These features arise from the ionization of heavy elements in the NS atmosphere and provide valuable insights into the physical and chemical properties of the NS environment. This approach will be discussed in detail in Sect.~\ref{Sec:burst_edge}.

Burst oscillations, hot spots generated by an as yet unknown mechanism during Type I X-ray bursts \citep[see][for a review]{2012ARA&A..50..609W}, may also provide information on the mass and radius. PPM for burst oscillations is challenging (given the intrinsic variability, the role of ongoing accretion, and uncertainty in the underlying surface pattern) \citep{Kini23,Kini24a,Kini24b}. However eXTP will observe burst oscillations with greater sensitivity in a softer waveband than existing observations, which may provide valuable insight into the mechanism and help to resolve surface pattern uncertainty. 

Millihertz quasi-periodic oscillations (mHz QPOs) in NS-LMXBs are linked to nuclear burning on the NS surface. The thermal emission from these QPOs can be modeled as a blackbody, allowing the size of the burning region to be estimated spectroscopically. Since the NS radius must exceed this size, mHz QPOs provide a novel method to constrain the NS radius. This approach will be discussed in detail in Sect.~\ref{Sec:mHzQPO}.

During accretion outbursts, nuclear reactions heat the crust and raising its temperature. Post-outburst, the crust cools via thermal emission, detectable as a decaying X-ray flux. By modeling this cooling, the thermal conductivity of the crust and core neutrino emissivity are linked to the NS composition. This approach will be discussed in detail in Sect.~\ref{Sec:Crustcool} and Sect.~\ref{Sec:Nscool}.

Kilohertz quasi-periodic oscillations (kHz QPOs) are high-frequency variability features observed in the X-ray emission of accreting NS-LMXBs. Iron (Fe) lines are emission features observed in the X-ray spectra of these systems. Both kHz QPOs and Fe lines are believed to originate from the inner regions of the accretion disk, close to the NS. These features provide additional methods to constrain the EOS for accreting NSs, which will be discussed in detail in Sect.~\ref{Sec:Disk}.

\subsection{Spectroscopy of NS thermal emission}
\label{Quiescent}

\subsubsection{Quiescent LMXBs 
} 
\label{Sec:qLMXB}

Most NS-LMXBs are transient systems. During accretion outbursts, a series of nuclear reactions deposit heat in the NS crust. In the quiescent state of LMXBs (qLMXBs) the accretion process almost halts, the disk dims, and the radiation from the NS surface dominates the emission in the soft X-ray band. Usually, the thermal emission is well described by a NS atmosphere model~\citep{Zavlin1996, Heinke2006}. This approach is based on the understanding that the observed thermal radiation is modified by the effects of the NS strong gravitational field, which causes gravitational redshift. By fitting the observed X-ray spectra with a model that accounts for the NS atmosphere, one can extract key parameters such as the effective temperature, the apparent radius, and — with the aid of relativistic corrections — even place constraints on the true radius and mass of the NS~\citep{Zavlin1996, Heinke2006,Guillot2011,Bogdanov2016}, provided the distance to the source is known precisely enough (order $\sim$5--10\% uncertainties). 

\begin{figure}[H]
\centering
\includegraphics[width=0.5\textwidth]{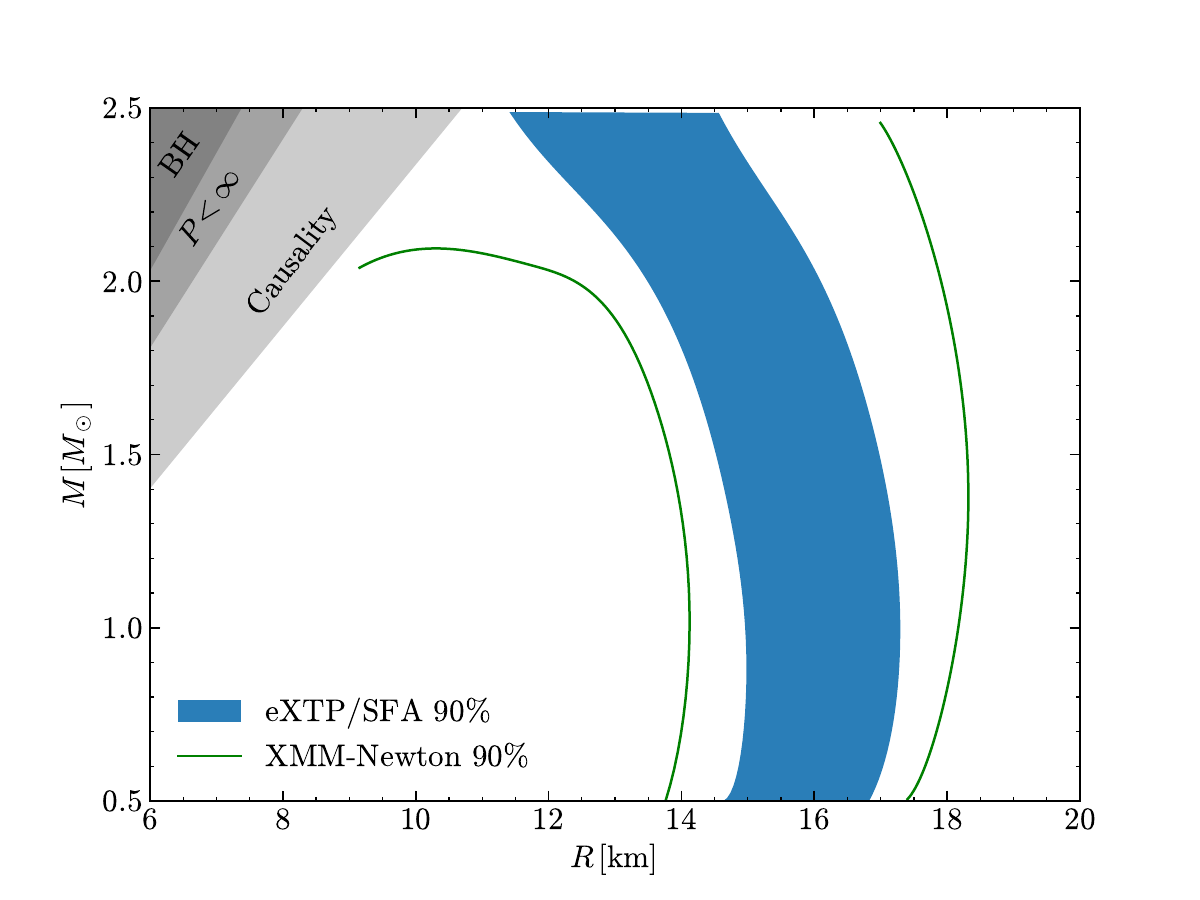}
\caption{A comparison of the fit results for NS mass and radius obtainable by eXTP with those obtained by XMM-Newton for the NS-LMXB EXO~0748-676. 
}
\label{fig:lmxb_exo_0748}
\end{figure}

Moreover, since the quiescent state exhibits low-level and relatively stable emission, it minimizes the complications that arise from variable accretion, thereby offering a clearer window into the thermal properties of the NS surface~\citep{Zhanggb2011, Degenaar2009}. In Fig. \ref{fig:lmxb_exo_0748} we present a comparison of the fit results for NS mass and radius obtained from XMM-Newton observations of the NS-LMXB EXO~0748$-$676 to those that would be obtained using eXTP. The simulations indicate that eXTP can significantly improve the precision of NS radius measurements using the same method, although systematics still exist in the form of uncertainty about the atmospheric composition and the emission uniformity across the stellar surface. 

Overall, the use of X-ray spectral analysis in quiescent NS-LMXBs has emerged as a powerful tool for probing the fundamental properties of NSs, although some systematics remain~\cite[e.g.,][]{Echiburu2020}. By combining high quality observational data with sophisticated spectral models – which account for both relativistic effects and the detailed microphysics of a NS atmosphere – researchers are able to extract robust estimates of a NS temperature, radius, and mass.

\subsubsection{Isolated NSs, Central Compact Objects 
}
\label{Sec:CCO}

CCOs emit predominantly thermal X-rays ($kT \sim 0.2-0.5$~keV) from their surfaces, likely due to residual heat from their formation or early-stage cooling \citep{Pavlov2002}. CCOs exhibit weak magnetic fields \citep{Gotthelf2013}, show no evidence of magnetospheric activity or pulsar wind nebulae, and are not surrounded by accreting material. This makes them unique laboratories for studying the intrinsic properties of NSs, such as their mass and radius \citep{Zavlin1998,Pavlov2009}. By modeling their thermal X-ray emission, researchers can infer the surface temperature distribution, emission area, and gravitational redshift, all of which depend on the NS compactness (M/R ratio). The absence of strong magnetic fields and accretion-related complexities simplifies the interpretation of their spectra, reducing uncertainties tied to radiative transfer in extreme magnetic environments or accretion physics.

A key advantage of CCOs lies in their thermal simplicity. Their X-ray emission is thought to originate from a large fraction of the NS surface, and the lack of detected pulsations in most CCOs suggests a relatively uniform temperature distribution. This uniformity minimizes geometric uncertainties, such as hotspot configurations or viewing angles, which complicate mass-radius estimates in pulsars or accreting NSs. Additionally, the weak magnetic fields of CCOs allow for the use of non-magnetic atmosphere models (e.g., hydrogen, carbon, or heavy-element atmospheres) to fit their spectra. For instance, surface emission from a light-element atmosphere produces distinct spectral features that can be compared to observations to infer the NS radius and gravitational field. The CCOs with detected pulsations further enable timing analysis to constrain spin parameters, though their low spin-down rates still imply minimal magnetic interference \citep{Gotthelf2013}.

However, significant challenges remain. CCOs are faint X-ray sources, and their distances — often inferred from associated SNRs — are poorly constrained, leading to large uncertainties in luminosity and emission area estimates. This directly impacts the precision of radius measurements \citep{Doroshenko2022}. Furthermore, the assumption of uniform surface emission is not definitive; the absence of pulsations could result from unfavorable alignment of the NS rotational axis 
with the observer line of sight, masking temperature anisotropies. The composition of the atmosphere also introduces degeneracies: heavier elements (e.g., carbon or oxygen) produce spectra that mimic larger emission areas at lower temperatures, skewing radius estimates. For example, a carbon atmosphere model may yield a radius 30--50\% larger than a hydrogen model for the same observational data. 
Finally,  
their youth ($\leq10^4$ years) means that their thermal emission reflects early-stage cooling, which depends on uncertain supernova fallback processes and crustal properties. Despite these limitations, CCOs remain promising targets for advancing our understanding of the interiors of the NS. 
With its large effective area capabilities, the eXTP-SFA will significantly improve the precision of NS radius measurements.

\subsection{Thermonuclear bursts 
} \label{sec:thermbursts}

Thermonuclear (type I) X-ray bursts are powered by unstable burning of material accreted on the NS surface from its low-mass companion. The burning process depends on the composition of the accreted material of the material, the accretion rate, and the ignition depth. For normal type I X-ray bursts, the ignition occurs at shallow depths of densities $\sim10^8~{\rm g~cm^{-2}}$, where pure helium or a mixture of hydrogen and helium burn explosively, releasing a significant amount of energy of $\sim10^{39}~{\rm erg}$. These bursts last only a few seconds to minutes and recur within several hours. More than 7000 X-ray bursts have been reported in a comprehensive catalog 
\citep{2020ApJS..249...32G}.

In some sources, carbon ashes accumulated during numerous normal X-ray bursts ignite at a depth of $\sim10^{12}~{\rm g~cm^{-2}}$, resulting in energetic long X-ray bursts (superbursts) with a total energy release of $\sim10^{42}~{\rm erg}$ and recurrence times from days to years. 
When a thick layer of helium can form in the envelope of accreting NSs, its explosion produces the so-called intermediate duration/long bursts, which have intermediate durations between normal frequent bursts, and superbursts. Unlike the superbursts,
they are associated with LMXBs with very low accretion rate, typically $\lesssim1\%\dot{M}_{\rm Edd}$, where $\dot{M}_{\rm Edd}$ is the Eddington-limited accretion rate \footnote{$\dot{M}_{\rm Edd}=1.75\times10^{-8}~M_{\odot}~{\rm yr}^{-1}$ has been widely used as the canonical value assuming $M=1.4~M_{\odot}, R=10~{\rm km}$, and 70\% of hydrogen mass fraction~\cite{JOHNSTON2020}.}. So far, 84 long X-ray bursts have been reported by a recent catalog \citep{2023MNRAS.521.3608A}. eXTP is projected to observe over 10 superbursts or intermediate duration bursts per year from well-studied persistent sources, such as 4U~1820$-$30, 4U~1636$-$536, 4U~0614$+$09, 4U~1254$-$69, 4U~1705$-$44, 4U~1735$-$44, GX~3$+$1, GX~17$+$2 or Ser~X$-$1, alongside active transient systems including Aql~X$-$1, 4U~1608$-$52, EXO~1745$-$248 and KS~1731$-$260 among others. eXTP/W2C will be able to monitor thermonuclear X-ray bursts from these sources in high-cadence and detect burst oscillations and their evolution, e.g., at $\sim$ 413 Hz from 4U~0614$+$09.

The spectral-timing properties of the burst lightcurves encode information about NS mass and radius that can be extracted in different ways, some of which we describe below (see also Fig. \ref{fig:structure})\footnote{Note that emission from the accretion disk will have an effect on the efforts to constrain mass and radius for all of the methods described in Sect. \ref{sec:thermbursts}. Disk reflection can contribute approximately 20-30\% of the observed flux from a burst, and bursts can also heat the disk, shifting its thermal emission \citep[see e.g.][]{Guver22,Speicher22}. Careful modeling of the burst-disk interaction will be needed to address these potential systematics.}

\subsubsection{Burst cooling tails 
}

\label{Sec:burst_cooling}

Over the past three decades, the thermal emission from NSs during thermonuclear X-ray bursts has been extensively used in efforts to determine NS masses and radii~\citep[e.g.,][]{vanparadijs1978, Damen1990, Guver2010, Zhanggb2011,Suleimanov2011, Guver2012, Suleimanov2020A&A}.
The most energetic bursts can expand the photosphere of the NS (photospheric radius-expansion, PRE). During their cooling phase, the photosphere contracts towards the NS surface, and the X-ray emission encodes imprints of the NS surface gravity and gravitational redshift. Atmospheric models tailored to NS environments account for relativistic effects and composition-dependent opacities, enabling precise extraction of the NS mass and radius~\citep{Suleimanov2012,Suleimanov2017}.

The methodology relies on time-resolved spectroscopy of the cooling tail, where the spectral evolution is modeled using relativistic atmosphere codes. These models self-consistently compute the emergent flux as a function of surface gravity ($g$), redshift ($1+z$), and atmospheric composition (e.g., hydrogen or helium). 
A key challenge arises from the degeneracy between the spectral parameters, uncertainties in the source distance and system inclination, and how all these relate to the mass and radius of the star. One mitigation comes when an independent estimate of the distance is known (e.g., when the source is in a globular cluster), but the main improvements come from the possibility of jointly fitting various spectra. This fitting is done within a Bayesian framework that eventually maps the posteriors from the fitting process onto the mass and radius parameter space\citep[e.g., ][]{Suleimanov2017,Nattila2017}. 

Recent advances in cooling tail spectroscopy have significantly reduced systematic uncertainties. Time-dependent modeling of photospheric contraction and temperature gradients across multiple burst phases has improved parameter accuracy, while multi-burst analyses (e.g., combining 5 bursts from 4U~1702$-$429) enhanced statistical robustness \citep[e.g.,][]{Nattila2017}. 
The high timing resolution and large collecting area of the eXTP-SFA will enable the cooling tail method to achieve even tighter constraints on key parameters by delivering high-quality spectra.

\subsubsection{Photoionization edges 
}
\label{Sec:burst_edge}

\begin{figure}[H]
\centering
\includegraphics[width=0.45\textwidth]{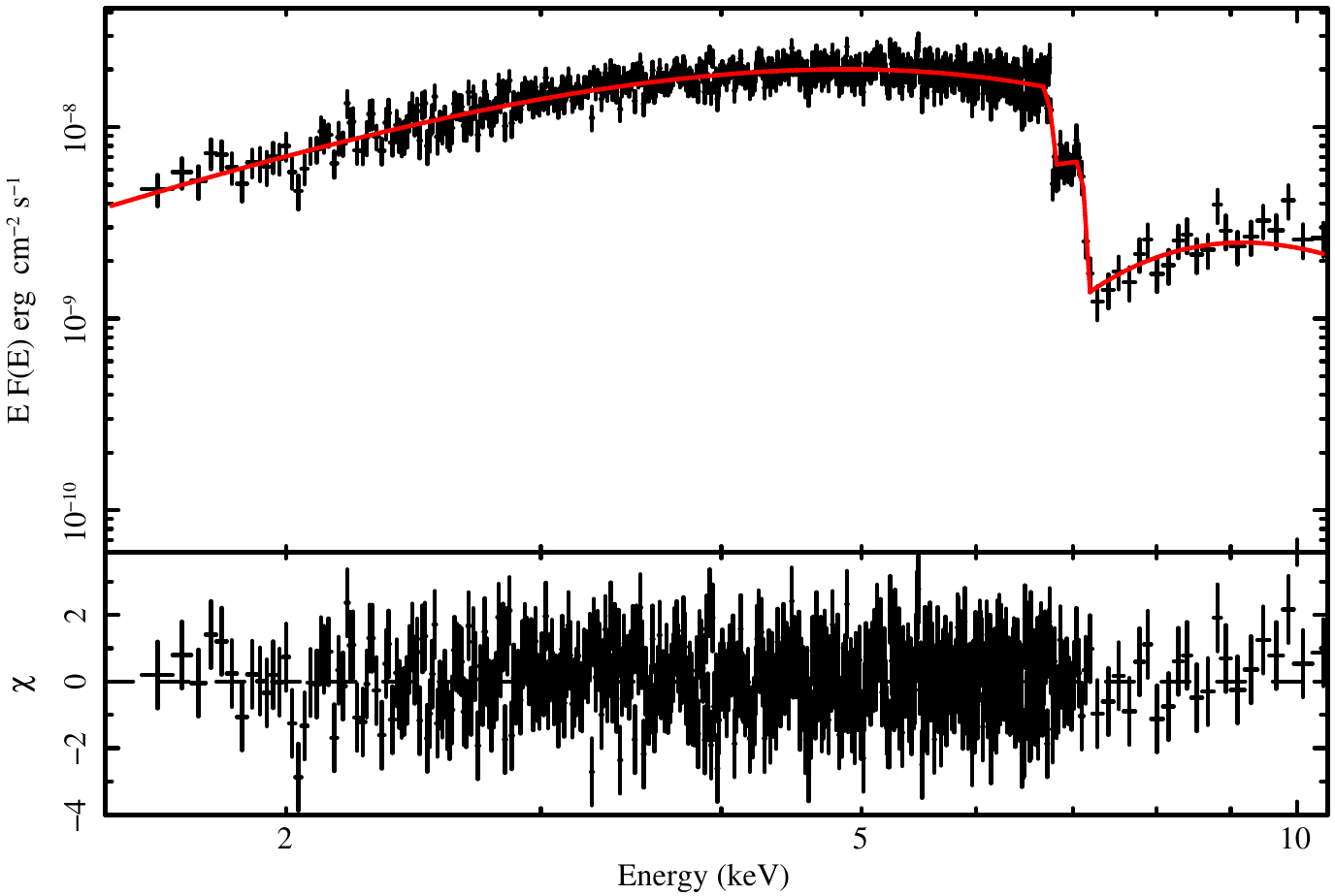}
\caption{Two photoionization edges from thermonuclear X-ray burst. The spectrum is simulated for the eXTP-SFA with an exposure time of 5 s. The photosphere is locates on the NS surface and the edges are redshifted by strong gravity.
Two edges in a single burst spectrum are shown: H-like and He-like Iron edges, at $7.13\pm0.03$ and $6.76\pm0.03$ keV, corresponding to a gravitational redshift of $1+z=1.3$. 
}
\label{fig:edge}
\end{figure}

Thermonuclear X-ray bursts can produce heavy elements via the rapid proton capture (rp-) process. During bursts, the heavy elements are ionized due to the high temperature of the NS atmosphere, and only one or two electrons remain. When the burst radiation passes through the NS photosphere, the elements can be fully ionized, causing a shortage of the burst radiation above the ionization energy. Photoionization absorption features in the range 6--11 keV have been identified in the sources 4U~0614+091, 4U~1722--30, and 4U~1820--30 \citep{int_Zand2010A&A}, HETE~J1900.1--2455 \citep{Kajava2017}, and GRS~1747--312 \citep{Li2018}.
They have been explained as bound-bound and bound-free transitions in the heavy elements produced during the bursts and transported in the photosphere by convection. The observed spectral features can be associated with the hydrogen-like Fe edge at 9.278 keV, or hydrogen/helium-like Ni edges at 10.8/10.3 keV, respectively. The presence of spectral features potentially allows us to determine the gravitational redshift on the NS surface \citep{Lewin93,Yu2018, Li2018} giving a direct measurement of $M/R$. So far, edge detections have been reported only from observations with the {\it Rossi X-ray Timing Explorer}\/ ({\it RXTE}) Proportional Counter Array (PCA). Due to the limited energy resolution of the PCA onboard RXTE, it was difficult to identify the origin of the observed edges, since the spectral states from Fe and/or Ni could not be disentangled effectively. The high resolution X-ray spectroscopy possible with the eXTP-SFA will provide the necessary sensitivity and resolution to identify these edges. In Fig.~\ref{fig:edge}, we show the potential to detect multiple edges corresponding to different ionized states such as hydrogen-like and helium-like Fe. This advancement is crucial for accurately measuring the gravitational redshift and thus determine the ratio of NS mass and radius.

\begin{figure*}
\centering
\includegraphics[width=0.85\textwidth,]{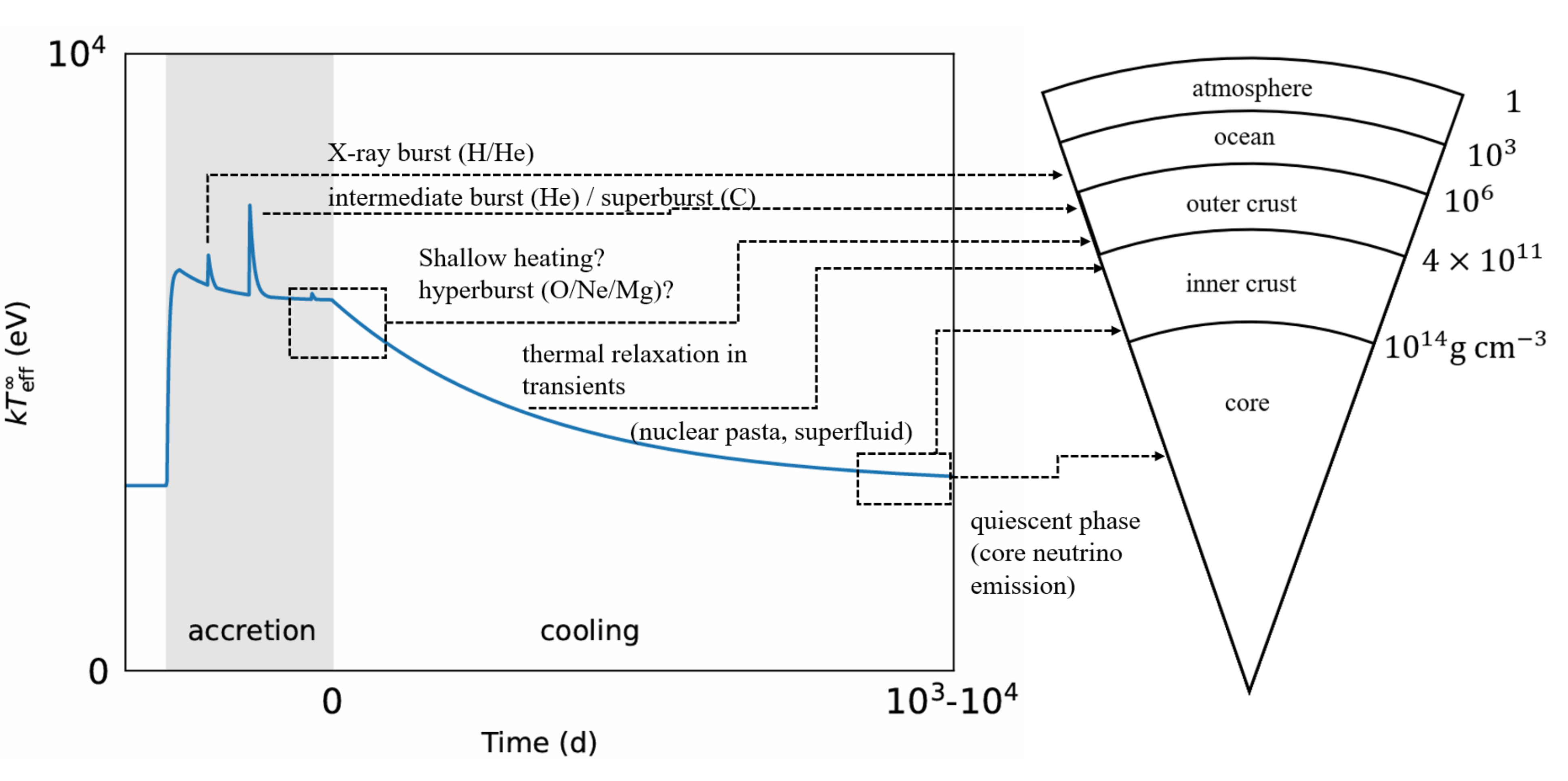}
\caption{The light curve serves as a probe for exploring the internal structure of NSs. Here the effective temperature is proportional to the fourth root of flux or luminosity. The gray shaded region indicates the period of accretion, with the end of accretion setting the starting point of the crustal thermal relaxation. Arrows connect structures at different depths in the crust to possible astrophysical phenomena (e.g., bursts and crustal cooling) with explorable crust properties (e.g., composition or superfluidity) in parentheses at various stages of the light curve. Note that hyperbursts triggered at significant depths do not exhibit any noticeable signature in the light curve. Nevertheless, we have marked the triggering time with a minor kink in the light curve.
}
\label{fig:structure}
\end{figure*}

\subsubsection{Millihertz quasi-periodic oscillations 
} 

\label{Sec:mHzQPO}

Millihertz quasi-periodic oscillations (mHz QPOs) have been discovered in 9 NS-LMXBs~\citep{2001A&A...372..138R,2011ATel.3258....1S,2019MNRAS.486L..74M,2012ApJ...748...82L,2018ApJ...865...63S,2021MNRAS.500...34T,2023MNRAS.521.5616M}. Currently, both observational and theoretical studies indicate that the mHz QPOs originate from nuclear burning on the NS surface \citep[][]{2001A&A...372..138R,2007ApJ...665.1311H,2009A&A...502..871K,2012ApJ...748...82L}. 

Since the thermal emission of the mHz QPOs could be described by a blackbody component~\citep{2016ApJ...831...34S}, the size of the burning region could be estimated from the spectroscopy of the QPOs. This provides a new way to put constraints on the NS radius, since the derived size of the burning regions should provide a lower limit to the NS radius. Previous work on 4U~1636$-$53 ~\citep{2016ApJ...831...34S} has constrained the NS radius to be larger than 11.0 km (2$\sigma$ lower limit) with XMM-Newton observations, thus ruling out EOSs with smaller NS radii.

The main limitation faced when studying the mHz QPOs is the number of detected photons, which determines the uncertainties in the derived spectral parameters: crucially, the size of the emitting region. The eXTP-SFA will introduce the advantage of a large collecting area, thus allowing to reduce the degeneracy between the fitted parameters. For example, the estimated error of the blackbody normalization is $\sim$ 1/3 of the one currently obtainable from NICER data (see WG5 white paper~\cite{WP-WG5}). 
Phase-resolved spectroscopy of the mHz QPOs across multiple epochs should provide a reliable lower limit for the radius of different sources.

\subsubsection{``Clocked'' bursters 
}
Most X-ray bursters exhibit bursts that vary significantly in profile from burst to burst. Remarkably, several X-ray bursters have been observed to produce light curves with extraordinarily regular shapes in an epoch with regular recurrence times; 
the best-studied such example is GS~1826$-$24~\cite{1999ApJ...514L..27U,2001AdSpR..28..375C,2003A&A...405.1033C,2004ApJ...601..466G} (see \cite{2017PASA...34...19G} for summarized information). Such bursters are termed ``textbook bursters'' due to the alignment with the macroscopic theory, or ``clocked bursters'' due to the almost constant recurrence time ($\Delta t$) for a series of bursts of an epoch. This regularity implies that the accretion rate ($\dot{M}$) of an epoch must be constant, the ignition occurs always at the same column depth, and the available fuel is completely burnt during the burst. Repeat observations of such bursters at different accretion rates allow comparisons against numerical models, for example by measuring the empirical relation, $ F_{\rm per}\propto\dot{M}\propto\Delta t^{-\eta} $, where $F_{\rm per}$ denotes the persistent flux that stems from the gravitational energy release, and $\eta$ is a power-law index close to 1. If $\eta=1$, which implies a constant persistent fluence, the amount of fuel needed for the ignition of light elements is constant and $\simeq\!\dot{M}\Delta t$ for every burst. 

Because of the simplicity of their light-curve behavior, clocked bursters are very useful to probe the properties of LMXBs, such as the accretion rate, the composition of the accreted matter, nuclear reaction rate uncertainties and the EOS, mass and heating, and cooling processes inside accreting NSs. Additionally, such objects are ideal targets for numerical light-curve modeling, which has been systematically explored for GS~1826$-$24~\cite{2007ApJ...671L.141H,2016ApJ...819...46L,2018ApJ...860..147M,2020PTEP.2020c3E02D, 2020MNRAS.494.4576J, 2021PhRvL.127q2701H, 2021ApJ...923...64D, 2022ApJ...937..124D, 2022ApJ...929...72L, 2022ApJ...929...73L, 2023NatPh..19.1091Z}.

Other examples of ``clocked'' bursting have recently been observed in several sources: GS~0836$-$429~\citep{2016A&A...586A.142A}, IGR~J17480$-$2446~\citep{2011MNRAS.418..490C}, MAXI~J1816$-$195~\citep{2022ApJ...935L..32B,2024A&A...689A..47W}, 1RXS~J180408.9$-$342058~\citep{2017MNRAS.472..559W,2019ApJ...887...30F} and SRGA~J144459.2$-$604207~\cite{Papitto2025}\footnote{Note that the first report of SRGA~J144459.2$-$604207 being a clocked burster was from\cite{2024ATel16485....1S} using INTEGRAL ToO observations.}. That is, among $\sim$120 X-ray bursters~\cite{2020ApJS..249...32G}, there are six clocked bursters identified so far. The long-term monitoring of new clocked bursters 
is a high priority as it results in large numbers of bursts observed and high-fidelity datasets that can be used for comparisons with numerical models, and hence constraints the host NS properties, and could enable us to constrain the nature of accreting NSs: For instance, the determination of $\eta$ value in the clocked bursters could lead to the constraint on NS mass and radius structure~\cite{2024ApJ...960...14D} as well as the compositions of accreted matter~\cite{2016ApJ...819...46L} (see also Fig. 1 in \cite{2006ApJ...652..559G}). In case of the latest clocked burster, SRGA~J144459.2$-$604207, it may be a He-enhanced massive accreting NS according to recent studies~\cite{2025ApJ...980..161F,2025PASJ..tmp...40D,2025PASJ..tmp...50T}. 

To monitor sequences of many bursts, a long exposure time is needed.
For example, to observe a series of 20 bursts with an average recurrence time of $\overline{\Delta t }=5~{\rm h}$, about 4 days would be needed for complete monitoring. The eXTP-PFA is suitable for monitoring clocked bursters because it will operate with long exposure times ($10^4~{\rm s}$) as well as having a large collecting area (900 cm$^2$ at $T=2~{\rm keV}$) in the energy range of 2--10 keV. In particular, long continuous monitoring helps prevent missing bursts. Thus, future long-term eXTP observations may unveil the mysteries of clocked bursters, such as \textit{why some X-ray bursters can behave very regularly and what is the physical origin of the $\eta$ relation}. These insights could, in turn, provide more reliable constraints on NS structure based on burst observations.

\subsubsection{Crustal relaxation after outburst 
} 
\label{Sec:Crustcool}

During accretion, the NS crust is compressed by the accreted matter, leading to a series of nonequilibrium reactions, e.g., electron capture, neutron emission, and pycnonuclear fusions. The heating mechanisms resulting from these reactions, namely deep crustal heating~\cite{1998ApJ...504L..95B}, will balance with the cooling mechanisms dominated by photon and neutrino emissions. In a long and strong enough accretion outburst, the crust can be driven away from thermal equilibrium with the core due to the burning of light elements in the accreted layer, resulting in subsequent cooling of the crust~\cite{2002ApJ...580..413R}. 
A schematic connection between the observed light curve at different stages and the NS structure is shown in Fig. \ref{fig:structure}. 
Comparisons between simulated light curves and observations thus reveal the structure and composition of the crust. 
With typical temperature of 115\,eV and flux of $6\times10^{-13}\,{\rm erg\,cm^{-2}\, s^{-1}}$ in the quiescent state, twenty observations with a net exposure of 50\,ks are sufficient to measure cooling curves after outburst over 3--5 years \cite{2020A&A...638L...2P}.

Generally, the longer the time after an outburst, the deeper the crust layer being probed. For KS~1731$-$260~\cite{2016ApJ...833..186M}, the unexpectedly high surface temperature after the outburst suggests the presence of an additional heating mechanism in the shallow crust, known as shallow heating~\cite{2009ApJ...698.1020B}. Currently, the origin and depth of this heating mechanism remain unclear. 
In the case of MXB~1659$-$29, its delayed cooling is linked to thermal relaxation at the bottom of the crust, specifically in the pasta layer~\cite{2013ApJ...774..131C}. The cause of this delay is under debate. One possibility is that the pasta layer has low thermal conductivity, with free neutrons remaining in the normal state~\cite{2017ApJ...839...95D}. Another explanation suggests that the delay results from the increased specific heat due to gapless neutron superfluidity~\cite{2024PhRvL.132r1001A}.
The accumulation of more observations from eXTP can improve our understanding of the crust structure and composition, continuing to refine related nuclear models. 

\subsection{Accretion flows in the disks: Fe lines, kilohertz QPOs and reflection polarization 
} \label{Sec:Disk}

Relativistic Fe K${\alpha}$ spectral emission lines and kilohertz quasi-periodic oscillations (kHz QPOs) are believed to be associated with the inner regions of the accretion disk, serving as potential probes into the physics of strong gravity and dense matter. However, their interpretation remains debated. For a more detailed discussion on the methods of constraining NS masses, radii, and the EOS using these two techniques, we refer readers to~\cite{2019SCPMA..6229503W}.

The Fe K$\alpha$ line is a key spectral feature observed in NS-LMXBs, typically appearing between 6.4 and 6.97 keV. It is believed to originate from the reflection of hard X-rays off the accretion disk~\citep{Fabian2000, Cackett2009, Miller2007, Miller2013}.
The line profile is shaped by various physical effects, including Doppler broadening, special relativistic beaming, gravitational redshifting, and general relativistic light bending. 
Fitting the iron reflection spectrum provides direct measurements of the inner radius of the disk. 
Since the disk inner radius $r_{\rm in}$ must lie outside the NS, the inferred $r_{\rm in}$ provides an upper limit on the stellar radius, thus additionally constraining EOS models \citep{Cackett2010, Bhattacharyya2011}. If one obtains only the ratio $r_{\rm in}c^2/GM$, then this transforms to a constraint in the $M$--$r_{\rm in}c^2/GM$ space which is nevertheless informative \citep{Cackett2010}.

kHz QPOs have been detected in several dozen NS-LMXBs~\citep[see reviews,][]{2000ARA&A..38..717V,2021ASSL..461..263M}. Due to their millisecond timescales, kHz QPOs are thought to be linked to dynamical timescales in the accretion flow near the NS. In numerous instances, twin kHz QPOs appear concurrently, and the correlations between their frequencies have been extensively studied.

If the frequency of one of the kHz QPOs reflects Keplerian orbital motions at a specific radial distance in the accretion disk, the stable orbit must lie outside the NS or the innermost stable circular orbit (ISCO). For an observed QPO frequency $\nu_{\rm QPO}$ and a dimensionless rotation parameter $j=cJ/GM^2$ for a NS (where $c$, $J$, $G$, and $M$ are the speed of light, the NS angular momentum, the gravitational constant and the NS mass respectively), these conditions limit the maximum allowed mass and radius to $M_{\rm max} = 2.2M_\odot (1+0.75j)(1000\,{\rm Hz}/\nu_{\rm QPO})$ and $R_{\rm max} = 19.5 \,{\rm km}(1+0.2j) (1000\,{\rm Hz}/\nu_{\rm QPO})$, respectively~\citep{1999NuPhS..69..123M}. The highest QPO frequency observed by RXTE is $1288\pm 8$ Hz in 4U~0614$+$09~\citep{2018MNRAS.479..426V}, placing an upper limit on the NS mass at $2.1M_\odot$. If this QPO reflects the Keplerian orbital frequency at the ISCO, the NS mass in 4U~0614+09 is estimated to be $2.0\pm 0.1M_\odot$.

There is ongoing debate regarding which of the upper or lower kHz QPO frequencies reflects the Keplerian orbital frequency at the inner edge of the accretion disk, although most models predict that it is the upper QPO. Nonetheless, increasingly high frequencies of the kHz QPOs give increasingly small radii, giving more stringent constraints on $r_{in}$ and therefore on the NS radius. However, the QPOs become more difficult to resolve as they rise in frequency, because they decrease in amplitude and coherence. 
Recent observations by NICER have now extended the detection of kHz QPOs to the soft X-ray band\cite{Bult2018ApJ}. With its substantially larger effective area, the eXTP mission is expected not only to detect kHz QPOs at higher frequencies and fainter amplitudes but also to investigate their spectral-timing properties specifically in the 0.5-10 keV range.

Simultaneous measurements of kHz QPO frequencies and relativistic Fe emission lines can potentially provide independent measures of the inner radius of the accretion disk and help distinguish between QPO models. In particular, detecting kHz QPOs in the power spectrum and a broad Fe line in the energy spectrum of the same object would offer separate ways to constrain the NS parameters. Currently, only one source, 4U~1636$-$53~\citep{2014MNRAS.440.3275S}, has been observed to exhibit both a broad Fe line in the energy spectrum and kHz QPOs in the power spectrum simultaneously on four separate occasions. Unfortunately, these observations covered different spectral states of the source, and the combined results from the kHz QPOs and the iron line did not yield a consistent NS mass value. 

\begin{figure}[H]
\centering
\includegraphics[width=0.475\textwidth]{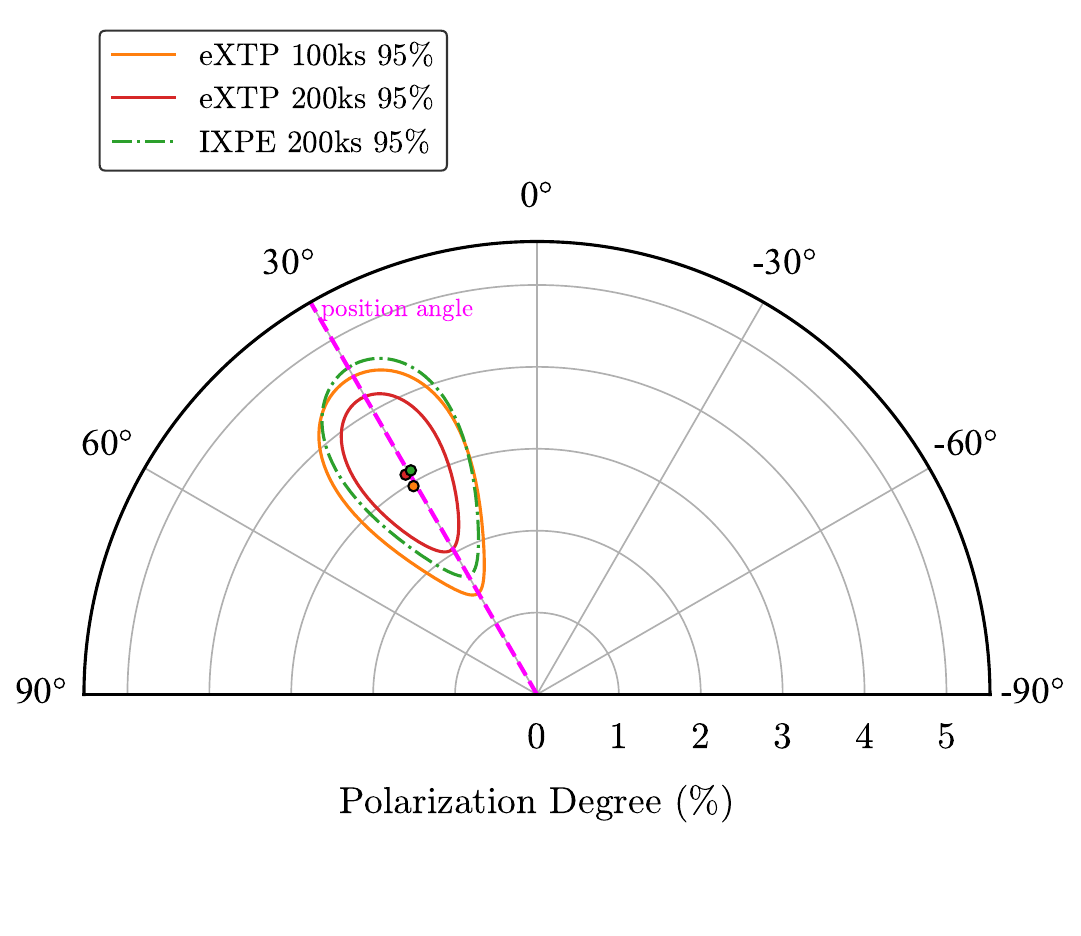}
\vspace{-0.5cm}
\caption{Polar coordinate plot of the polarization degree (PD) and polarization angle (PA) for 4U~1636$-$53 at 95\% confidence level, derived from IXPE simulations with 200 ks (green dash-dotted line), along with eXTP simulations using 100 ks (Orange solid line) and 200 ks (red solid line). The magenta dashed line indicates the position angle.
}
\label{fig:Feline}
\end{figure}

Even more stringent constraints on the inner radius can be obtained by the inclusion of simultaneous X-ray polarimetric observations. The X-ray polarization of the reflection is sensitive to the geometrical properties of the disc, especially the inclination angle \citep[e.g.,][]{Li2009ApJ, Podgorny2023MNRAS}. Therefore, with X-ray polarimetric measurements we can put tighter constraints on the inclination, and break the degeneracy between inner radius and inclination when modeling the reflection component in the energy spectrum. To demonstrate the ability of eXTP, we perform PFA simulations of 4U~1636$-$53, assuming an exposure time of 100\,ks. As for the polarization, we utilize the \textsc{xsstokes\_disc} (\cite{Podgorny2023MNRAS}) to model the polarization properties of the reflection component. The results are presented in Fig.~\ref{fig:Feline}, in which the PD and PA extracted from simulated IXPE data are also shown for comparison. These simulations show that eXTP can reach the same accuracy as IXPE in half the time, or give an error smaller by $\sim 40\%$ than IXPE with the same exposure time. It is therefore clear that eXTP/PFA will be able to significantly detect the polarization signature from the reflection component.

\subsection{Thermal evolution of NSs 
} 
\label{Sec:Nscool}

With the ever-increasing accuracy of observational instruments, more details of the signals emitted by NSs can be quantitatively monitored.
Apart from the measurements of NS masses and radii, high-density NS models can be constrained by the surface temperatures of isolated NS~\citep{Yakovlev2004, Page2006, Tsuruta1998}. Known or estimated ages and the thermal photon luminosity of the X-ray transients in quiescence are especially useful for this purpose. 
Since the first detection of surface radiation from cooling NSs by ROSAT in 1990, numerous subsequent detections have been made using Chandra, XMM-Newton, and NICER, along with the establishment of upper limits on NS temperatures.\footnote{https://www.ioffe.ru/astro/NSG/thermal/cooldat.html}
As noted in Sect.~\ref{Sec:Crustcool}, the NS EOS determines the stellar structure, as well as the effective masses and superfluid gaps of baryons, and is therefore crucial for the heat capacity and neutrino emission rate.
Understanding thermal evolution requires considering the complex interplay of cooling and heating processes. This, in turn, provides a valuable opportunity to probe the physics of dense matter within the stellar interior.
In this section, we discuss recent progress and highlight key cases where further progress is expected, while also outlining the necessary ingredients for interpreting observations.

For newborn NSs (ages $<$ 10-100 years), the surface temperature reflects the thermal coupling of the crust to the core. 
Similarly, in qLMXBs (e.g., Cen~X-4, Aql~X-1), the observed thermal luminosity is a direct consequence of accretion heating during outbursts, and modeling the crust thermal response (i.e., how the crust temperature and heat transport properties change in response to heating) allows us to probe its composition, thermal conductivity, and heat capacity, as discussed in Sect.~\ref{Sec:Crustcool}.

For NSs with ages $<10^5$ years, their cooling is primarily dominated by neutrino emission from the core, and the surface temperature can provide information about the core thermal structure. At this age the magnetic field also plays a crucial role in the surface temperature map, and in certain neutrino processes \cite{Vigano2013, Gourgouliatos2016, Dehman2023, Ascenzi2024}.
Several neutrino emission processes, such as dUrca, modified Urca (mUrca), Cooper Pair Breaking and Formation (PBF), and Bremsstrahlung processes, may dominate this stage of NS cooling \cite{1996PhRvL..76.1994E,2004ApJS..155..623P}.

The standard cooling scenario is dominated by the mUrca process~\cite{Friman1979_ApJ232-541}, and its cooling curve is insensitive to the stellar mass.
The observed rapid cooling of the Cassiopeia A NS challenges standard cooling models, which primarily consider neutrino emission from the core. This discrepancy suggests that either enhanced neutrino emission (e.g., due to exotic particles or Cooper pair breaking) is occurring, or that our understanding of the core heat capacity (e.g., non-Fermi liquid effects) is incomplete (see also \cite{Page2011_PRL106-081101,2011MNRAS.412L.108S}). 

The minimal cooling scenario extends standard cooling by incorporating the effects of superfluidity on cooling~\cite{2004ApJS..155..623P,Page2009_ApJ707-1131}. 
Well-studied examples of cooling NSs in this age range include PSR~J0205+6449 in 3C 58 \citep{Slane2002ApJ,Livingstone2009ApJ} and RX~J0822$-$4300 in Puppis A \citep{Petre1996ApJ,Gotthelf2009ApJ}.
Recent studies of these isolated NSs in terms of secular cooling, and considering young cold and relatively low-magnetic NSs, has confirmed the need of fast cooling processes in dense matter EoS for at least certain mass ranges. Deep X-ray observations of PSR~J0205+6449 confirmed that it is no longer consistent with the minimal cooling scenario \cite{marino2024}.
Superfluidity plays a dual role in cooling: on the one hand, superfluidity suppresses the specific heat and the mUrca process~\cite{2004ApJS..155..623P}; on the other hand, superfluidity opens a neutrino emission mechanism more efficient than that of the mUrca process, known as PBF process~\cite{Flowers1976}. 

The enhanced cooling scenario includes any dUrca processes involving nucleons and possibly non-nucleonic particles~\cite{Pethick1992_RMP64-1133,Prakash1992}. 
Both the minimal and enhanced cooling scenarios can lead to rapid cooling of the NSs. 
In the minimal cooling scenario, the trigger time of the PBF process is closely related to the superfluid critical temperature, and several studies have constrained the critical temperature of $^3P_2$ neutron superfluidity to $\sim0.5\times10^9$ K~\cite{Page2011_PRL106-081101,Beloin2018_PRC97-015804}. However, recent theoretical progress suggests that the mUrca rate is comparable with the PBF process~\cite{Alford2024_PRC110-L052801}, indicating that the characterization of cooling scenarios still requires joint constraints from theory and observations.

For older NSs (ages greater than $\sim 10^5$ years), the surface photon emission becomes the dominant cooling mechanism. The effective surface temperature ($T_{\rm s}$) is related to the interior temperature ($T$) through the properties of the NS envelope, a thin layer that insulates the hot core~\citep{Gudmundsson1983, Potekhin1997}. The resulting photon luminosity scales as $T_{\rm s}^4$, and this thermal radiation is directly observable in the X-ray band, making X-ray telescopes essential tools for probing NS cooling.
The composition of the envelope plays an important role in the detection of the thermal emission. 
Lighter element (H, He) envelopes are more thermally conductive and result in higher surface temperatures for a given core temperature compared to heavier element (Fe) envelopes \citep{Potekhin1997}. 
Surface thermal emission has been detected in several middle-aged pulsars, including PSR~B0656+14, PSR~B1055$–$52, and Geminga \citep{DeLuca2005ApJ}. 

Since XDINSs are free from the complexities of accretion disks or strong magnetospheric activity, their surface thermal radiation provides a relatively clean window into NS cooling, particularly during the photon-dominated cooling stage \citep{Haberl2007ApSS}. Their surface temperatures and luminosities directly reflect their internal thermal structure and the heat transport properties of the envelope \citep{2015SSRv..191..239P}. By fitting their high-quality X-ray spectra with different atmosphere models, one can infer the apparent radius and surface temperature \citep{Potekhin2014}. Combined with distance information, this allows constraints to be placed on the NS mass and radius, thereby constraining the EOS \citep{Ozel2016}. Broad absorption features have been observed in the spectra of some XDINSs, whose origin is still under discussion, offering clues about physics under extreme conditions \citep{Haberl2007ApSS, 2009ASSL..357...91B}.

Among older NSs, MSPs form a distinct and particularly interesting class. These are rapidly rotating, old NSs that have been spun up through accretion in binary systems. While they are expected to be cool due to their age, many MSPs exhibit higher surface temperatures than predicted by standard cooling models \citep{Zavlin1996,Bogdanov2009ApJ}. 
A prominent example is the nearby MSP, PSR~J0437$–$4715, whose thermal emission has been extensively studied in both the X-ray and ultraviolet bands, providing important constraints on its surface temperature, mass, and radius~\citep{Durant2012, Guillot2016, Gonzalez2019, Choudhury24}.

The composition of the NS core has a profound impact on its cooling processes. For a purely nucleonic star (composed of neutrons, protons, electrons, and muons), the $np$ dUrca process is only possible if the proton fraction exceeds a critical value. This threshold is sensitive to the nuclear symmetry energy, particularly its slope $L$ at saturation density. A larger $L$ generally favors a higher proton fraction and a lower threshold mass for the dUrca process (e.g.,~\citep{2011PhRvC..84f5810C,2017IJMPE..2650015L,2019PTEP.2019k3E01D,2024PhRvC.110a5805L,Mendes:2021tos}). Current constraints ($L \lesssim~60 \rm{MeV}$) suggest that the $np$ dUrca process may be unlikely in many NSs~\citep{Lattimer2013, Oertel2017}.
However, the presence of non-nucleonic particles, such as hyperons or $\Delta$ isobars, can significantly alter the theoretical framework. These particles can participate in dUrca processes, often with lower threshold densities than the $np$ dUrca process \cite{Providencia:2018ywl,Fortin:2020qin,Fortin:2021umb}. Therefore, the observation of rapid cooling (i.e., a low X-ray luminosity for a given age) in a relatively low-mass NS could be a strong indication of presence of exotic matter. 
For NSs containing non-nucleonic particles, the threshold mass for the dUrca process is related to the interaction between these non-nucleonic particles. Similarly, if present in NSs, quark matter would also participate in dUrca processes and their cooling signature could be observable and even constrain the quark-hadron phase transition density~\citep{2006NuPhA.774..815B, 2012PhRvC..85c5805N, 2020MNRAS.498..344W, Mendes:2024hbn}. 

While cooling processes tend to decrease the temperature of a NS, the relatively high temperatures observed in old NSs indicate the presence of significant heating mechanisms \cite[e.g.,][]{Rangelov2017}.
Theoretical models must therefore incorporate not only cooling processes but also heating mechanisms, such as rotational energy dissipation (particularly important for MSPs), magnetic field decay (dominant in magnetars like SGR~1806-20), and vortex creep (relevant for older, slower pulsars). 

More observational data from diverse sources (isolated cooling NSs, magnetars, MSPs, XDINs and qLMXBs), including new objects, could be observed by eXTP-SFA thanks to its high flux sensitivity \cite{Beloin2018_PRC97-015804}. By comparing detailed theoretical models, which incorporate a range of cooling and heating scenarios, we can constrain the microphysics of dense matter, including the EOS, the presence of exotic particles, superfluidity, magnetic field evolution, and the properties of the crust and envelope. 

As previously noted, in the study of NS thermal evolution, the distance to NSs remains one of the most uncertain parameters, presenting significant challenges in distinguishing between competing theoretical models. We further discuss distance determination in Sect.~\ref{distance}.

\section{Spin measurement 
}
\label{Sec:Spinmeas}

NSs with the fastest spins constrain the EOS since the limiting spin rate, at which the equatorial surface velocity is comparable to the local orbital velocity and mass-shedding occurs, is a function of mass and radius. Softer EOS have smaller radius for a given mass, and hence have higher limiting spin rates. More rapidly spinning NSs - especially a sub-millisecond spin period if one could be found - would place increasingly stringent constraints on the EOS \citep{Haensel90}.
However, the current record holder (the MSP PSR~J1748$–$2446ad in the Globular Cluster Terzan 5), which spins at 716 Hz~\citep{hessel2006}, does not rotate rapidly enough to rule out any EOS models.
More rapidly spinning NSs could very well be discovered in future radio surveys~\citep{2015aska.confE..43W}. However, since the standard formation scenario for MSPs involves accretion \cite{Alpar82,Radhakrishnan82,Bhattacharya91}, searches for fast-rotating NSs among accretion-powered compact objects, i.e., in the X-ray domain, is an crucial avenue for the discovery of the next record holder. 

The theory of the origin of MSP has long suggested that accretion could spin stars up close to the break-up limit~\citep{1994ApJ...423L.117C}. Until now, interestingly, the spin distribution of MSPs does not shown the cutoff that is seen in the current sample of accreting NSs~\citep{Watts:2016uzu}. Since eXTP would have a larger effective area than other soft X-ray timing missions, it is well-suited to measure more NS spins, using both accretion-powered pulsations and perhaps also burst oscillations \citep{2008ApJS..179..360G,2012ARA&A..50..609W}. From RXTE observations, only some accreting NSs show accretion-powered pulsations, perhaps due to either the geometric alignment \citep{2009ApJ...705L..36L} or a complex magnetic field geometry \citep{Das22,Das24}. Searches for weak pulsations can exploit the blind search techniques used for the \textit{Fermi} pulsar surveys \cite{2006ApJ...652L..49A,2011PhRvD..84h3003M,2012ApJ...744..105P}, which compensate for orbital Doppler smearing. eXTP will also be able to detect burst oscillations in individual type I X-ray bursts to amplitudes of 1\% rms in the burst tail (rise) assuming a 1\,s integration time; by stacking bursts, the sensitivity improves.

Another approach is to search for weak accretion-powered pulsations using semi-coherent search techniques. These methods divide the data into short segments (typically a few hundred seconds) and perform a coherent search within each segment \cite{2015ApJ...806..261M, 2018ApJ...859..112P}. The results from longer time intervals (of tens to hundreds kiloseconds) are then combined to improve the signal-to-noise ratio. In this way, the sensitivity is greatly enhanced compared to incoherent search methods (such as power spectra), while keeping the computational cost manageable.

The sensitivity of a single 10-ks observation taken by eXTP-SFA can be calculated in three cases: a faint source (0.5–10~keV flux of approximately $2\times10^{-10}\,\rm erg\, s^{-1}\,cm^{-2}$), a moderately bright one ($3\times10^{-9}\,\rm erg\,s^{-1}\,cm^{-2}$), and a bright source ($2\times10^{-7}\,\rm erg\,s^{-1}\,cm^{-2}$). These values correspond approximately to the fluxes observed in sources such as the AMXP XTE~J1807–294 (faint), Aql~X-1 (moderate), and Sco~X-1 (bright).
We adopted a coherent segment length of 512 seconds and targeted a signal-to-noise ratio of 5 for the pulsations. Under these conditions, we find that it is possible to place upper limits on the pulsed fraction of 0.9\%, 0.3\%, and 0.03\% for faint, moderate, and bright sources, respectively.

\section{Timing studies of pulsar interiors 
}
\label{Sec:Timing}

Apart from the X-ray lightcurve timing analyses with eXTP (see Sect.\ref{Sec:Disk} and Sect.\ref{Sec:Spinmeas}), this section explores timing studies of pulsars. A detailed discussion of the eXTP timing capabilities is provided in the WG3 White Paper~\cite{WP-WG3}. Continued timing observations will help identify more glitching and precessing pulsars, expanding the sample size and diversity, and enabling studies on shorter timescales. This enriched dataset will offer valuable insights into superfluid dynamics, crust-core coupling, and internal torques, thereby advancing our understanding of the internal structure and viscosity of NSs.

\subsection{Timing analyses of pulsars}

Pulsar timing is a technique allowing precise measurements of the pulsar rotational and astrometric parameters. For RPPs, the emitted radiation originates from its rotational kinetic energy, thus the pulsar spins down gradually due to rotational energy loss~\cite{1979Natur.277..437T}. To obtain a timing solution for a pulsar, a monitoring program with dozens of sessions over several years permits us to obtain the times of arrival (ToAs) of the pulsations. Subsequently, fitting a timing model to the ToAs yields rotational and astrometric parameters, including the pulsar spin frequency, spin-down rate, position and proper motion. Assuming magnetic dipole braking, the surface magnetic field strength and characteristic age can also be deduced. Combined with surface temperature measurements, the characteristic age can be connected to NS interior physics through NS thermal evolution. Therefore, timing techniques are very important for research on pulsar spin evolution and NS interior physics.

The standard procedure for pulsar timing is as follows. Using software such as \textsc{Tempo2}~\cite{2006MNRAS.369..655H} or \textsc{PINT}~\cite{Luo2021}, one can perform a simple fitting of the ToAs to obtain the rotational parameters and the timing residuals. The model adopted for timing fitting is a Taylor expansion of the rotation phase $\phi(t)$, i.e.,
\begin{equation}
   \phi(t) = \phi_0 + \nu_0(t - t_0) + \frac{1}{2}\dot{\nu}_0(t - t_0)^2 + \frac{1}{6}\ddot{\nu}_0(t - t_0)^3 + \cdots, 
\label{Eq:Tming_model}
\end{equation}
where $\nu$ is the rotational frequency, $\dot{\nu}$ and $\ddot{\nu}$ are its first and second derivatives. 
$\phi(t_{\rm 0})$ is the pulse phase at the reference time $t_{\rm 0}$, generally we assume $\phi(t_{\rm 0}) = 0$. The rotational frequency changes with time in the form:
\begin{equation}
\nu(t)=\nu_{0}+\dot\nu_{0}(t-t_{0})+\frac{1}{2}\ddot\nu_{0}(t-t_{0})^{2}+\frac{1}{6}\dddot\nu_{0}(t-t_{0})^{3}+\delta\nu,
\end{equation}
where $\delta\nu$ is the residual in spin frequency. If no timing irregularities occur during the data span time, the timing residual will fluctuate around zero. However, there exist two kinds of timing irregularities in pulsar timing solutions, timing noise and the glitch phenomenon~\cite{1972ApJ...175..217B}.

Timing noise manifests as white noise when the power is distributed uniformly across all frequencies, or as red noise processes where the timing residuals are dominated by slow, long-timescale structures. Red noise is primarily observed in young pulsars and is characterized by a continuous, low-frequency power spectrum. Over long timescales, the main sources of timing red noise can be described as random walk processes of pulse phase, spin frequency, and its first derivative~\cite{1972ApJ...175..217B}. Although some studies suggest that timing noise may be related to changes in the magnetosphere of pulsars~\cite{2010Sci...329..408L}, the physical mechanisms underlying most timing noise remain largely unexplained. The second type of timing irregularities, glitches, is the focus of the next subsection.

\begin{figure*}
\centering
\vspace{-1cm}
\includegraphics[width=0.95\textwidth]{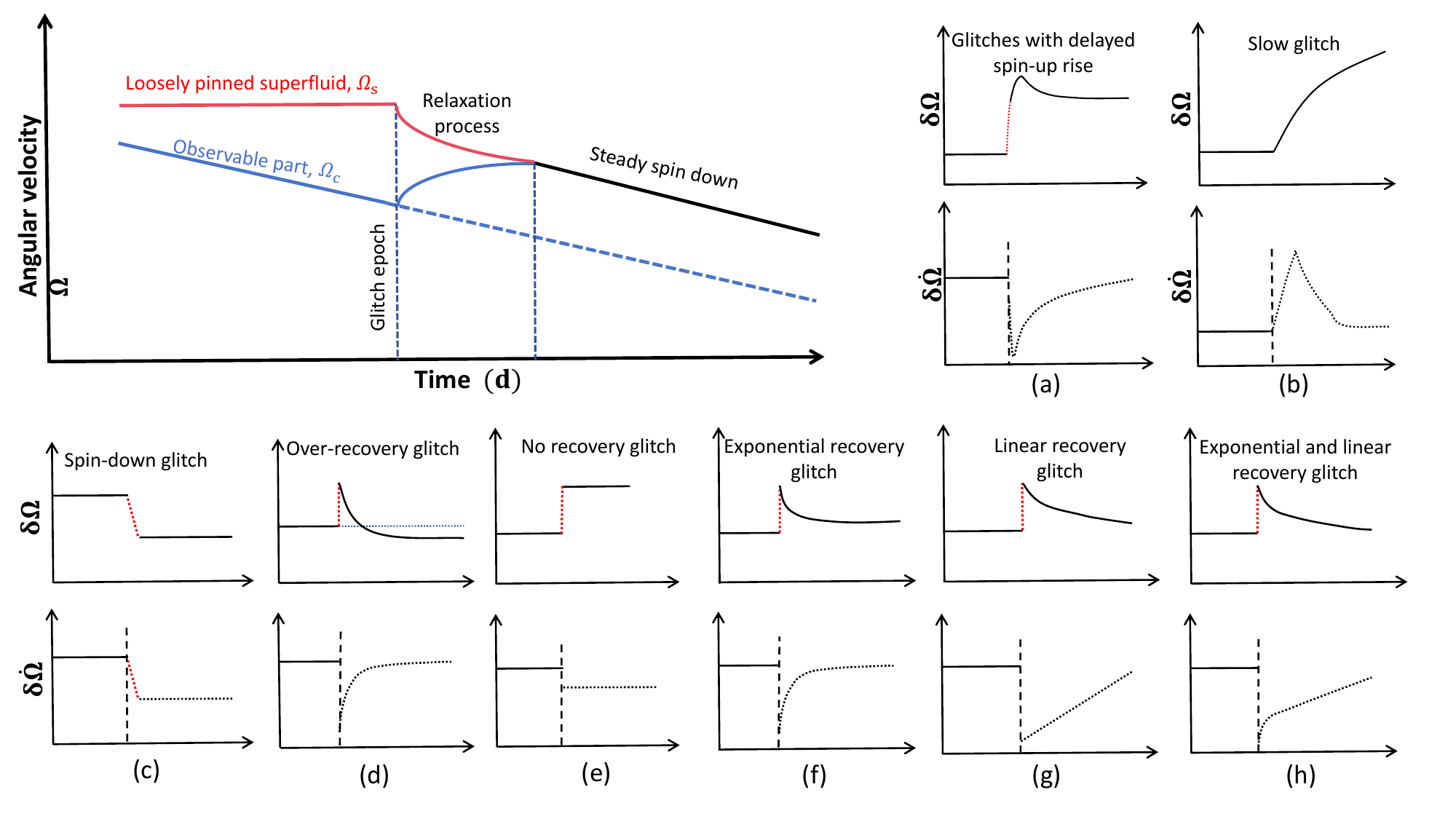}
\vspace{-0.75cm}
\caption{A sketch of various glitch patterns. Panel (a): the glitch with a delayed spin-up is characterized by a fast rise followed by a day-long timescale extended component. Panel (b): the slow glitch is characterized by a slow increase in spin frequency and a decrease in spin down rate over a timescale of several months to years. Panel (c): the spin-down glitch is usually characterized by a step decrease in spin frequency, but recently, the rapid spin down process of a spin-down glitch from SGR~J1935+2154 has been observed~\cite{2024Natur.626..500H}. Panel (d): the over-recovery glitch is usually characterized by an exponential increase in spin down rate and a decrease in spin frequency following the fast glitch rise, resulting in a net spin down eventually. Panel (e): subtracting a mean value of pre-glitch spin down rate, the no recovery glitch shows no obvious trend towards its pre-glitch value. Panel (f): the post-glitch spin frequency and spin down rate could be fitted by one or several exponential decaying component(s) and the permanent change component. Panel (g): the post-glitch spin down rate recovers linearly towards the pre-glitch value. Panel (h): the post-glitch spin down rate recovers exponentially and then linearly towards the pre-glitch value. For panels (a)-(h), the black solid lines and dotted lines represent what we have observed in spin frequency and spin down rate, the vertical dashed lines represent the glitch epoch, while the red dashed lines represent the unresolved spin up process. 
}
\label{fig:glitch}
\end{figure*}

\subsection{Mechanisms of glitch trigger and recovery}
\label{sec:glitches}

The glitch phenomenon is characterized as an abrupt increase or decrease in the pulsar rotational frequency, generally accompanied by an increase in the absolute value of the pulsar spin down rate. Glitches are mainly classified into four categories according to their different manifestations~\cite{2022Univ....8..641Z}, namely, glitches with unresolved spin-ups (including no recovery glitches, exponential recovery glitches, and linear recovery glitches), glitches with delayed spin-ups, slow glitches, and spin-down glitches (sometimes named anti-glitches), as shown in Fig.~\ref{fig:glitch}. 

The detection of glitch events within an observing period is achieved through timing residual analysis. A typical glitch is confirmed when the spin period exhibits abrupt discontinuities accompanied by persistent spin-up signatures.
The change in pulse phase caused by glitches can usually be fitted by the following model~\cite{2006MNRAS.372.1549E}:
\begin{equation}
\label{equ:1} 
\phi_{\rm g}(t)=\Delta\phi+\Delta\nu_{{\rm p}}(t-t_{\rm g})+\frac{1}{2}\Delta\dot\nu_{{\rm p}}(t-t_{\rm g})^{2}[1-e^{-(t-t_{0})/\tau_{\rm d}}]\Delta\nu_{\rm d}\tau_{\rm d},
\end{equation}
where $\Delta\phi$ is the increment of pulse phase after the glitch, $t_{\rm g}$ is the glitch epoch, 
$\Delta\nu_{{\rm p}}$ and $\Delta\dot\nu_{{\rm p}}$ are the permanent changes in rotational frequency and spin down rate, while $\Delta\nu_{\rm d}$ is the transient change in frequency which will be recovered on the timescale $\tau_{\rm d}$. Therefore, the glitch size is given by $\Delta\nu=\Delta\nu_{\rm p}+\Delta\nu_{\rm d}$, the spin down rate increment will be $\Delta\dot\nu=\Delta\dot\nu_{\rm p}-\Delta\nu_{\rm d}/\tau_{\rm d}$. $\Delta\nu/\nu$ and $\Delta\dot\nu/\dot\nu$ are their fractional sizes respectively. $Q=\Delta\nu_{\rm d}/\Delta\nu$ is the recovery factor. The time interval between two nearby glitches is called waiting time. Currently, more than $700$ glitches have been found in about $240$ pulsars~\footnote{https://www.jb.man.ac.uk/pulsar/glitches/gTable.html~\cite{2022MNRAS.510.4049B}, and https://www.atnf.csiro.au/research/pulsar/psrcat/glitchTbl.html.}, most of which are relatively young with $\tau_{\rm c}\lesssim 20~\rm{Myr}$~\cite{2011MNRAS.414.1679E}.

There is still no consensus on the triggering mechanisms or physical origins of the glitch phenomenon. Our understanding of the overall picture remains incomplete; for example, it is unclear whether there is a connection between the stellar interior and the magnetosphere, or how the diverse types of post-glitch behavior can be consistently explained. Quantitatively describing glitch behavior remains a major challenge due to the complexity of the underlying physical processes. 
Over the years, two main models have been proposed: the crustquake model~\cite{1969Natur.223..597R, 1971AnPhy..66..816B} and the superfluid model~\cite{1969Natur.224..673B, 1975Natur.256...25A}.

The crustquake resembles earthquake on the Earth. NSs are born with high temperature and spin frequency; the equilibrium shape of a rotating liquid NS is described by a Maclaurin ellipsoid, so the NS is actually oblate. When the crustal temperature drops below the melting temperature of the crustal components, the outer crust will solidify. The solid phase of the crust hinders the adjustment of the shape and oblateness of the NS, making its oblateness changes slower with time than a liquid NS would have, therefore, stress appears and accumulates during long-term spin down process. The crust will break at one critical point~\cite{2018MNRAS.480.5511B}, resulting in an abrupt decrease in NS oblateness and moment of inertia, and a sudden increase in its spin frequency according to angular momentum conservation. 

Note that, crustquakes can account for the small glitches with $\Delta\nu/\nu\sim 10^{-8}$ in the Crab pulsar (PSR B0531+21) but not the frequent and large glitches with $\Delta\nu/\nu\sim 10^{-6}$ in the Vela pulsar (PSR J0835$–$4510) given the short waiting time of about three years~\cite{1971AnPhy..66..816B}. 
For example, the glitch size is proportional to the rigidity parameter~$b=B/(A+B)$~\cite{1971AnPhy..66..816B}. 
$B=\mu V/2$, $\mu$ is the shear modulus, $V$ is the star volume.
$A=3GM^{2}/(25R)$, $M$ and $R$ are stellar mass and radius respectively, $G$ is the gravitational constant.
However, it was found that $b$ was overestimated by a factor of about 40 when a more realistic NS structure was used~\cite{2003ApJ...588..975C}. The dependence of parameters 
$A$ and $B$ on the NS EOS and mass was also studied, revealing that $A$ had been severely underestimated~\cite{2008A&A...491..489Z}. Additionally, it was determined that the strain accumulated between two successive glitches in the Vela pulsar is insufficient to trigger a crustquake unless the occurrence of crustquakes is considered a history-dependent process~\cite{2020MNRAS.491.1064G}.
Although the crustquake model may not be supported by theoretical calculations, the possibility that crustquake serves as the trigger of glitches is not excluded~\cite{2018MNRAS.473..621A}.

The superfluid model divides the NS into two parts, the crust which couples with the external torque, including the crustal component and what is coupled to it, and the faster rotating superfluid component which is loosely pinned to the crust. The crust spins down gradually as a result of the external braking torque, while the superfluid component imitates the spin down through outward migration of the quantum vortices. The spin of these two parts is not synchronous, resulting in an angular velocity difference or {\it spin lag} between them. A glitch may be triggered when the Magnus force exerted on vortices is strong enough to overcome the pinning between vortices and pinning sites. Angular momentum will be transferred to the crust when the outward migrated vortices repin with the crust in new pinning sites, resulting in the glitch rises. Besides this, a fraction of the NS moment of inertia decouples from the charged component on which the external torque acts, resulting in an increase in the spin down rate according to braking torque conservation. In some cases, the enhanced spin down rate will relax back to a predicted value had the glitch not occurred within a timescale of days to hundreds of days. The vortex creep model attributes the post-glitch relaxation to responses of vortex creep to rotational changes~\citep{1984ApJ...276..325A}. This model has been viewed as the standard model for post-glitch description due to its success in describing Vela pulsar glitches, whose post-glitch spin down rate relaxation is characterized by at least one exponential component and a linear component~\citep{1993ApJ...409..345A}.

The vortex creep model has the capability to simultaneously explain the triggering and post-glitch relaxation behaviors of pulsar glitches~\citep{1984ApJ...276..325A,1993ApJ...409..345A,1989ApJ...346..823A}. In this model, vortices are pinned in the inner crust due to their interaction with the nuclear lattice, and the pinning energy depends on the properties of the inner crust. The pinning of vortices leads to a local lag $\omega$ between the superfluid and the normal components. Vortices can be thermal activated to overcome the pinning barriers and creep radially outward, and the continuous vortex creep results in the angular momentum transfer from the crustal superfluid to the crust. Under an external torque, the pulsar can reach a steady-state of spin-down, where the superfluid and normal components share the same spin-down rate. A local fluctuation raises the lag $\omega$ beyond a critical value $\omega_{\rm cr}$, which would trigger the vortex avalanches resulting in a glitch, because the lag $\omega$ deviates from its steady-state value. During the glitch recovery, the linear or non-linear responses of different distinct superfluid regions in the inner crust to the lag deviation determine the post-glitch relaxation behavior. The relaxation time-scales of these response regions, as well as their corresponding moments of inertia, are strongly dependent on the internal temperature and local pinning energy. Therefore, within the framework of the vortex creep model, comparisons between theoretical predictions and observations can reveal the internal structure and temperature of pulsars.

In addition, glitches can impact the pulsar emission properties, including its pulse profile and spectral characteristics. Studying how glitches affect emission can provide insights into the internal dynamic, as well as the radiation processes near the pulsar surface. Recent studies have explored the connection between pulsar glitches and emission changes. While glitches are observed in various pulsars, 
only a few show clear correlations between spin behavior and emission changes. Notable examples include glitches in PSR~J1119$-$6227, PSR~J0742$-$2822, PSR~B2035+36, PSR~J2021+4026, and PSR~J1048$-$5832, linked to shifts in radiation modes, pulse profile variations, and changes in flux. For magnetars, glitches are more strongly associated with radiation changes, often linked to outbursts and shifts in pulse profiles, believed to result from magnetic energy release or field reconfiguration in the crust. These topics are central to WG3 discussions~\cite{WP-WG3}.

\subsection{Glitching pulsars} 

\subsubsection{Statistical study}

Since the first detection of the glitch event in Vela pulsar in 1969, this phenomenon has been detected in various types of pulsars~\cite{1969Natur.222..229R,1969Natur.222..228R}, including isolated normal pulsars, MSPs, binary pulsars, magnetars, and accretion-powered pulsars~\cite{SasmazAG2014,SerimSDS2017,WeisbergNT2010}. It is worth noting that most glitches occur in relatively young pulsars, with characteristic ages ranging from $10^{3}$ to $10^{5}$ yr~\cite{2013MNRAS.429..688Y}.
Exceptions include MSPs J0613$-$0200 and J1824$-$2452 (B1821$-$24), which have exhibited minor glitches with fractional sizes around $\sim10^{-12}$~\cite{CognardB2004, McKeeSLC2016}, magnetars such as 1E 2259+586~\cite{SasmazAG2014}, the accretion-powered pulsar SXP 1062~\cite{SerimSDS2017}, the ultraluminous X-ray pulsar NGC 300~X-1~\cite{Ray2019}, and binary pulsars like PSR~J1915+1606~\cite{WeisbergNT2010}. 
Studies have shown that there seems to be a certain correlation between the amplitude of a glitch and the characteristic age of pulsars: larger glitches often occur in younger pulsars, while smaller glitch events are more associated with older pulsars, but this association is not yet fully clear and more observational data is needed to confirm it~\cite{2022MNRAS.510.4049B}.

The range of relative glitch sizes ($\Delta{\nu}/\nu$) distribution is relatively wide, from a minimum of $2.5 \times 10^{-12}$~\cite{McKeeSLC2016} to a maximum of $1.37 \times 10^{-3}$~\cite{SerimSDS2017}. Statistical results indicate that both the amplitude ($\Delta{\nu}$) and relative amplitude of the glitch exhibit a bimodal structure, which can be well fitted by the double Gaussian model~\cite{2021MNRAS.508.3251L}. 
A study of 543 glitch sizes found that the peaks of the two Gaussian components occur at 0.032 $\mu$Hz and 18 $\mu$Hz, respectively~\cite{2022MNRAS.510.4049B}. 
According to the distribution characteristics of their amplitudes, glitches can be categorized into large glitches ($\sim 10^{-6}$) and small glitches ($\sim 10^{-9}$)~\cite{2017A&A...608A.131F}. 
The differing distributions of large and small glitch sizes suggest that these two types of glitches may involve different physical mechanisms~\cite{2017A&A...608A.131F,2022MNRAS.510.4049B}.

The typical glitch recovery process usually displays an exponential decay, with a timescale ranging from 10 to 300 days. In some cases, the spin-down rate may present a linear recovery process~\cite{2010MNRAS.404..289Y,2022MNRAS.510.4049B}. Note that there may be more than one exponential recovery process following the glitch. For example, the Vela pulsar glitch 16 (MJD $\sim$ 51559) featured four exponential recovery phases, marking the highest number observed in a single glitch~\cite{2002ApJ...564L..85D,2013MNRAS.429..688Y}.

Except for the two special cases of Vela pulsar and PSR~J0537$-$6910, the inter-glitch intervals of other pulsars seem to be random. 
A study based on the self-organizing criticality model conducted a statistical analysis of glitch sizes and waiting times of nine pulsars~\cite{2008ApJ...672.1103M}. The results indicated that the glitch sizes of seven of these pulsars followed a power-law distribution, while the waiting times adhered to an exponential distribution. However, for Vela and PSR~J0537$-$6910, which exhibit quasi-periodic glitch behaviors, both glitch sizes and waiting times were better characterized by a Gaussian distribution. 
Reanalysis of eight prolific pulsars with the latest data showed that, aside from Vela and PSR~J0537$-$6910, the size distributions of the glitches of the remaining six pulsars were best described by power laws, exponential or log normal functions, while waiting-time distributions were best fitted an exponential function~\cite{2019A&A...630A.115F}.
When considering glitch phenomena over a longer timescale, often referred to as `glitch clusters', it is assumed that the angular momentum accumulated in the superfluid reservoir is not fully released following a glitch. For different pulsars, both glitch size and waiting time follow a unified Gaussian distribution, and the long timescale period of glitches exhibits a significant linear relationship with the characteristic age of the pulsar. This indicates a statistical basis for a common, unique physical mechanism underlying glitches across various pulsars, and that all pulsars exhibit long timescale periods for glitches rather than a purely random occurrence~\cite{2025ApJ...978...49Z}. 

\subsubsection{Constraining EOS with fractional moment of inertia}

Theoretical research on the interior physics of NSs through glitch phenomenon centers mainly around two aspects, the dynamical evolution of post-glitch recovery process, and the fractional moment of inertia (FMoI) of crustal superfluid, determined from frequently glitching pulsars.

The FMoI of crustal superfluid is tightly connected with the parameter $\dot\nu_{\rm g}/|\dot\nu|$, where $\dot\nu_{\rm g}$ is defined as the glitch activity. For one specific pulsar, glitch activity has the form $\dot\nu_{\rm g}=(\sum_{i}\Delta\nu_{i})/\Delta T$~\citep{2000MNRAS.315..534L}, where $\Delta T$ is the total observation time over which this pulsar has been searched for glitches. $\dot\nu_{\rm g}/|\dot\nu|$ represents the fractional spin down rate of the pulsar reversed by glitches. There are two ways to constrain the NS EOS with FMoI. 

First, the crustal superfluid exists in the highest density region of the crust, so the FMoI of crustal superfluid is almost equal to the FMoI of the NS crust. Since the latter is determined by NS EOS~\citep{1994ApJ...424..846R}, the FMoI of crustal superfluid determined by pulsar glitches can be used to place constraints on NS EOS.

Assuming the angular momentum reservoir is the crustal superfluid that coexists with the inner crust lattice, a connection can be made between the FMoI of the crustal superfluid (for Vela and six other pulsars) and the NS EOS through the relation~\citep{1999PhRvL..83.3362L}:
\begin{equation}
\label{FMoI lower limit}
I_{\rm c}/I\geq I_{\rm res}/I\geq \dot\nu_{\rm g}/|\dot\nu|, 
\end{equation}
where $I_{\rm c}$ and $I$ are moments of inertia of NS crust and the whole star, respectively, $I_{\rm res}$ is moment of inertia of the angular momentum reservoir.
Based on the structure of a NS, the ratio $I_{\rm c}/I$ depends uniquely on the mass and radius for any given EOS, and provides a lower limit on the radius for a given mass.
This method is only suitable for pulsars which glitch regularly and frequently and transfer angular momentum steadily, while not for others such as the young Crab pulsar. In addition, it is highly dependent on the assumption of the crustal angular momentum reservoir. Whether the core superfluid gets involved in glitches is still an open question~\citep{2012PhRvL.109x1103A, 2012PhRvC..85c5801C}.

Second, FMoI could constrain the NS EOS when combined with NS thermal evolution~\citep{2015SciA....1E0578H}. Thermal evolution of NSs are influenced by many NS properties, including the superfluid models, the internal heating mechanisms, the crustal composition, as well as its mass and the EOS, which are critically important. The FMoI of the angular momentum reservoir considers only the parts whose temperature is lower than the local superfluid critical temperature (density dependent), so it is also determined by the EOS and superfluid models. Therefore, for those frequently and regularly glitching pulsars with measured $\dot\nu_{\rm g}/|\dot\nu|$ (equal to the definition $G=2\tau_{\rm c}A$ in~\citep{2015SciA....1E0578H}) and precise ages, there will be connections between $\dot\nu_{\rm g}/|\dot\nu|$ and surface blackbody temperature. To match the observations of FMoI of the crustal superfluid and surface blackbody temperature at their current ages simultaneously, the NS mass, EOS and superfluid models should be constrained to some extent. The involvement of multiple such NSs with measured $\dot\nu_{\rm g}/|\dot\nu|$ and surface blackbody temperature will place more tight constraints on these internal physics parameters.

Based on the above idea, the masses of nine glitching pulsars, including the Vela pulsar and the prolific PSR~J0537$–$6910, were measured by comparing the observed surface temperatures and FMoI of the crustal superfluid at their current ages~\citep{2015SciA....1E0578H}.
The superfluid models considered include those confined to the crust as well as those extending into the core. 
The surface temperatures of these pulsars were taken from spectral fitting.

Third, there is an obvious difference between the post-glitch behaviors of the Crab pulsar and other young RPPs, i.e., the existence of a large persistent shift, which is an increase in the spin down rate of the pulsar compared with the linear trend of the pre-glitch spin down rate~\cite{1977AJ.....82..309G,1981A&AS...44....1L,1983MNRAS.202..437D}. 
This phenomenon has been observed in all relatively large glitches of the Crab pulsar~\cite{2015MNRAS.446..857L}, perhaps in all of its glitches. Nevertheless, large persistent shifts have not been observed in the frequent glitching Vela pulsar~\cite{2019Ap&SS.364...11X} or other RPPs. Theoretically, starquake induced external torque variation has been proposed as an explanation~\cite{1992ApJ...390L..21L, 1996ApJ...459..706A, 1997ApJ...478L..91L}. However, if this theory is correct, it will be hard to understand why starquakes do not happen in the other more than 200 glitching pulsars~\footnote{https://www.jb.man.ac.uk/pulsar/glitches/gTable.html~\cite{2022MNRAS.510.4049B}}. Therefore, the persistent shift in spin down rate of the Crab pulsar could be unique to Crab-like young pulsars and correspond to one specific physical process, such as superfluid decoupling~\cite{2012NatPh...8..787H} or anything else. Systematic analysis on the physical origin of this phenomenon could reveal structural differences between Crab-like young pulsars and Vela-like mature pulsars and further inspire new ideas to constrain the interior physics of NSs.

\subsubsection{Constraining the interior physics of NS with the dynamical evolution of post-glitch recovery process}

As stated above, the standard superfluid model includes only the contribution of crustal angular momentum reservoir~\cite{1984ApJ...282..533A}, and the following post-glitch relaxation process represents the responses of vortex creep to glitch-induced rotational changes~\cite{1984ApJ...276..325A}. According to the vortex creep model, vortex creep responds to changes in the rotation rate of the crust, the different relaxation timescales characterize physically distinct pinning regions~\cite{2019MNRAS.488.2275G}. The relaxation timescale is proportional to the internal temperature of NSs, so studies of post-glitch behaviors offer another way to determine the internal temperature of NSs~\cite{2021arXiv210603341A}. On the other hand, the additional negative torque exerted on the crust and plasma by vortex creep is proportional to the moment of inertia that got involved in the glitch~\cite{1984ApJ...276..325A, 2019MNRAS.488.2275G}, therefore, studies of post-glitch evolution provide the lower limit of the total moment of inertia needed for the glitch. Note that the total moment of inertia that gets involved in glitches is generally consistent with the FMoI of crustal superfluid of prolific glitching pulsars discussed above, except for that of the young Crab pulsar~\cite{2019MNRAS.488.2275G}. Nevertheless, although this method is physically meaningful, it probably does not provide more information about NS structure or the EOS.

In recent years, one aspect of the glitch phenomenon which attracted much attention is spin evolution during the earliest stage of the glitch, including the delayed spin up phenomenon of the Crab pulsar and the overshoot phenomenon of the Vela pulsar glitches. The fast rise of the glitch process is still not resolved, but the rise timescale has most recently been constrained to be within $12.6~{\rm s}$, through single pulse observations of the Vela pulsar~\cite{2019NatAs...3.1143A}. The overshoot phenomenon is characterized by a rapid rotational frequency increase which is higher than that of the steady state, the timescale from the overshoot to a new steady state one could be about hundreds of seconds~\cite{2019NatAs...3.1143A}. Nevertheless, for the young Crab pulsar, no overshoot has ever been observed, and six out of 30 Crab pulsar glitches have been partially resolved~\cite{1992Natur.359..706L,2001ApJ...548..447W,2020ApJ...896...55G,2018MNRAS.478.3832S,2021MNRAS.505L...6S}, i.e., the fast glitch rise process is followed by an extended component of days-long timescale. This phenomenon is named the delayed spin-up glitch (DSU glitch). There are also two DSU glitch candidates from 1E 2259+586~\cite{2004ApJ...605..378W} and the soft gamma-ray repeater (SGR) J1935+2154~\cite{2024RAA....24a5016G} respectively. The peculiar evolutions of pulsar spin frequency during the early state of a glitch in the Crab and Vela pulsars may shed light upon the vortex interaction and mutual friction~\cite{2018ApJ...865...23G,2020MNRAS.493L..98S,2020A&A...636A.101P}, which may further reflect the triggering location of a glitch, in the crust or in the core of a NS~\cite{2018MNRAS.481L.146H,2020A&A...636A.101P}.

Another aspect of the post-glitch recovery process which may be promising to constrain NS interior physics is the exponential relaxation timescale. As stated above, the relaxation timescale reflects the response of vortex creep to 
rotational change of the crust in different pinning regions in the vortex creep model~\cite{1989ApJ...346..823A}. As of today, the entrainment effect requires the core superfluid to contribute to the glitch in at least some pulsars, such as the Vela pulsar~\citep{2012PhRvL.109x1103A}. Accordingly, recent works have provided the basis of how core superfluid would contribute to glitch and post-glitch relaxation~\cite{2014ApJ...788L..11G,2016MNRAS.462.1453G}. A subsequent study~\cite{Gugercinoglu2017} showed that timescales based on vortex lines-flux tubes interaction in the outer core of NSs agree well with observed exponential relaxation timescales in certain NS EOS. Therefore, the post-glitch exponential relaxation could be a probe of NS structure. 

Linear recovery, with an approximately constant $\ddot\nu$, generally dominates the long-term recovery of a large glitch after the exponential recovery is over and persists until the next glitch event~\cite{2013MNRAS.429..688Y}. The non-linear response regime in the vortex creep model may be responsible for the linear recovery, therefore the post-glitch linear recovery has the potential to probe the interior physics of NSs, particularly the outer layer of NSs. 
Recently, a comprehensive analysis of glitches in gamma-ray pulsars was conducted~\cite{Liu2025_MNRAS537-1720} by combining Fermi-LAT and Parkes timing data, unveiling several novel phenomena within the glitch recovery process. This multi-messenger study not only enhances the effective observation cadence of the pulsars but more importantly, has the potential to fill the observational gap after the glitch occurs, thereby providing a more accurate understanding of the exponential recovery process on the short time scale following a glitch. 

While the continuously accumulating glitch observations challenge our understanding of the phenomenon, they also offer a solid foundation for developing a well-established interpretation.
The triggering, rise, and relaxation of a glitch are three sequential phases. A comprehensive interpretation should be able to account for these phenomena in a consistent manner (see e.g.,~\cite{2021ApJ...923..108S,2025ApJ...984..200T} for recent efforts). 
The detection capabilities of eXTP will enable detailed investigations into whether pulsar glitches originate from internal or external mechanisms within NSs. Moreover, eXTP is expected to place quantitative constraints on both established and emerging theoretical models of glitches, thereby contributing to more precise determinations of the NS EOS through X-ray observations. As demonstrated in simulations of X-ray pulsars presented in the WG3 White Paper~\cite{WP-WG3}, eXTP is capable of detecting glitch amplitudes as small as $\Delta \nu / \nu \sim 10^{-9}$ under typical exposure conditions. These studies will be further enhanced by accurate measurements of key NS properties, including distance, age, and surface temperature.

\subsection{Precessing pulsars 
}

Omitting higher-order effects, the total angular momentum $\bm{J} = {\bf I} \cdot \bm{\omega}$ for an isolated NS is conserved in an inertial frame, where ${\bf I}$ is the inertial tensor and $\bm{\omega}$ is the angular velocity. For an isolated asymmetric NS, the eigenvalues of the inertial tensor ${\bf I}$, corresponding to three principal axes (presented by three eigenvectors of ${\bf I}$), are different. If $\bm{\omega}$ is not parallel to any of the three principal axes of the body, free precession might happen and the spin direction rotates around the angular momentum $\bm{J}$~\cite{1969mech.book.....L}. Eventually, the body of a NS, as well as its deformation, is related to the EOS of supranuclear dense matter. Therefore, probing the precession of NSs provides a unique way in studying the EOS of dense matter~\cite{Gao:2020zcd}. This is further facilitated by the existence of quasi-universal relations between the moment of inertia and the stellar compactness~\cite{Lattimer2013, Yagi2013a, Breu2016}. While these relations may be broken by the presence of strong magnetic fields or rapid rotation~\cite{Haskell2014}, they are expected to be valid for the typical magnetizations and rotation rates in pulsars. 

The precession behaviour heavily depends on the EOS of the NS. If the interior is fluid, the precession will be dissipated quickly, as was realized early on in the hint of a possible precession observed in PSR~B1828$-$11 using different characterization hyper-parameters of its pulse profiles~\cite{Stairs:2000zz, Jones:2000ud}. Instead, if the NS is a solid star (e.g., made from strange quark clusters~\cite{Xu:2003xe}) and the dissipation is minimal, the precession can last for a long time. Therefore, the presence of precession (at its timescale) provides valuable information about NS interiors. Recent high-cadence radio observations of the magnetar XTE J1810$-$197, following an X-ray outburst, suggest a possible precessional behavior on a timescale of months~\cite{Desvignes:2024vle}. It is worth noting that whether the ${\cal O}~({\rm month})$ damping originated in the interior of the magnetar or was caused by magnetospheric changes needs in-depth investigation. Nevertheless, in such a scenario, the eXTP satellite would be helpful to capture magnetar burst and characterize their behaviour, thus informing subsequent radio observations and reducing the parameter space. 

On the other hand, eXTP might also contribute to the study of precession itself if the X-ray pulse profiles of a precessing magnetar are observed in detail~\cite{Gao:2022hzd}. Depending on the geometry and the strength of the magnetic fields, the emission geometry, as well as the interior of the magnetar, free precession behaviour encodes information onto the X-ray pulse profile shapes and, mostly importantly and bearing high observational significance, their polarization details. The profile shapes and polarization properties will change in a characteristic way as a function of time. Therefore, the eXTP satellite is uniquely placed to offer both X-ray pulse profile and X-ray polarization observations, decoding the precession behaviour. 
In data analysis, how to map parameters that describe the free precession to parameters that describe the EOS of dense matters is intricate. One may firstly aim to extract the shape parameters of the NS (e.g., whether biaxially or triaxially deformed, as well as how large is the deformation), and then combine them with theoretical studies or simulations of the NS structure (e.g., its breaking strain for the solid crust, etc). Some preliminary theoretical models are constructed~\cite{Gao:2020zcd}, but more details need to be incorporated in the future.

Finally, the damping of a pulsar free precession due to internal viscosity may leave imprints in its timing data since its magnetic tilt angle evolution is probably affected~\cite{2019PhRvD..99h3011C}. X-ray polarization observations of pulsars using eXTP may help measuring their magnetic tilt angles and the dependence of the precession on the structure of their internal magnetic fields. It has been shown \cite{Hamil2016} that more than one dipole component could explain the observed change in the angle with time. Using the measured timing data (including the period $P$, its first derivative $\dot{P}$, and the braking index $n$) and magnetic tilt angle, we can set constraints on the number of precession cycles, a quantity characterizing the mutual friction between superfluid neutrons and other particles in the NS interior~\cite{2019PhRvD..99h3011C,2002PhRvD..66h4025C,2009MNRAS.398.1869D}. The analyses have been performed for several young pulsars with measured braking indices, however, roughly determined magnetic tilt angles in various physical scenarios~\cite{2019PhRvD..99h3011C,2023RAA....23e5020H,2024JHEAp..41...13Y,2024SCPMA..6729513Y}. The results can be compared to that obtained from modeling of pulsar glitches, and they are generally consistent with each other~\cite{2018MNRAS.481L.146H,2019NatAs...3.1143A,2023RAA....23e5020H,2024JHEAp..41...13Y,2024SCPMA..6729513Y}. Therefore, by virtue of eXTP we could study the damping mechanisms of free precession and the recovery mechanisms of pulsar glitches. Moreover, if the deformation of the pulsar is caused by its internal magnetic fields, the configuration of internal fields may also be probed by using the measured timing data and tilt angle~\cite{2019PhRvD..99h3011C,2023RAA....23e5020H}. For freely precessing magnetars, measurements of the precession period and surface thermal emission could be used to probe the strength and configuration of internal fields~\cite{2025PhRvD.111b3038S}, which cannot be directly determined through observations.

\section{Multi-wavelength studies} 

\subsection{Mass} \label{mass}

\subsubsection{Radio timing of post-Keplerian parameters
}

Pulsars offer exceptional opportunities for precise mass measurements. The remarkable stability of pulsar rotation (aside from occasional glitches in some pulsars, see Sect.~\ref{sec:glitches}), combined with precise measurements of pulse times-of-arrival (ToAs), allows for the detection of subtle effects caused by gravitational interactions in binary systems~\citep{Lorimer2012}.

The foundation for mass measurements in binary systems is the binary mass function ($f_{\rm{bin}}$): 
\begin{equation}
f_{\rm{bin}}(M_{\rm c}, M_{\rm NS}) = \frac{(M_{\rm c} \sin i)^3}{(M_{\rm NS} + M_{\rm c})^2} = \frac{K_{\rm c}^3 P_{\rm{orb}}}{2\pi G} (1 - e^2)^{3/2} = \frac{4\pi^2}{GP_{\rm orb}^2}x^3
\label{eq:mass_function}
\end{equation}
where $M_{\rm NS}$ is the pulsar mass, $M_{\rm c}$ is the companion mass, $i$ is the orbital inclination, $K_{\rm c}$ is the companion radial velocity semi-amplitude, $P_{\rm{orb}}$ is the orbital period, $e$ is the orbital eccentricity, $G$ is the gravitational constant, and $x = a_{\rm p} \sin i / c$ is the projected semi-major axis of the pulsar orbit ($a_{\rm p}$ is the pulsar semi-major axis and $c$ is the speed of light).

The observable quantity from pulsar timing is $x$. While $K_{\rm c}$ is not directly measurable via radio timing, the pulsar radial velocity semi-amplitude, $K_{\rm NS} = 2\pi x / P_{\rm{orb}}$, is. The mass function alone cannot independently determine $M_{\rm NS}$, $M_{\rm c}$, and $i$; it provides only a lower limit on the pulsar mass. If both $K_{\rm NS}$ and $K_{\rm c}$ are measurable (e.g., $K_{\rm c}$ from optical spectroscopy of the companion), the mass ratio $q = M_{\rm NS}/M_{\rm c} = K_{\rm c}/K_{\rm NS}$ becomes known. Combining this with the mass function allows for the determination of both masses.

To overcome the limitations of the mass function, we use post-Keplerian (PK) parameters. The five classical PK parameters measurable through pulsar timing are: the orbital period derivative ($\dot{P}{\rm{orb}}$), the advance of periastron ($\dot{\omega}$), the gravitational redshift and time dilation parameter ($\gamma$), the Shapiro delay range ($r$), and the Shapiro delay shape ($s$). Measuring two or more PK parameters, along with the mass function, enables the determination of $M_{\rm NS}$, $M_{\rm c}$, and $i$.

The Shapiro delay is particularly valuable. Its shape parameter, $s = \sin i$, directly provides the orbital inclination, while the range parameter, $r$, is proportional to the companion mass. The Shapiro delay is most pronounced in systems with massive companions and edge-on orbits~\citep{Lorimer2012}. Moreover, measuring multiple PK parameters also provides a test of general relativity, with any discrepancies potentially indicating deviations from general relativity (GR).

Through Shapiro delay measurements, we have discovered some of the most massive NSs known, including PSR~J1614$–$2230 and PSR~J0740+6620~\citep{Demorest2010, Arzoumanian2018, Cromartie2020, Fonseca2016, Fonseca21}. These measurements offer crucial constraints on the EOS of dense matter.

These discoveries of massive NSs have profound implications for nuclear physics. The EOS of dense matter at supranuclear densities remains one of the most significant open questions in modern astrophysics. NSs with masses approaching (or potentially exceeding) $2 M_{\odot}$ provide strong constraints, ruling out many ``softer" EOS models that predict lower maximum NS masses~\citep{Lattimer2012,Ozel2016}. The existence of these pulsars necessitates relatively stiff EOS models that can support such high masses against gravitational collapse.

Beyond individual mass measurements, the population of measured NS masses is also crucial. A statistically significant sample of precise mass measurements allows for investigations into NS formation mechanisms and evolutionary pathways. For instance, different supernova core-collapse scenarios, or subsequent accretion processes in binary systems, might lead to distinct NS mass distributions~\citep{Antoniadis2016, Tauris2017ApJ, Shao2020}.

\subsubsection{Combined radio timing and optical (dynamical) observations 
} 

While radio timing provides the mass function $f_{\rm{bin}}$ and thereby constrains a combination of masses and orbital inclination, many NS binaries also contain an optically visible companion star whose dynamical signals can decisively break the degeneracy. In particular, the companion star can often be observed, supplying two key measurements, $K_c$, and orbital inclination $i$ \citep[][ and references therein]{Kennedy22}. Time‐resolved spectroscopy of the companion yields its radial‐velocity semi‐amplitude $K_c$. One cross‐correlates absorption lines in each orbital phase with template stellar spectra, thereby measuring the Doppler shifts caused by the companion orbital motion \cite{2016ApJ...828....7R,2018ApJ...864...15K,Kennedy22}. Multi-band photometry of the companion over a full orbit also provides its brightness and temperature variations—crucial for determining the orbital inclination $i$. In tight binaries such as ``spider" pulsars (with a low-mass companion nearly filling its Roche lobe), the pulsar wind strongly heats one side of the companion, producing large brightness contrasts from day side to night side. Modeling these phase‐resolved fluxes (and accounting for effects like Roche‐lobe deformation, gravity darkening, and atmospheric limb darkening) tightly constrains $i$ \cite{2016ApJ...828....7R,2018ApJ...864...15K,Kennedy22}.

Once $K_c$ and $i$ are known, they can be combined with the mass function (Eq.~ \ref{eq:mass_function}) to deduce both the NS mass $M_{\rm{NS}}$ and companion mass $M_{\rm{c}}$. This method has been applied to several ``spider" pulsars~\citep{Linares18,Kennedy22,romani2021psr,Sen24}. Such precision optical measurements therefore complement the radio approach by removing degeneracies inherent in the mass function alone and have revealed a growing sample of massive NSs.

\subsection{Distance 
} \label{distance}

Knowledge of the distance to a NS may be necessary to infer the NS properties and provide constraints on the EOS, but precise distance measurements are often challenging to obtain from observation. The most reliable method is parallax measurement, which uses the Earth's annual motion around the Sun to detect shifts in the apparent position of a source. This can be done at optical wavelengths or, more precisely, at radio frequencies using Very Long Baseline Interferometry (VLBI). 

For radio pulsars, a second type of parallax can be measured through timing observations, where distances are derived from variations in pulse arrival times caused by the curvature of the incoming wavefront as Earth moves in its orbit. Unlike imaging parallax, the precision of timing parallax is highest for pulsars at low ecliptic latitudes and lowest near the ecliptic pole \citep[][ and references therein]{2011A&A...528A.108S}. 

In addition, binary pulsar systems offer another method for distance determination through secular or annual variations in orbital parameters. However, these methods are typically limited to relatively nearby sources. Advances in telescope baseline lengths and sensitivity, such as those provided by the Square Kilometre Array (SKA), promise significant improvements in precision, enabling more accurate distance measurements~\cite{2011A&A...528A.108S,2009IEEEP..97.1482D}. Another proposed method is the neutral hydrogren (H I) kinematic distance technique, uses the absorption of pulsar emission by Galactic H I gas, with distances inferred from a Galactic rotation model \citep[][ and references therein]{2012ApJ...755...39V}. Finally, the methods of distance measurements could be combined, such the timing, optical image and VLBI to improve the accuracy utilizing different facilities~\cite{2011A&A...528A.108S,2018ApJ...864...26J}.

\section{X-ray study of dense matter in the multi-scale and multi-messenger era 
}
\label{Sec:MM} 

The study of dense matter is a challenging task that requires theoretical models to extrapolate from matter at nuclear densities to the largely unknown regimes of high-density, neutron-rich or exotic compositions. Additionally, it depends on identifying the relevant degrees of freedom, which may range from nucleons to exotic particles, and even dark matter particles. 
One generally assumes that there is one theoretical model that can correctly explain the nuclear matter data of different physical situations obtained in both laboratory nuclear experiments and astronomical observations. 
It is necessary to combine efforts from different communities and discuss mutual interests and problems~\cite{lattimer2021,schatz2022horizons}. Experimental nuclear physics provides critical data needed to benchmark theories of dense matter EOS residing in pulsars. The current and upcoming multimessenger observatories (e.g., LIGO/Virgo/KAGRA, FAST, SKA, LHAASO, HUBS, QTT)~\cite{eosbook} will continue improving detection of NSs together with the precise measurements of their global properties, as well as their dynamical evolution. Laboratory experiments~\cite{okuno2020present,tamura2012strangeness,guan2021compact,hong2023status,lovato2022long,2025arXiv250315575N} will provide an emerging understanding of dense matter EOS, together with the properties of strong interactions and the transition to deconfined quark matter. 
In this section, we outline recent progress and future perspectives along this line.

\subsection{Study from binary mergers and their electromagnetic counterparts} 
The detection of GW170817~\cite{LIGO2017} by LIGO/Virgo firmly established gravitational wave astronomy as a revolutionary tool in nuclear astrophysics, while its accompanying short GRB and kilonova observations~\cite{LIGO-P1700294} ushered in a new era of multi-messenger astronomy. Together, these breakthroughs marked a watershed moment in probing matter under extreme densities. 
With the upcoming upgraded and new facilities, including third-generation gravitational wave observatories like Cosmic Explorer~\cite{CE2019} and Einstein Telescope~\cite{ET2020}, many more multi-messenger observations of NS mergers are expected in the coming decades. These observations promise to provide deeper insights into dense matter and place unprecedented constraints on the EOS and the underlying nuclear interaction models~\cite{Somasundaram:2024ykk}. 
In this section, we briefly outline several perspectives on this aspect.

First, we focus on the gravitational waveform during the inspiral phase, which provides a direct measurement of tidal deformability. 
In a coalescing system containing a NS, the NS is tidally deformed by its companion, transferring part of the orbital energy into the stellar deformation which accelerates the orbital decay. 
This effect leads to a correction in the evolution of the gravitational wave phase, which depends on the quantity~\cite{Flanagan2008}
\begin{equation}
   \tilde \Lambda = \frac{16}{13}\frac{(M_1+12M_2)M_1^4}{(M_1+M_2)^5}\Lambda_1+(1\leftrightarrow2),
\end{equation}
where $M_1$ and $M_2$ are the component masses. $\Lambda_1$ and $\Lambda_2$ are the corresponding tidal deformabilities for each component, characterizing how easily the NS can be deformed. 
The tidal deformability depends on the internal structure of NS and is therefore closely linked to the EOS. 
The analysis of GW170817 by the LIGO-Virgo collaboration provided an estimate of $\tilde\Lambda \leq 800$ for a low-spin prior (assuming a dimensionless spin parameter $\chi<0.05$), which was subsequently translated into a constraint of $\Lambda_{1.4} \leq 800$ for a $1.4\,M_\odot$ NS~\cite{LIGO2017}. 
These results have been widely used to investigate the nature of matter at supranuclear densities~\citep{LIGO2018}. 
Multiple stacked events with advanced LIGO and third-generation interferometers will improve these NS EOS constraints.
Moreover, the improvement in detection precision is also expected to help distinguish higher-order tidal effects, such as resonant tides, facilitating the exploration of exotic states of matter inside NSs, including phase transition~\cite{Most2018b, Miao2024} and superfluidity~\cite{Yuhang2017}.

In the post-merger phase (not probed by current gravitational wave facilities), determining the fate of the remnant is a crucial question, as it is closely linked to the maximum mass of a NS, $M_{\rm TOV}$. 
Any EOS model that cannot support this mass will be ruled out. In very low-mass systems where the remnant mass is below this threshold (i.e., $M_{\rm rem} < M_{\rm TOV}$), a stable massive NS is expected to form. 
In contrast, sufficiently massive binaries will undergo a prompt collapse into a black hole after the merger, while intermediate-mass systems may go through a phase as either a long-lived supramassive NS (SMNS) or a short-lived hypermassive NS (HMNS).
While gravitational wave observations alone do not directly indicate which scenario applies to GW170817, its electromagnetic (EM) counterpart provides valuable clues about the post-merger fate. 
The detection of the short gamma ray burst (GRB170817A), suggests a delayed collapse of the merger remnant into a black hole, disfavoring a prompt collapse scenario~\cite{Ruiz2018,Gill2019}. 
However, it remains uncertain whether the remnant evolved into an SMNS or an HMNS. The prevailing interpretation favors a short-lived remnant, as a long-lived remnant would have injected around $10^{52}$ erg of rotational energy into the ejecta, which is inconsistent with observations~\cite{Margalit2017}. Based on this short-lived remnant scenario and the estimated total remnant mass, an upper limit on the maximum mass is suggested with $M_{\rm TOV} \lesssim 2.3 M_{\odot}$~\cite{Margalit2017,Ruiz2018,Shibata2019}. 
Nonetheless, some studies argue that a long-lived NS remnant remains compatible with multi-messenger observations within the magnetar model for short GRBs~\cite{Piro2019}. 
This alternative interpretation suggests a stiffer EOS, implying a lower limit on the maximum mass, $M_{\rm TOV} \gtrsim 2.2 M_{\odot}$~\cite{Most2020d}. 
Looking ahead, direct detections of post-merger GWs from the remnant can provide a definitive answer regarding its fate, offering a robust constraint on the maximum mass of NSs. 

If detected, a kilonova can provide complementary insights into the properties of dense matter. Kilonovae are powered by the radioactive decay of heavy elements synthesized in the ejected matter, which is expelled from the system by tidal forces, shocks, or disk winds. Their observational signatures, such as peak luminosity, temperature, and the time at which the emission peaks, are primarily determined by the properties of the ejecta, including mass, velocity, and electron fraction.
These ejecta properties are closely correlated with the binary parameters, particularly the total mass, mass ratio, and the NS EOS. This connection makes kilonova observations a valuable tool for inferring the properties of dense matter. Indeed, there have been studies with this goal utilizing the data from the AT2017gfo kilonova event~\cite{Margalit2017}, which was observed several hours after the merger associated with GW170817.
Moreover, kilonova observations can also offer clues about the fate of the merger remnant. For instance, some recent studies suggest that the late-time emission from AT2017gfo is more consistent with energy injection from a long-lived NS remnant~\cite{AiSK2018}. 
If this interpretation is correct, it would imply a rather stiff EOS capable of supporting a maximum NS mass of \( M_{\rm TOV} \gtrsim 2.4\,M_{\odot} \).
In summary, electromagnetic transients from binary mergers serve as a great probe of the unknown EOS of NSs. 

Thus, GW170817 is just the beginning of the multimessenger era. 
As an increasing number of gravitational wave signals and electromagnetic counterparts from BNS mergers are detected in the future~\cite{national2021pathways}, our knowledge of NS and even quark star EOS models will be enriched. 
Future observations, for example, of the time delay between the merger and the collapse to black hole, whether via post-merger gravitational wave signals or via more extensive electromagnetic counterpart observations, may allow a robust determination of the TOV maximum mass and provide independent constraints on the NS radius for a given mass (complementary to NICER and eXTP), bringing us closer to a comprehensive understanding of the EOS of dense matter in NSs. 
Moreover, eXTP is well suited to play a central role in the follow-up of binary NS mergers by detecting X-ray signatures from the post-merger remnant~\cite{WP-WG4}, thanks to its excellent spectral and temporal resolution. These observations serve as an important complement to gamma-ray and optical observations, offering additional means to identify the nature of the merger outcome and to further constrain the underlying EOS of dense matter.

\subsection{Physics-driven machine learning approach}

Recent advances in machine learning have opened new avenues for tackling physics exploration, especially of the inverse problems that are inherent in decoding dense matter. By embedding physical priors into machine learning framework, researchers are now able to directly map complex experimental observables back to the underlying dense matter EOS (see~\cite{Aarts:2025gyp,Zhou:2023pti,He:2023zin,Ma:2023zfj} for recent reviews).
Such a physics-driven machine learning approach leverages deep neural networks — not just as black-box regressors but as interpretable models that honor the constraints imposed by our physics knowledge. 
Machine learning models, especially those designed with physics-informed learning strategies, have been employed to invert NSs observations to decode dense matter EOS information~\cite{Li:2025obt}. 
Specifically, the automatic differentiation method is deployed in these works to evaluate the gradients (linear response) of the calculated NS properties (such as mass-radius) with respect to the dense matter EOS (represented as a deep neural network), based upon which the optimization can be performed via gradient descent to reconstruct the appropriate dense matter EOS which matches the astrophysical observations of NSs.

Furthermore, recent studies have demonstrated that hybrid approaches — combining deep learning in the format of automatic differentiation and deep neural network with Bayesian statistical methods — can substantially speed up the reconstruction of EOSs with the observations~\cite{Li:2025obt}. 
A key advantage of this physics-driven machine learning framework is its ability to capture subtle features in the dense matter EOS, most notably, variations in the speed of sound squared, especially those related to first-order phase transition and estimate its strength.

These approaches not only quantify the uncertainties in the extracted parameters including information for first-order phase transitions, but also allow for a systematic update of the model as new data become available. 
This means that as future observatories and gravitational wave measurements further refine the precision of mass, radius, and tidal deformability NS observations, the developed machine learning framework can adapt readily and quickly, ensuring that the dense matter properties are continuously improved. 
Overall, machine learning represents a paradigm shift in the study of dense nuclear matter. 

\subsection{Interdisciplinary synthesis} 

The study of the EOS relies on a continuous and iterative interplay between experiment and theory, with reciprocal feedback proving essential for progress. To obtain precise quantitative constraints and qualitative insights, it is crucial to integrate nuclear experiments, multi-messenger astrophysics, and EOS theory.

Experimental data from the FOPI (symmetric nuclear matter) and ASY-EOS (symmetry energy) experiments at GSI have been incorporated into dense matter EOS inference alongside NS observations~\cite{Huth:2021bsp}. The inclusion of HIC data increases the pressure in the range of one to two times the nuclear saturation density by approximately, favoring stiffer EOSs. 
Astrophysical data dominate constraints at densities exceeding twice the nuclear saturation density, excluding overly stiff EOSs inconsistent with the GW170817 tidal deformability constraint. 
The ASY-EOS constraints on the symmetry energy slope $L$ at saturation agree with neutron skin thickness measurements, which helps reduce uncertainties in NS radius predictions.
Moreover, a Bayesian odds ratio of $0.4\pm0.1$ disfavors a strong first-order phase transition in NS cores under current constraints~\cite{Huth:2021bsp}. 

Nevertheless, opposing results were obtained in a recent Bayesian analysis \cite{OmanaKuttan:2022aml} that utilized experimental data on mid-rapidity proton observables — specifically, the elliptic flow ($v_2$) and the mean transverse kinetic energy ($<m_T>-m_0$) — in HICs with beam energies $\sqrt{s_{NN}}\approx 2$--$10$~GeV. 
Their sensitivity analysis for when all 15 available experimental data points are used inferred 
a broad peak structure in the squared speed of sound, indicating a rather stiff EOS up to four times the nuclear saturation density. 
However, if two key data points for $<m_T>-m_0$ at $\sqrt{s_{NN}}=3.83$ and $4.29$ GeV are excluded from the Bayesian inference — reducing the input to 13 data points — the extracted squared speed of sound drops drastically at high densities, consistent with a strong first-order phase transition. 
The phase transition parameters derived from beta-stable NS matter can be tested in nearly symmetric nuclear matter produced in HIC~\cite{2023PhRvD.107d3005L}. However, to fully understand their implications, a deeper understanding of the connection between phase transitions in symmetric and asymmetric nuclear matter is required, and further investigations are needed on the roles played by temperature and magnetic field.

Future high-statistics measurements at existing and upcoming HIC facilities, particularly over the relevant intermediate beam energy range~\cite{friese2006cbm,musulmanbekov2011nica,guo2024studies,sun2020huizhou}, will be essential for robustly constraining the high-density EOS. 
Since the probed densities can reach up to four or five times the nuclear saturation density, these experiments will also be invaluable for testing the potential relevance of hyperons or quarks in dense matter. Such findings will further refine our understanding of the dense matter EOS.

\section{Summary 
}

eXTP will play a key role in the global effort to uncover the properties of dense matter in NSs. From 2030, the eXTP mission will deliver new constraints on the properties of dense matter through precision X-ray observations.

The synergy of eXTP instrumentation — combining large collecting area, polarimetry, and timing capabilities — allows for the deployment of multiple independent techniques. These include PPM of MSPs, spectral modeling of thermonuclear bursts and their cooling tails, analysis of burst oscillations, and study of timing irregularities. In addition, eXTP will probe phenomena associated with nuclear burning on the NS surface (such as mHz QPOs) and accretion disk dynamics (such as kHz QPOs and relativistic Fe line profiles).

We anticipate that the comprehensive observational strategy of eXTP will open a new quantitative era for NS radius measurements.
These measurements offer sensitivity to the nuclear symmetry energy, its density dependence, and possible phase transitions to exotic degrees of freedom (hyperons, quark matter) in the core.
Additionally, a single SDD array in the SFA telescopes can be replaced by one pn-CCD, which offers combined spectral, timing, and imaging capabilities. The pn-CCD imaging performance (W50 =$30^{\prime\prime}$) enables tighter constraints on the phase-invariant component in energy-resolved pulse profiles through simultaneous observations of target sources. Furthermore, joint analysis of the SDD and pn-CCD data—for example in non-accreting MSPs—can enhance the precision of NS mass and radius measurements. Similar approaches have been employed in the analysis of NICER and XMM data (though not using simultaneous observations) using a joint log-likelihood function in PPM. 

Simultaneously, eXTP will build an extensive sample of high-precision data for exploring the thermal and rotational evolution of NSs. 
Evolution studies of NSs observed by eXTP provide complementary probes of superfluidity, viscosity, crustal elasticity and neutrino emission mechanisms from the core. 

Together with gravitational wave constraints and laboratory experiments, eXTP will play a critical role in establishing a unified and consistent description of dense matter from the crust to the core of NSs, advancing our understanding of the EOS and the possible presence of exotic matter in the Universe.


\emph{Acknowledgements.}  
This work is supported by China's Space Origins Exploration Program. 
AL is supported by the National Natural Science Foundation of China (grant No. 12273028).
ALW, BD and TS acknowledge support from ERC Consolidator grant No.~865768 AEONS. ALW also acknowledges support from NWO grant ENW-XL OCENW.XL21.XL21.038.
SG acknowledges the support of the CNES.
S-NZ is supported by the National Natural Science Foundation of China (No. 12333007), the International Partnership Program of Chinese Academy of Sciences (No.113111KYSB20190020) and the Strategic Priority Research Program of the Chinese Academy of Sciences (No. XDA15020100).
ZM is supported by the China national postdoctoral program for innovation talents (No.~BX20240223) and the China Postdoctoral Science Foundation funded project (No.~2024M761948).
YC acknowledges support from a Ramon y Cajal fellowship (RYC2021-032718-I) ﬁnanced by MCIN/AEI/10.13039/501100011033 and the European Union NextGenerationEU/PRTR.
AP acknowledges support from grant  PID2021-124581OB-I0, PID2024-155316NB-I00 and 2021SGR00426. 
XZ is supported by the Natural Science Foundation of Xinjiang Uygur Autonomous Region (No. 2023D01E20)
National SKA Program of China (Grant No. 2020SKA0120300).
XPZ and XZL are supported by the National Natural Science Foundation of China (grant No.12033001,12473039) and the 
WHW is supported by Zhejiang Provincial Natural Science Foundation of China under grant No. LQ24A030002.
QC is supported by the National Natural Science Foundation of China (Grant No. 12003009).
WWZ is supported by the National SKA Program of China No. 2020SKA0120200, the National Nature Science Foundation (grant No. 12041303).
ZL is supported by the National Natural Science Foundation of China (grant No. 1227303).
LS was supported by the National SKA Program
of China (2020SKA0120300), the Beijing Natural Science Foundation (1242018), and the Max Planck Partner Group
Program funded by the Max Planck Society.
AD is supported by JSPS KAKENHI (the Japan Society for the Promotion of Science, Grants-in-Aid for Scientific Research) under grant No. 23K19056 and No. 25K17403.
HS is supported by the National Natural Science Foundation of China (grant No. 12175109).
CP acknowledges support from FCT (Fundação para a Ciência e a Tecnologia, I.P, Portugal) under projects 
UIDB/04564/2020 and 2022.06460.PTDC.
LT is supported from the program Unidad de Excelencia María de Maeztu CEX2020-001058-M, from the project PID2022-139427NB-I00 financed by the Spanish MCIN/AEI/10.13039/501100011033/FEDER, UE (FSE+), and by the CRC-TR 211 'Strong-interaction matter under extreme conditions'- project Nr. 315477589 - TRR 211.
LL is supported by the National Natural Science Foundation of China (grant No. 12122513).
The work of KH, MM, AS and IS was supported in part by the European Research Council (ERC) under the European Union’s Horizon 2020 research and innovation programme (Grant Agreement No.~101020842) and by the Deutsche Forschungsgemeinschaft (DFG, German Research Foundation) -- Project-ID 279384907 -- SFB 1245.
ML acknowledges support from the ERC Consolidator grant No. 101002352 (LOVE-NEST).
NR is supported by the European Research Council (ERC) via the
Consolidator Grant “MAGNESIA” (No. 817661) and the Proof of Concept
``DeepSpacePulse" (No. 101189496), by the Catalan grant SGR2021-01269, the Spanish grant ID2023-153099NA-I00, and by the program Unidad de Excelencia Maria de Maeztu
CEX2020-001058-M.
AV is supported by the Research Council of Finland under grant No.~354533.


\emph{Conflict of interest.} 
The authors declare that they have no conflict of interest.



\bibliographystyle{apsrev4-1}  
\bibliography{refs}

\end{multicols}
\end{document}